\documentstyle[aps,twocolumn,psfig]{revtex}
\newcommand{\th}{\vartheta}
\newcommand{\al}{\alpha}
\newcommand{\De}{\Delta}
\newcommand{\ga}{\gamma}
\newcommand{\la}{\lambda}

\newcommand{\f}{\varphi}
\newcommand{\eps}{\varepsilon}
\newcommand{\om}{\omega}

\newcommand{\qp}{quasi particle\ }
\newcommand{\qps}{quasi particles\ }
\newcommand{\qpr}{quasi proton\ }
\newcommand{\qprs}{quasi protons\ }
\newcommand{\qn}{quasi neutron\ }
\newcommand{\qns}{quasi neutrons\ }
\newcommand{\conf}{configuration\ }
\newcommand{\confs}{configurations\ }
\newcommand{\barr}{\begin{array}}
\newcommand{\bea}{\begin{eqnarray}}
\newcommand{\beq}{\begin{equation}}
\newcommand{\ear}{\end{array}}
\newcommand{\eea}{\end{eqnarray}}
\newcommand{\eeq}{\end{equation}}
\newcommand{\am}{angular momentum\ }
\newcommand{\bef}{\begin{figure}}
\newcommand{\enf}{\end{figure}}

\begin{document}

\title{Description of Multi Quasi Particle Bands by the Tilted Axis Cranking
Model }
\author{S.Frauendorf}

\address { Department of Physics, University of Notre Dame, IN  , USA\\ and\\
 Institute for Nuclear and Hadronic Physics,
Research Center Rossendorf, PB 51 01 19,  01314 Dresden, Germany }

\maketitle
\begin{abstract}
The selfconsistent cranking approach is extended to the case of rotation about
an axis which is tilted with respect to the principal axes of 
the deformed potential (Tilted Axis Cranking). 
Expressions for the energies and the intra
 bands electromagnetic transition probabilities are given.
  The mean field solutions  are interpreted in terms of
quantal rotational states. The 
construction of the 
\qp configurations and the elimination of spurious states is discussed.
The application of the theory to high spin data
is demonstrated by analyzing the
multi \qp  bands in the nuclide-s with $N=102,103$ and $Z=71,72,73$.  

\end{abstract}


\section{Introduction}

Since its introduction in ref. \cite{tac}, the Tilted Axis Cranking (TAC)
approach has turned out to be quite successful in describing $\De I=1$ rotational
bands \cite{er163,yb168,re181,tm164,ba128,onishi1,os182,w178,hybrid}. 
In particular it has led to
the understanding of     
the appearance of regular magnetic dipole bands in nearly
spherical nuclei 
\cite{mr1,mr2,mr3,pblt,sn105,sn106,sb108,cd110,sn106l,cd109,cd108l,sm139,rn205,rb74}.
 The physical aspects of these investigations
have been reviewed in \cite{mr,rmp}.
 Though different aspects of the actual calculations
 were touched in these papers, a comprehensive presentation of the theory,
the calculational methods, and the practical application of the 
TAC approach is still missing.
In the  present paper we  provide it by using the
rotational bands in the nuclides with $N=102,103$ and $Z=71,72,73$
as examples.
The TAC approach   has been applied so far only for potentials of
the Nilsson type, which are
 combined with  a pairing plus quadrupole model interaction
or the shell correction method for finding the deformation.
A systematic exposure of the applied techniques, the experiences gathered as 
well as the successes and  limitations of the TAC approach as it stands seems
to be timely for two reasons. On the one hand hand it is meant as guideline
for application of the existing program system, which has turned out 
quite useful in the data analysis. One the other hand it may serve as a 
starting point for more sophisticated versions of the TAC
mean field  approximation, as the up to date  versions of the Hartree-Fock
approximation or the Relativistic Mean Field approach.

The earliest invocations of cranking about a non-principal
axis were in the context of wobbling motion
\cite{thouless,wob,mar79,kerman}. Kerman and Onishi \cite{kerman} pointed out 
the possibility of uniform rotation about   a non-principal
axis. Frisk and  Bengtsson \cite{frisk} demonstrated the existence of such
solutions for realistic nuclei and discussed
conditions where to expect them \cite{friskpol,bengtsson}. 
Goodman \cite{goodmomi} demonstrated that the  moments of inertia may
strongly depend on the orientation of the rotational axis, which implies the
possibility of uniform rotation about a tilted axis.
However, these studies did not give the physical
interpretation of the TAC solutions and
left open the question
if taking into account the self-consistency with respect to the shape
degrees of freedom would not result in rotation about a principal axis.
In fact, the investigation of the rotating harmonic oscillator by
Cuypers \cite{cyupers} and a few level model by Nazarewicz and Szymanski 
\cite{szym}
seemed to support the latter possibility. Frauendorf \cite{tac}
found the first fully self-consistent TAC
solutions and gave their interpretation in terms of $\Delta I=1$ rotational
bands. This marked the origin of the fully fledged Tilted Axis Cranking
(TAC) approach.

Marshalek \cite{mar87,mar91} studied the
possibility of tilted rotation generated by superpositions of collective
vibrations, while Alhassid and Bush \cite{alhassid}, Goodman \cite{good92}, 
and Dodaro and
Goodman \cite{good94,good96} included the tilt of the rotational axis in 
their analysis of nuclei at nonzero
temperature. A recent reinvestigation of the 
rotating harmonic oscillator by Heiss and Nazmitdinov
\cite{nasmi} claims the existence
of TAC solutions within this model, in contrast to \cite{cyupers}.
Horibata and Onishi \cite{onishi1}, Horibata et al. \cite{onishi2} and 
D\"onau et al. \cite{doenau}
 have started to investigate the
dynamics of the orientation angles in the frame of the Generator Coordinate 
Method.

 Section \ref{s:tac} presents the relevant expressions
for the energies and electro-magnetic transition matrix elements. It
discusses the  interpretation of the cranking solutions,  important technical
aspects and approximations that help to find  the solution of the TAC
equations in an efficient way. It investigates the relation of TAC 
to the treatment
of $\De I =1 $ bands in the framework of the standard Principal
Axis Cranking (PAC) approach.  It explains how to read the \qp diagrams.
Section \ref{s:mqpc} analyses the   rotational bands in the yrast region of
the nuclides with $N=102,103$ and $Z=71,72,73$. The main purpose is 
to illustrate how to construct the multi \qp configurations  and how 
to relate them to the experimental  rotational
bands. Merits and limitations of the method will be exposed and
compared with the 
standard Cranking approach. 
We are not going to optimize all parameters of
the mean field for each configuration. In the spirit of the Cranked Shell
Model approach \cite{bf} only semi quantitative agreement with the data
is sought, the focus being the qualitative structure of the band spectrum. 
Well deformed nuclei are taken as examples because the assumption of
one and the same deformation for the various \qp configurations is
realistic. The specific nucleon numbers are chosen because 
 a large number of high $K$ bands and low $K$ 
bands have been found in these nuclides. This makes  them an appropriate
test ground for the TAC model. 
This paper is restricted to the HFB approximation for pairing. 
A more sophisticated
version of TAC based on particle number projection will be presented 
separately \cite{almehet}. 
Since the change of the pair field is not in the concern  of this paper but
rather an unwanted complication, 
the self-consistency  of this degree of freedom is treated in a schematic way.
 
Section \ref{s:rules} provides a set of rules for the analysis
 of rotational bands in terms TAC, which is meant as a reference for
potential users of the method. The conclusion are given in
section \ref{s:conc}.

                     \section{ Tilted axis cranking} \label{s:tac}

                       \subsection{ General layout}\label{s:tacg}

     Two versions of the TAC have been developed and applied 

 i) The pairing plus quadrupole model (PQTAC)\cite{tac}

ii) Shell correction method (SCTAC)

In the subsections \ref{s:PQQ} - \ref{s:kappa}
 we present the PQTAC  in detail. 
Section \ref{s:sc} describes      the differences between SCTAC and PQTAC.
The PQTAC it more appropriate for small deformations, whereas SCTAC is better
suited for large deformation. Subsection \ref{s:dela} discusses 
the schematic  treatment of pairing and subsection \ref{s:read} explains the
how to read the \qp diagrams. 
\subsection{      The pairing plus quadrupole model (PQTAC)}
\label{s:PQQ}
We assume that the rotational axis is the z-axis and start 
with the two-body Routhian
\beq \label{H'}
H'=H-\omega \hat J_z. 
\eeq
It consists of the rotationally invariant two-body Hamiltonian $H$ 
and the constraint $\omega \hat J_z$ which
ensures that the low-lying states have a 
finite angular momentum projection $J=<\hat J_z>$. As a two
body Hamiltonian, the pairing plus quadrupole interaction is used, 
\beq\label{HPQQ} 
H=H_{sph}-{\chi\over2}\sum\limits^2_{\mu=-2}Q^+_{\mu}Q_{\mu}-GP^+P-\lambda N.
\eeq 
The model and its justification are described in the textbooks (see, for
example,  Ring 
and Schuck \cite{ring}). We use
a slightly modified version, which is constructed such 
that the derived mean field Hamiltonian
coincides with the popular Nilsson Hamiltonian (see, for example,
 Ring and Schuck \cite{ring} and 
Nilsson and
Ragnarsson \cite{ragnarsson}). 
The motivation is that the parameters of the Nilsson
Hamiltonian are  carefully adjusted and that it is useful 
to have the same standard mean field for nuclei with
large deformation, where the shell correction method \cite{strutinsky} is more 
appropriate. Thus, the spherical part 
\beq 
H_{sph}=\sum\limits_k \eps_k c_k^+c_k
\eeq
is parameterized in the same way as the Nilsson Hamiltonian. For the
calculations in this paper we use the set of parameters given 
 in \cite{oldparameter}. 

The pairing interaction is defined by the   
monopole  pair field 
\beq \label{P}
P^+=\sum\limits_{k>0}c^+_k c_{\bar k}^+.
\eeq
Here $\bar k$ is the time reversed state of $k$.  The quadrupole
 interaction is defined by the operators
\footnote{This definition of the quadrupole operators corresponds to
$Q_0=r^2P_2(\cos\th)$, with $P_2$ being the Legendre polynomial.}
\beq \label{Q}
Q_{\mu}=\sum\limits_{kk'}\sqrt{\frac{4\pi}{5}}<k|r^2Y_{2\mu}|k'>c_k^+c_{k'}.
\eeq
In order to simplify the notation  all expressions are written  
 only for one kind of particles. They are  understood as sums of a proton
and a neutron part. 

     The wave function is approximated by the Hartree - Fock - Bogoljubov (HFB) mean
field expression $|>$. Neglecting exchange terms, the HFB -
Routhian becomes
\beq \label{h'}
h'=h_{sph}-\sum\limits^2_{\mu=-2}q_{\mu}(Q^+_{\mu}+Q_{\mu})-\Delta(P^++P)-
\lambda N- \omega \hat J_z.
\eeq
The self-consistency equations determine the deformed part of the potential
\beq \label{scQ}
q_{\mu}=\chi<Q_{\mu}>
\eeq
and the pair potential
\beq \label{scP}
\Delta=G<P>.
\eeq
 The chemical potential $\la$ is fixed by the standard condition
\beq \label{scN}
N=<\hat N>.
\eeq
The quasi particle operators
\beq \label{alpha}
\alpha_i^+=\sum\limits_k (U_{ki}c_k^++V_{ki}c_k)
\eeq
obey the equations of motion
\beq \label{eom}
[h',\,\alpha_i^+]=e_i'\alpha_i^+,
\eeq
which define the well known HFB eigenvalue equations for the quasi 
particle amplitudes $U_{ki}$
and $V_{ki}$. The explicit form of these equations can be found 
in \cite{ring}. They have the familiar symmetry under particle hole
conjugation, which has the consequence that for each \qp solution $i$
there is a conjugate $ i^+$ with 
\beq\label{phconj}
e'_{i^+}=-e'_i ,~~U_{ k i^+}=V_{ki},~~V_{k i^+}=U_{ki}. 
\eeq 
The \qps have good parity, but in general no good signature.
 The consequences of
 good or broken signature will be discussed in subsection \ref{s:ts}.

The quasi particle operators
refer to the vacuum state $|0>$,
which is defined by the condition
\beq\label{vac}
\alpha_i\,|\,0>=0\;\forall\;i.
\eeq
They define the excited quasi particle configurations
\beq\label{conf}
|i_1,\,i_2,\,\cdots>=\alpha_{i_1}^+\alpha_{i_2}^+\,\cdots|0>.
\eeq
The rules and strategies for constructing 
\qp \confs from them  will be be discussed below by means of concrete
examples. 

The set of HFB eq. (\ref{h'}) - (\ref{vac}) can be solved for 
any configuration $|i_1,\,i_2\,\cdots>$. For the self-consistent 
solution, the total Routhian
\beq\label{E'}
E'=<H'>
\eeq
has an extremum
\beq\label{extrE'}
{\partial E'\over\partial_{q_{\mu}}}|_{\omega}=0,\,{\partial 
E'\over\partial\Delta}|_{\omega}=0.
\eeq
The total energy as function of the angular momentum 
\beq\label{J}
J=<\hat J_z>
\eeq
is given by
\beq\label{E}
E(J)=E'(\omega)+\omega J(\omega)
\eeq
where the eq. (\ref{J}) implicitly fixes $\omega(J)$. 
The total energy is also extremal for the self-consistent solution
\beq\label{extrE}
{\partial E\over\partial_{q_{\mu}}}|_J=0,\,{\partial
E\over\partial\Delta}|_J=0,
\eeq
where the derivatives have to be taken at a fixed value of $J$.
For a family of self-consistent solutions $|\omega>$, found for different values of $\omega$, there hold the
canonical relations
\beq\label{can}
{dE'\over d\omega}=-J,\,{dE\over dJ}=\omega.
\eeq

In ref. \cite{kerman} it was  shown that for a self-consistent solution the vector
of the angular velocity
\beq
\vec \om =(\om_x,\om_y,\om_z)=(0,0,\om)
\eeq
and the vector of the expectation values of the \am components
\beq
\vec J=(\langle \hat J_x\rangle,\langle \hat J_y\rangle,\langle \hat
J_z\rangle)
\eeq
must be parallel. The argument is as follows.
 Since the interaction is rotational invariant, one has
\beq
\langle [H',\hat J_x]\rangle =i\om \langle \hat J_y\rangle,
\eeq
\beq
\langle [H',\hat J_y]\rangle =-i\om \langle \hat J_x\rangle.
\eeq
Since the left-hand sides are small variations of $E'$, the
stationarity of $E'$ implies
\beq
<\hat J_x>=<\hat J_y>=0,
\eeq
i. e.
\beq\label{scJ}
\vec \om || \vec J.
\eeq
This holds also in the intrinsic frame of reference, which will be 
discussed in sect. \ref{s:intr}.

\subsection{Tilted Solutions}\label{s:ts}

The formalism   presented above is the well 
known "Cranking Model" as laid out in the
textbooks \cite{ring,ragnarsson}. 
The "Tilted-Axis Cranking" (TAC) version \cite{tac} accounts 
for the possibility
that  the principal axes  (PA) of the quadrupole tensor $q_{\mu}$ 
need not to coincide with the rotational axis (z). Hence, one has to 
distinguish between two possibilities:
\begin{itemize}
\item Principal Axis Cranking (PAC).   The z - axis
(rotational or cranking axis)  coincides with one of the PA.
Then, the signature $r$ is a good quantum number, i. e. 
\beq \label{sig}
 e^{-i\pi \hat J_z}|\pi, \alpha, \om>=r|\pi, \alpha, \om>.
\eeq
Following \cite{bf} we indicate the signature $r=e^{-i\pi \alpha}$ by
the signature exponent $\alpha$.
The \qp configuration $|\pi, \alpha, \om>$  describes a $\De I=2$
 rotational band, the spins of which take the values \cite{bf} 
\beq
I=\alpha + even~ number. 
\eeq
\item Tilted Axis Cranking (TAC). 
The z - axis does not
 coincides with one of the PA, i. e. it is {\em tilted} away from the PA.
Then, 
\beq 
 e^{-i\pi \hat J_z}|\pi, \om>\not=e^{-i\pi \alpha}|\pi \om>.
\eeq
The signature is no longer a good quantum number. The \qp configuration
$|\pi,  \om>$  describes a $\De I=1$
 rotational band of given parity. 

\end{itemize}
 The different interpretation~  of solutions with different
symmetry is characteristic for spontaneous symmetry breaking
in the   mean field approximation. It makes it necessary to eliminate
spurious configurations and   will lead to discontinuities when 
the symmetry changes as a function of the frequency $\om$.
 These problems which will be
discussed in the subsections \ref{s:spur} by means of  concrete examples.

\subsection{                    Intrinsic coordinates}
\label{s:intr}
     It is useful to reformulate the TAC approach in the frame 
of the PA of the quadrupole tensor. This "intrinsic" coordinate system is 
defined by demanding that 
the components of the quadrupole tensor satisfy the conditions 
\beq \label{q'}
q'_{-1}=q'_1=0,\;q'_{-2}=q'_2.
\eeq
The orientation of the PA, which are denoted by 1, 2, and 3,
 with respect to the lab frame is fixed by the
three Euler angles $\psi$, $\th$ and $\f$, the meaning of which is
illustrated in fig. \ref{f:angles}. 
The two intrinsic quadrupole moments $q'_0$ and $q'_2$
specify the deformation of the potential.
The quadrupole moments in the lab
frame are related to them   by
\beq\label{eq:intlab}
q_{\mu}=D^{2}_{\mu 0}(\psi,\th,\f)q'_{0}+(D^{2}_{\mu 2}(\psi,\th,\f)+
D^{2}_{\mu -2}(\psi,\th,\f))q'_{2},
\eeq
where $D^{2}_{\nu \mu}(\psi,\th,\f)$ are the Wigner $D$-functions
\footnote{The convention of \cite{bm2} is used.}.

The different angles $\psi$ correspond to one and the same
intrinsic state. They are degenerate. We choose the one with $\psi=0$.
We restrict the consideration to $planar$ TAC solutions, 
which is the 
case when the z-axis lies in one of the three principal planes 
defined by the PA. We
assume that it lies in the 1 - 3 plane, i. e.   choose $\f=0$. 
In the case of axial shapes,   
this is one choice from the equivalent
solutions differing by the angle $\f$. 
For triaxial shapes one may always relabel the PA 
by means of the  shape
parameterization, letting the triaxiality parameter $\ga$
vary within an interval of 180$^o$ \cite{ragnarsson}. 
Which axes of the triaxial potential lie in the 1-3 plane can be found in  
table \ref{table}. It is seen that
all three possibilities appear in the half-plane $-60^o\le\ga\le 120^o$. 
The other half-plane is a repetition with the axes 1 and 3 exchanged.
 
With the above mentioned restrictions and conventions 
the deformed potential is fixed by the 
two intrinsic quadrupole
moments $q'_0$ and $q'_2$ and the
orientation ("tilt") angle
 $\vartheta$ between the 3 - and the z - axis, which is the direction of
the rotational axis. In the intrinsic frame the HFB Routhian reads
\bea \label{h'intr}
h'=h_{sph}-q'_0Q'_0-q'_2(Q'_2+Q'_{-2})-\Delta(P^++P)\nonumber \\
-\lambda N-\omega(\sin\,\vartheta\,J_1+\cos\,\vartheta\,J_3).
\eea
Figs. \ref{f:qp} and \ref{f:qn} show examples of the the \qp levels
$e'_i$ as functions of the rotational frequency $\om$ and the orientation
angle $\th$. 

The shape is fixed by the two equations
\beq \label{scQintr}
q'_0=\kappa<Q'_0>,\,q'_2=\kappa<Q'_2>
\eeq
and the orientation angle $\vartheta$ by the condition that the expectation 
value of  the angular momentum  and the
angular velocity must have the same direction, i. e. 
\beq \label{scJintr}
\vec J=(<\hat J_1>,\, 0,\, <\hat J_3>)~~||~~\vec \om=
(\omega\,\sin\,\vartheta,\,0,\,\omega\,\cos\,\vartheta),
\eeq
respectively. These parameters correspond to extrema of total Routhian,
that is
\beq
{\partial E'\over\partial_{q'_{0}}}|_{\omega}=0,\,{\partial
E'\over\partial_{q'_{2}}}|_{\omega}=0,\,{\partial 
E'\over\partial\vartheta}|_{\omega}=0.
\eeq
Of course only the minima  are interpreted as bands. 

In praxis it is convenient to solve 
the equation (\ref{scJintr}) for each combination of $q'_0$ and
$q'_2$ which is needed to obtain the shape from the 
equations (\ref{scQintr})
 with the desired accuracy. Very often
it is enough to determine the shape for one value of 
$\omega$ and then keep it fixed for other values, only
calculating the 
orientation  angle $\vartheta$ by means of the condition (\ref{scJintr}).

Using the Cartesian representation of the quadrupole moments, the HFB Routhian (21)
becomes the modified oscillator potential \cite{ring,ragnarsson}
\bea\label{h'nilsd}
h'=\sum\limits^3_{\nu=1}\frac{(p^2_{\nu}+\omega^2_{\nu}x^2_{\nu})}{2M}+
\kappa l_{\nu}s_{\nu}+\mu (l^2_{\nu}-<l^2_{\nu}>) \nonumber \\-
\Delta(P^++P)-\lambda N-\omega(sin\,\vartheta\,J_1+\cos\,\vartheta\,J_3),
\eea
where the oscillator frequencies are parameterized by means of Nilsson's 
 deformation parameters $\delta$  and $\gamma$ (cf. e. g. \cite{ragnarsson}), 
\beq\label{omdel}
\omega^2_{\nu}=\omega^2_{00}\,\left[1-{2\over3}\delta\,
\cos(\gamma-{2\pi\nu\over3})\right].
\eeq
The only difference to the standard modified oscillator model 
is that there is no volume
conservation in the pairing plus quadrupole model. Since we are only interested  in small 
deformation the coupling
between the oscillator shells is not taken into account when diagonalizing the HFB Routhian
(\ref{h'nilsd}). Solving the self-consistency equation (\ref{scQintr}) 
and calculating the total energy, the coupling
between the oscillator shells is also neglected. 

 \subsection{              Strutinsky Renormalization (SCTAC)}
     \label{s:sc}                           
     An alternative version of TAC starts with the modified oscillator 
Routhian (\ref{h'nilsd}). As e. g. 
described in \cite{ragnarsson}, stretched coordinates are introduced
and  the matrix elements $<N|J_{\mu}|N\pm2>$ are neglected in the 
stretched basis. This is a standard procedure which takes into account  
most of the couplings between the oscillator shells.
The oscillator frequencies are parameterized by means of Nilsson's
alternative  set of  deformation parameters $\eps$  and $\gamma$  , 
\beq\label{omeps}
\omega_{\nu}=\omega_0\,\left[1-{2\over3}\varepsilon\,
\cos(\gamma-{2\pi\nu\over3})\right],
\eeq
where the condition of volume conservation $\omega^3_0=\omega_1\omega_2\omega_3$
 fixes $\omega_0$.  The total Routhian is obtained by applying the 
Strutinsky renormalization to the energy
of the non-rotating system $E_0$.
This kind of approach has turned out to be a quite reliable calculation
scheme in the case of standard PAC \cite{tbengtsson}.
One minimizes the total Routhian\footnote{For the treatment of the term
$\la <\hat N>$ see sect. \ref{s:dela}.} 
\bea\label{E'str}
E'(\omega,\th,\eps,\eps_4,\ga,\De,\la)=E_{LD}(\eps,\eps_4,\ga)-\tilde
E(\eps,\eps_4,\ga)\nonumber \\
+<h'> +(2\De-G<P>)<P>,
\eea
where $|>=|\omega,\th,\eps,\eps_4,\ga,\De,\la>$ is a \qp \conf belonging to 
the mean field Routhian $h'(\omega,\th,\eps,\eps_4,\ga,\De,\la)$ as defined
above.  The smooth energy 
$\tilde E$ is calculated  from the single-particle energies,
 which are the
eigenvalues of $h'(\omega=0,\th=0,\eps,\eps_4,\ga,\De=0)$ by means of Strutinsky 
averaging \cite{strutinsky}. The expressions for the liquid drop energy
$E_{LD}(\eps,\eps_4,\ga)$ are given, for example, in \cite{ragnarsson},
where also the averaging procedure is described.     
For given $\eps, ~\eps_4$  and $\ga$,
the tilt angle is determined by means of  the condition (\ref{scJintr}). 
Then, the minimum of $E'(\omega,\eps, \eps_4, \gamma)$ with respect
to the deformation parameters is found.
Since $|\omega>$ is an eigenfunction of $h'(\omega, \th)$ the Routhian 
$<\omega,\th |h'|\omega,\th >$ is stationary at the angle where  the
condition (\ref{scJintr}) is fulfilled and so is  
$E'$ because the other terms do not depend on $\th$. Hence, the procedure 
determines a stationary point with respect to the mean field parameters
and the canonical relations (\ref{can}) are satisfied. 

The SCTAC approach is preferred to the PQTAC version  for well 
deformed nuclei, because it is a reliable standard method for determining
large deformations. In the calculations of well deformed nuclei
it is usually a good approximation to keep the deformations fixed within
a rotational band.  However this is a matter of the needed accuracy and 
of how much effort one is willing to invest. 

\subsection{               Electro - magnetic matrix elements}
\label{s:em}
     The intra band M1 - transition matrix element is calculated by 
means of the semiclassical expression 
\bea
<I-1I-1|{\cal M}_{-1}(M1)|II>&=
<{\cal M}_{-1}(M1)>=&\nonumber\\
=\sqrt{\frac{3}{8\pi }}
\left[\mu_3\sin\,\vartheta -\mu_1\cos\,\vartheta \right].&&
\eea
The components of the transition operator ${\cal M}_{\nu}$
refer the the lab system. The expectation value is taken with the 
TAC \conf $|>$. In the second line  ${\cal M}_{\nu}$ is expressed by
the components of the magnetic moment in the intrinsic frame.
The reduced M1-transition probability becomes  
\beq\label{bm1}
B(M1,\,I\rightarrow I-1)=<{\cal M}_{-1}(M1)>^2.
\eeq
The spectroscopic magnetic moment is given by
\bea\label{mu}
\mu=<II|\mu_z|II>=\frac{I}{I+1/2}<\mu_z>\nonumber \\
=\frac{I}{I+1/2}[\mu_1\sin\,\vartheta+\mu_3\cos\,\vartheta].
\eea
The factor $\frac{I}{I+1/2}$ is a quantal correction which is close to one
for high spin. 
The components of the magnetic moment with respect to the PA are calculated by
means of
\bea
\mu_1=\mu_N(J_{1,p}+(\eta 5.58-1)\,S_{1,p}-\eta 3.82\,S_{1,n}),\nonumber\\
\mu_3=\mu_N(J_{3,p}+(\eta 5.58-1)\,S_{3,p}-
\eta 3.82 \,S_{3,n}),
\eea     
where the components 
of the vectors of angular momentum $\vec J$ and of the spin 
$\vec S=<\vec s>$ are the expectation values  with the TAC
\conf $|>$. The free spin magnetic moments are attenuated by a factor
$\eta = 0.7$. For mass  other mass regions a somewhat different 
attenuation may be
taken, which needs not be the same for protons and neutrons.

     The intra band E2 - transition matrix elements are calculated by 
means of the semiclassical expressions\footnote{Refs.
\cite{berk94,berk95,crete96} contain some unfortunate 
inconsistency  between   the quadrupole moments of the
quadrupole interaction and the electric transition matrix elements. 
These concern only the written formulae, the results of the calculations quoted 
are correct and consistent with the ones given here.}  
 \bea
&<I-2I-2|{\cal M}_{-2}(E2)|II>=
<{\cal M}_{-2}(E2)>=&\nonumber\\
&=\sqrt{{5\over4\pi}}\left(\frac{eZ}{A}\right)
\Bigl[\sqrt{3\over8}<Q'_0>(\sin\,\vartheta)^2 &\nonumber \\ 
&+{1\over4}<Q'_2+Q'_{-2}>\left( 1+(\cos\th)^2\right) \Bigr],&\\
&<I-1I-1|{\cal M}_{-1}(E2)|II>=
<{\cal M}_{-1}(E2)>=&\nonumber\\
&=\sqrt{{5\over4\pi}}\left(\frac{eZ}{A}\right)
\Bigl[\sin\,\vartheta \cos\th( \sqrt{3\over2}<Q'_0>&\nonumber \\
&-{1\over2}<Q'_2+Q'_{-2}>)\Bigr],&
\eea
and the spectroscopic quadrupole moment by
\bea\label{Qstat}
&Q=<II|Q_0^{(BM)}|II>=\frac{I}{I+2/3}<Q_0^{(BM)}>=&
\nonumber\\
&\frac{I}{I+2/3} \frac{2eZ}{A}\Bigl[<Q'_0>\left((\cos\th)^2
-{1\over2}(\sin\,\vartheta)^2\right)&\nonumber\\
&+\sqrt{3\over8}<Q'_2+Q'_{-2}>(\sin\,\vartheta)^2\Bigr].&
\eea
We use the conventional definition of the static quadrupole 
moment as  given in ref.  \cite{bm2}, which differs by a
factor of 2 from our quadrupole moments in the lab frame. There is a similar
quantal correction factor as for the magnetic moment.
 
The reduced E2-transition probabilities are
\beq\label{be22}
B(E2,\,I\rightarrow I-2)=<{\cal M}_{-2}(E2)>^2
\eeq
and
\beq\label{be21}
B(E2,\,I\rightarrow I-1)=<{\cal M}_{-1}(E2)>^2.
\eeq
The mixing ratio is
\beq
\delta=\frac{<{\cal M}_{-1}(E2)>}{<{\cal M}_{-1}(M1)>}.
\eeq
     The mass quadrupole moments consist two terms. The first one 
contains the microscopic expectation 
values \mbox{$<Q'_0>_N$} and \mbox{$<Q'_{\pm 2}>_N$}, where the
subscript $N$ indicates that only the $\Delta N=0$ matrix elements of the 
quadrupole operator are taken. The second term takes care of the  
 coupling between the oscillator shells. 

In the case of SCTAC the 
 stretched coordinates are introduced to approximately 
take the coupling between the oscillator shells into account.
The expectation values needed in eqs. (\ref{be22}, \ref{be21},
\ref{Qstat}), are the quadrupole moments
in unstretched coordinates, which are given by 
\bea
&<Q'_0>=&\nonumber\\ 
&\frac{1}{6}(\frac{2\om_o}{\om_3}-\frac{\om_o}{\om_1}-
\frac{\om_o}{\om_2})<r^2>_N&\nonumber\\ 
&+\frac{1}{6}(\frac{4\om_o}{\om_3}+\frac{\om_o}{\om_1}+\frac{\om_o}{\om_2})
<Q'_0>_N&\nonumber\\ 
&-\frac{1}{2\sqrt{6}}(\frac{\om_o}{\om_1}-\frac{\om_o}{\om_2})
<Q'_2+Q'_{-2}>_N&\label{Q0uns}
\eea
\bea\label{Q2uns}
&<Q'_2+Q'_{-2}>=&\nonumber\\ 
&\frac{1}{\sqrt{6}}(\frac{\om_o}{\om_1}-\frac{\om_o}{\om_2})
<r^2>_N&\nonumber\\ 
&-\frac{1}{\sqrt{6}}(\frac{\om_o}{\om_1}
-\frac{\om_o}{\om_2})<Q'_0>_N&\nonumber\\ 
&+\frac{1}{2}(\frac{\om_o}{\om_1}+\frac{\om_o}{\om_2})
<Q'_2+Q'_{-2}>_N&
\eea
where the semiclassical value $<r^2>_N\approx1.2\,ZA^{-2/3}fm$ is used.

     In the case of the PQTAC the coupling between the oscillator 
shells is neglected.  This is a reasonable approximation for 
the rotational response of the valence particles. However,
when calculating electric quadrupole moments it cannot not be neglected,
because it accounts for the polarization of the core 
by the valence nucleons. We describe the polarization by means of 
eqs (\ref{Q0uns},\ref{Q2uns}),
 setting $\varepsilon=\delta$.  This prescription satisfies 
 the consistency condition that the deformations of 
the potential and the
density should be the same \cite{bm2}. It corresponds to a 
 polarization charge  
close to 1, as estimated for the
isoscalar quadrupole mode \cite{bm2}. This choice of the polarization 
charge makes to PQTAC and
the SCTAC as similar as possible. 

In the above described methods  one could also use  
the proton part of the quadrupole moments 
instead of $Z/A$ times the mass quadrupole moments.

   \subsection{                    Quantization}
\label{s:quant}
     Due to leading quantal correction (cf. e. g. \cite{bf,fm}) 
one must associate 
the total angular
momentum $J$ calculated in TAC with $I+1/2$, where $I$ is the quantum number of the angular
momentum. This prescription permits us to compare the TAC calculations with the experimental
energies and the static moments. 
     Genuine TAC solutions $(\vartheta\neq0^o,\,90^o)$ represent $\Delta I=1$ bands. In this case, the experimental
rotational frequency $\omega$ is introduced by 
\beq\label{omexp1}
J=I,\,\omega(I\rightarrow I-1)=E(I)-E(I-1),
\eeq
and the experimental Routhian by 
\beq\label{E'exp1}
E'(I\rightarrow I-1)={1\over2}[E(I)+E(I-1)]-\omega(I\rightarrow I-1) J.
\eeq
Here, the canonical relations (\ref{can}) are approximated by quotients 
of finite differences. The data  define a discrete sets of points $J(\om)$
and $E'(\om)$,
which are connected by interpolation.  
     If the axis of rotation coincides with one of the principal
 axes $(\vartheta = 0^o,\,90^o)$, states
differing by two units of angular momentum arrange into a $\Delta I=2$ band of 
given signature $\alpha$. 
In this case the frequency is calculated by 
\beq\label{omexp2}
J=I-1/2,\,\omega(I\rightarrow I-2)={1\over2}[E(I)-E(I-2)],
\eeq
and the experimental Routhian by
\beq\label{E'exp2}
E'(I\rightarrow I-2)={1\over2}[E(I)+E(I-2)]-\omega(I\rightarrow I-2) J.
\eeq

For the transition probabilities, $J$ is associated with the mean value of 
$I+1/2$
of the transition, i. e.
\bea\label{bexp}
B(M1,I\rightarrow I-1)=B(M1,J=I)\\
B(E2,I\rightarrow I-1)=B(E2,J=I)\\
B(E2,I\rightarrow I-2)=B(E2,J=I-1/2),
\eea
where the rhs denotes the result of the 
TAC calculation taken for the indicated value of
$J$.
Another possibility is to compare the experimental transition 
probabilities with the ones
calculated at the experimental  frequency of the transition
(\ref{omexp1},\ref{omexp2}). As long as the 
experimental and
calculated functions $J(\om)$ agree well, both ways will give about the 
same result. 

Of course, one may also use the relations (\ref{omexp2}) and
(\ref{E'exp2}) for a $\De I =1$ band. Then, the two signature branches
will lie nearly on top of each other if the discrete points are connected by
smooth interpolation. This choice has the disadvantage that the distance 
between the discrete frequency points is doubled. It has the advantage
to give smooth curves when the splitting between the two signature
branches gradually develops with increasing frequency. In such a case 
(\ref{omexp2}) and
(\ref{E'exp2}) should be used.

 \subsection{Diabatic tracing }
\label{s:diab}
     The goal of the calculation is to describe a rotational band, 
which corresponds to the
"same quasi particle configuration" for a set of increasing values of 
$\omega$. This means that one
should keep fixed  the occupation of the quasi particle states 
with similar structure.  Usually one band does 
not correspond to the same configuration if 
the quasi particle levels are labeled according to their energy, because
the \qp trajectories cross each other as functions of $\om$ and $\th$. 
In order to find the
equilibrium angle, one has to calculate the functions 
$J_1(\vartheta)$ and $J_3(\vartheta)$. This  
 becomes very tedious  if the configurations are assigned
 manually by identifying the crossings from quasi particle diagrams like
figs. \ref{f:qp} and \ref{f:qn}. The task is greatly
facilitated by tracing  the structure of the quasi particle wave functions. 
The calculations are  run changing $\vartheta$ or $\om$ in finite steps. 
For a given grid point the overlaps of 
each quasi particle state with all states of the previous 
grid point are calculated. The pair with
the maximal overlap continues one quasi particle 
level from the previous to the present grid point.
The pair with the next lower overlap 
continues the   second \qp trajectory. This procedure is repeated 
 until all quasi particle trajectories are continued. For 
all the single particle and \qp
diagrams shown in this paper  the grid points are
connected by means of this {\em diabatic tracing}. 

In a practical calculation, 
the configurations are assigned manually for the first
grid point in a loop. The following strategy has turned out to be quite
efficient: First a typical angle $\th_s$ is chosen and the \qp diagram
$e'(\om_i,\th_s)$ is generated. The step size $\De \om= 0.05MeV$ has turned
out to be a good choice. 
Configurations are assigned for a typical frequency.
The occupation numbers for the other  grid points $\om_i$
are found by means 
of the \qp diagrams or, if the crossing pattern is complex, using
the tracing facility of the code. 
These occupation numbers are used to set the configurations in a 
$\th$-loop starting at $\om_i$ and $\th_s$. The configurations of the 
other grid points in the loop are determined by means of
 diabatic tracing. Then, 
the code finds the orientation angle $\th$ for each $\om_i$ 
by means of the self-consistency
condition (\ref{scJintr}) and calculates  the interesting  quantities.
 The step size $\De \th =5^o$
 has turned out to be a good choice.
At which $\th_s$ the loop is started depends on the type of the band  and
will be discussed below. 

 Problems are encountered when the \qp levels do not cross sharply when
$\vartheta$ or $\omega$ are changing. If the grid point happens to be 
located in the middle of the region where the levels strongly 
mix and repel each 
other, the diabatic tracing  does not
always follow the desired structure. Such cases necessitate human
interference in order to continue the correct structure.
One reruns the calculation with the complementary configuration
and puts the parts with the correct configuration together. 
The grid point itself is problematic
because the cranking model becomes a bad approximation due to the 
unphysical mixing of states
with different angular momentum. These problems have been 
investigated for the
standard cranking model \cite{tbengtsson}. We restrict ourselves to the most 
simple solution advocated in \cite{bf}:
We discard such grid points and bridge the crossing region by 
means of interpolation. 

   The results of the diabatic tracing  depend 
on the step size. It should not be
too small. If the step size is much smaller than 
the mixing region, the procedure follows the
levels adiabatically, i. e. it connects the levels (of the same parity) 
according to their energy. On the other hand it should not be 
too large in order to preserve a reasonable precision. As mentioned above, 
step sizes of $\Delta\vartheta=5^o$  and 
$\Delta\omega=0.05\,MeV$ have turned out to be good choices.

For low $K$ bands it is usually convenient to choose $\th_s=85^o$ for
the manual assignment  of the configurations. The reason is
that with increasing $\om$ the  equilibrium angle $\th_o$
changes quickly  from zero to values close to $90^o$. 
As seen in figs. \ref{f:qp} and \ref{f:qn}, the number of avoided crossings,
which cause problems, is small at low frequency. Therefore the diabatic
tracing works well in most cases and permits  calculating the  interesting 
 range of $\om$  without human interference.
The \conf assignment should not be done at 90$^o$, where the signature is good and
the levels are often degenerated. 
The \confs discussed in subsections \ref{s:0qp}-\ref{s:2qn} are calculated 
by assigning \confs at $\th_s=85^o$.

For high $K$ bands, $\th_o$ remains relatively small up to rather large 
values of  $\om$. Then, starting at $\th_s=85^o$ becomes less efficient
because the number of avoided crossings increases. A smaller value
of $\th_s$ closer to the equilibrium angle is preferable.
For the \confs discussed in subsections \ref{s:p8m} we used   $\th_s=45^o$.
This choice has the disadvantage that one has to run the $\th$ loop
two times, for $\th<\th_s$ and  $\th>\th_s$. The choice $\th_s=0^o$
has turned out to be quite efficient in other applications of TAC to high 
$K$ bands.    

 Diabatic tracing  is also used when the other parameters of the 
mean field
Hamiltonian are changed in order to solve the complete
set of self-consistency equations. 
Approaching the minimum on the multi dimensional surface 
$E'(\om,\th,\eps,\eps_4,\gamma,\De)$ it is applied for each step in one
of the parameters.

 \subsection{             Choice of the QQ coupling constant}
\label{s:kappa}

   Using the PQTAC version, the 
coupling constant $\chi$ of the QQ - interaction must be fixed.
So far it has been adjusted such that 
the  quadrupole deformations 
$\eps$ and $\ga$  calculated 
for PAC solutions come as close as possible to the ones obtained
by means of the shell correction method, which  has a 
considerable predictive potential
concerning the nuclear shapes (cf. e.g. \cite{ragnarsson}). 
The adjustment has been carried out for selected nuclei.
The QQ coupling constant scales 
with $A^{-5/3}r^4_{osc}$, 
where $r_{osc}$ is the oscillator length \cite{ring}. 
This scaling has been used to determine $\chi$ in neighboring nuclei.

In the first  TAC calculations  \cite{tac}
the equilibrium shape was calculated 
for $A=170$ using the standard shell correction
method at $\om =0$.
The calculation was repeated
for PQTAC at the same deformation and $\om=0$. The coupling constant
 $\chi$ was chosen such that the
self-consistency equations (\ref{scQintr}) were fulfilled. 
This value of $\chi$  was kept constant in the full TAC calculation 
for all values of $\om$. 
The value  $\chi=0.0174~MeV~ r^{-4}_{osc}$ was found.  Scaling gives 
 $\chi=91~MeV~ A^{-5/3}r^{-4}_{osc}$ for the rare earth region.
When SCTAC became available, it turned out 
that  the results of PQTAC and SCTAC 
 were nearly identical  for the nuclei around $A=170$. 

Extrapolating by means of scaling  gives 
\mbox{$\chi= 0.0133~MeV~ r^{-4}_{osc}$}
for  $A=200$.  
This value (scaled locally) gives a good overall description 
of  the magnetic dipole bands in the Pb- isotopes \cite{mr1,mr2,mr3,pblt}.
A deformation of $\eps \approx -0.11$ is obtained. SCTAC gives   
larger deformations of $\eps \approx -0.15$, which 
account  less well for the data on the Pb - isotopes.

Scaling of the rare earth value
gives $\chi = 0.036~MeV~ r^{-4}_{osc}$ and $0.024~MeV~ r^{-4}_{osc}$ for
 $A=110$ and 140, respectively.
A new adjustment of $\chi$ was carried out for $^{110}$Cd and $^{139}$Sm. 
 The  respective values  $\chi = 0.036~MeV~ r^{-4}_{osc}$ and $0.022~
MeV~ r^{-4}_{osc}$ were 
determined by making equal the deformation obtained by means of  PQTAC and  
the shell correction method for zero pairing at finite
$\om$. The latter values 
  (including  local scaling) gave a good description of the magnetic dipole
bands in a number of nuclides of the two regions
 \cite{sn105,sn106,sb108,cd110,sn106l,cd109,cd108l,sm139}.  
 The data on electro-magnetic
transitions in $^{105,106,108}$Sn \cite{sn105,sn106l}, which
according to the calculations have a deformation of $\eps<0.11$, 
seem to point to a
smaller deformation than calculated. That is  a smaller value of $\chi$
leading to smaller deformation than predicted by the shell correction method 
appears to be 
more appropriate for $^{105,106,108}$Sn. This is similar to the Pb-isotopes. 

    The shell correction method accounts rather well for the overall tendencies 
of the shape, in
particular for the well deformed nuclei.
 However, it is not obvious that 
for  the small deformations encountered for magnetic rotation (typically $|\eps|
< 0.11$) the shell correction method provides a reliable gauge for $\chi$. 
In such cases it seems preferable to fix  the QQ coupling 
constant in a different way. 
Since $\chi$ controls the quadrupole polarizability, one may 
adjust it to 
the static quadrupole moments of high spin states  and 
the $B(E2)$ values of  transition between them, which  are particular sensitive
to the quadrupole polarizability.  It seems promising to use this experimental
information for a fine tuning of $\chi$. 
 This approach, which is discussed in more detail in the review \cite{rmp},
is being investigated \cite{chmel}.

\subsection{Approximate treatment of self-consistency }\label{s:dela}

The CHFB equations are a complex system of nonlinear equations.
TAC adds a new dimension to it, the orientation angle $\th$.
The fully self-consistent solution of the equations becomes rather
tedious, in particular if one tries to describe several 
non yrast bands.  The success of the CSM \cite{bf} shows that for a first 
analysis of the excitation spectrum  it is often 
sufficient, even preferable, to keep fixed the parameters of the mean field.
For selected bands, they may be determined self-consistently in 
subsequent calculations if one is interested in specific properties. But
very often  the additional effort does not pay off  the gain in
insight. We shall follow the CSM approach and carry out the
calculations assuming that the deformation and the parameters of
the pair field, $\De$ and $\la$ do not depend on the rotational
frequency $\om$. Only the tilt angle, which usually strongly changes,
is determined by means of (\ref{scJintr}) for each value of $\om$.

For the well deformed nuclei considered in this paper
the deformation changes turn out to be moderate. They are negligible for the 
more qualitative comparison with the data which we are aiming at. 

 The approximation   of a constant pair field
needs a more careful discussion. The original assumption of the CSM
\cite{bf} to keep $\De$ at 80\% of the experimental odd - even
mass difference $\De_{oe}$ becomes problematic, because modern data reach 
rotational frequencies where the static pair field disappears \cite{shimizu}.
Since the transition to the unpaired state
may substantially change rotational response, self-consistency must
be taken into account at least in some rough way.
We found the following compromise between accuracy and effort quite
satisfying. The TAC calculations are carried out at a few values
of $\De$, which do not depend on $\om$. 
The total Routhians $E'(\om,~\De)$ are plotted.
At each frequency one
can easily choose the best $\De$-value as the one that  has  the lowest
value of $E'$.
The upper panel of fig. \ref{f:dela} shows as an  
example the yrast band of $^{174}$Hf.
The  values $\De_n=0.95,~0.69, ~0~MeV$ and $\De_p=1.05,~0.75, ~0~MeV$
are used. The first point  corresponds to the self-consistent
ground state value and the corresponding curve is lowest
for small $\om$. At $\om=0.24 ~MeV$
the curve with the reduced values of  $\De_n$ and $\De_p$ takes over.
 Within the considered frequency
range the unpaired solution cannot compete, though the neutron 
correlation energy is rather small.

For paired \confs the proper Routhian is $E'(\om,\la)+\la N$, where $N$ is the 
exact particle number. The term
$\la N$ compensates the Lagrangian multiplier introduced in (\ref{HPQQ}).
However, exact compensation appears only if     
the self-consistency condition (\ref{scN}) for $\la$ is fulfilled.
If $\la$  is  approximately determined  
 one can proceed  in a similar way as for $\De$ by
plotting $E'(\om,\la)+\la N$.  Since
\beq
\frac{\partial}{\partial \la}
 E'(\om,\la )=-<\hat N>,
~ \frac{\partial}{\partial \la}
<\hat N> ~>0,
\eeq
the Routhian $E'(\om,\la)+\la N$  has a maximum at the self-consistent value of $\la$. Accordingly, TAC
calculations are carried out  for a few values of $\la$, which do not
depend on $\om$ and   $E'(\om,\la)+\la N$ is plotted. The highest curve
corresponds to the best value of $\la$. The lower panel of 
fig. \ref{f:dela} shows
the three points $\la_n=49.0, ~49.15, ~49.3~MeV$. The arrows
indicate the frequencies where the  self-consistency  condition (\ref{scN})
is fulfilled. For $\la=49.0~MeV$, the deviation in particle number is about
2 at $\om=0.5 ~MeV$. The upper envelop of the  curves represents
 the best  choice of $\De$ and $\la$ within the restricted set of grid points
investigated. For $\om > 0.3~MeV$, 
it behaves very similar to the 
unpaired curve.  
The small correlation energy of 0.1 - 0.2 $MeV$ indicates weak static
pairing.   We will show this optimized
Routhian of the yrast sequence in the figures as a reference.  
As long as the values of $\la$ are the same for all \confs one may leave
away the term $\la N$. It is however needed to correctly 
calculate the relative position of
\confs with different $\la$ or of paired and unpaired \confs.

This method is quite useful because it is   simple and it can easily 
be made as accurate as needed by adding more $\De$ and $\la$ values. 
At each stage one has a clear idea of the remaining  error of the
energy.
The simplest variant of considering only 
$ \De \approx 0.8 \De_{oe}$ and $ 0 $ and choosing $\la_p$ and
$\la_n$  such that the particle numbers are right for 
$\om=0$ turns out to be sufficient for a first orientation.
It shows the pair correlation energy directly.
The discussion of pairing  will be restricted to
this minimal variant. 
All figures showing total Routhians display the quantity $E'+\la_pZ+\la_nN$.
In order to keep the figures simple, the ordinate is labeled with $E'$ 
only. The energy of the ground state of the nucleus, 
which is not the concern of this paper, is not calculated correctly.
In all figures, only the Routhians relative to the ground state energy are of
relevance.    

\subsection{Reading the quasi particle diagrams}
\label{s:read}

Figs.  \ref{f:sp3} - \ref{f:sn6} 
show the single particle Routhians 
$e'_i(\om,\th)$ as functions of the 
frequency  $\om$ and of the tilt angle $\th$. In order to demonstrate
change of the particle response to with the magnitude and orientation of 
$\om$  two different 
frequencies   are presented for each kind of particles.
The figures with intermediate frequency are relevant for the present
day high spin data. The high frequency figures show territory 
yet to be explored. The side panels of each figure are added
for  helping the reader to connect to new middle panel with  the familiar 
single particle Routhians for rotation about the PA axes.  

 The slope of the 
trajectories gives negative projection of the \qp \am on the $\om$ - axis,
\beq
\frac{\partial e'_i}{\partial \om}=-j_{||}
=-(j_{1,i}\sin \vartheta+j_{3,i}\cos \vartheta ),
\eeq
and  its perpendicular component,   
\beq
\frac{\partial e'_i}{\partial \th}=-\om j_\perp
=-\om(j_{1,i}\cos\vartheta-j_{3,i}\sin\vartheta ), 
\eeq 
where $j_{1,i}$ and $j_{3,i}$ are the expectation values of the \am 
 in the single-particle or  \qp state $i$.

For $\th=0^o$, the cranking term $\om  \hat J_3 $ commutes
with the axial symmetric deformed potential and the projection of the 
\am on the symmetry axis $K$ is a good quantum number.
 In this case, the states
coincide with  the non-rotating Nilsson states, which are indicated by
the labels  in the
figure. For $\th=90^o$, the signature $\alpha$  is a
good quantum number which is also indicated in the figure. 
As the signature operator $e^{-i\pi \hat J_1 }$ and $ \hat J_3$ do not commute, there is 
a transition from one to the other type of symmetry when $\th $ changes
from 0$^o$ to $90^o$.

Discussing the features of the \qp diagrams, 
we will refer to the three types of coupling schemes that appear as a
consequence of the competition between the deformed potential, 
the inertial forces and the pair correlations. They are discussed in
\cite{fal}. Let  us start with moderate frequencies $\om < 0.3~MeV$, which are
illustrated in figs. \ref{f:sp3},  and \ref{f:sn3}.

The normal parity states with $K\ge 5/2$ obey the deformation aligned
coupling (DAL) scheme. 
 These orbitals are strongly coupled to the deformed potential. 
 In the $e'(\th)$ plot, they are recognized as 
the pairs of trajectories, which branch 
at $\th =90^o$.
They have a small component $j_{1,i}$ but a large
component $j_{3,i}\approx K$. They 
approximately behave like $-K\cos\vartheta$ in an extended region. 
Near $\th=90^o$ there is the
very narrow  transition region from good signature to almost good $K$,
where the slope changes from zero to approximately $K\sin\vartheta$.
The region is too narrow to be discerned in the figure, where it looks
like a kink.

There are pairs of parallel trajectories in the $e'(\th)$ plot,
originating from the [521]1/2 neutron and [411]1/2 proton
orbitals. These are  pseudo spin singlets. A discussion
of the pseudo spin symmetry in deformed potentials is given in
\cite{pseudospin}.
The projection $\tilde \Lambda_3$ of pseudo orbital momentum is 
zero. Thus,
the pseudo spin is decoupled from the orbital motion. It only reacts
to the cranking term $-\vec \om \cdot \vec  S$, where $ \vec S$ is the pseudo spin. 
The  two parallel, nearly
horizontal trajectories with the distance $\De e_{pso}=\om$
 correspond to the pseudo spin being
aligned or anti aligned with the rotational axis $\vec \om $. 
The pseudo spin vector follows the tilt of $\vec \om $, remaining parallel
to it. Since the pseudo orbital momentum remains small, the two trajectories
are almost horizontal.    
The signature $\alpha$ is gradually lost when the pseudo spin 
vector tilts away from the 1 - axis.

The states with the highest $K$ values of the $h_{11/2}$ 
and $i_{13/2}$ intruder orbitals 
obey the  DAL coupling. The states with lower $K$ 
 have  an extended
region around $90^o$ where the Routhians are relatively flat functions of
$\th$, what means that $j_\perp$ is small. In this region 
 the orbitals are rotational
aligned (RAL), precessing around the  rotational axis $\vec \om $, where the
precession cone follows the tilt of the axis. 
The signature
is gradually lost when  $\vec \om $ tilts away from the 1 - axis.
With    decreasing  $\th$, they make a quasi crossing
 with other members of the  same
intruder orbital. These crossings mark the  transition to 
the deformation aligned (DAL) coupling, which is 
shows up as the $-K\cos\vartheta$ behavior.

Fig. \ref{f:qn} shows the quasi neutron energies, which are relevant when
the pair correlations are important. What has been said about 
the single particle Routhians also applies to the quasi particle Routhians.
As a new type, the Fermi aligned (FAL) coupling 
\cite{fal} appears. It is realized
by the lowest $i_{13/2}$ trajectory, denoted by A. 
 The FAL coupling 
appears at
some distance from $90^o$. It corresponds to a substantial
component $j_3$ as well as to a substantial $j_1$.
It is most favored at the minimum  of
$e'_A(\th)$ at $\th = 38^o$, where $j_\perp=0$, that is $\vec j || \vec \om$.
With $\th \rightarrow 0$ the non rotating quasi particle state is
approached, i. e. $j_3 \rightarrow K$ and $j_1 \rightarrow 0$, corresponding
to the maximum. Overall, the lowest $i_{13/2}$ trajectory A is rather flat,
indicating that the orientation of
 $\vec j$ does never too strongly deviate from $\vec \om$.
At larger $\om$, where the negative and positive quasi particle states
strongly interact with each other, a  complex pattern of 
avoided crossings emerges, which we have not found a simple interpretation
for.

The high frequency regime is illustrated in figs. \ref{f:sp6} and
\ref{f:sn6}. It is characterized by many avoided crossing between the
orbitals. This indicates the progressive dissolution of the approximate
conservation of the $K$ quantum number for the DAL orbitals.
  
\subsection{Relation to the
PAC  treatment  of high-$K$ bands}\label{s:csm}

Bands with a finite value of $K$ have been studied by means of the standard
PAC scheme using the following prescription
\cite{bf}, which 
may be considered as an approximation to  TAC.  
A fixed $K$ value is ascribed to each band, which is the spin value
at the band head. It is taken from experiment or calculated by
means of the cranking model choosing $\vec \om$ parallel to the symmetry
axis. It is assumed that $J_3=K$, independent of $\om$
and  \mbox{$J_1=<J_1>$}. The \conf $|>$ is generated by the quasi particle 
Routhian (\ref{h'intr}) assuming $\th=90^o$.
Only  the reduced cranking term $-\om_1 J_1$ appears,
where $\om_1$ is the 1 - component of the angular velocity.

With these assumptions TAC goes over into  the CSM \cite{bf} scheme:
The constraint (\ref{J}) becomes
\beq 
J=I+1/2=\sqrt{J_1^2+J_3^2}=\sqrt{<J_1>^2+K^2},
\eeq
Solving for $<J_1>$,  the standard cranking constraint 
\beq\label{J1}
 <J_1>=J_1\sqrt{(I+1/2)^2-K^2}
\eeq 
to fix $\om_1$ is obtained.  In the  CSM 
one uses  $\om_1$ as the independent variable. Experimental values of 
$\om_1$ are derived by means of the expression \cite{bf}
\beq\label{om1exd}
\om_1=\frac{E(I)-E(I-2)}{J_1(I)-J_1(I-2)},
\eeq
with $J_1$ being the rhs. of expression (\ref{J1}).
The TAC condition $\vec J~||\,\vec\omega$ implies 
\beq\label{om1ex}
\om_1=\frac{J_1}{J}\om=\frac{\sqrt{(I-1/2)^2-K^2}}{2(I-1/2)}(E(I)-E(I-2)), 
\eeq
where the expression (\ref{omexp2}) for the experimental frequency is used.
The $\om_1$ values obtained by expressions (\ref{om1exd}) and (\ref{om1ex}) 
almost coincide, except 
near the band head.  As
demonstrated in the model study \cite{fm},  expression (\ref{om1ex})
reproduces the quantal  results slighly better.

With the fixed-$K$ assumption the expressions for the electro-magnetic
matrix elements given in sect. \ref{s:em}
become the ones of the semiclassical vector model of ref. \cite{df}.
For the magnetic moments, the vector model additionally assumes that each quasi  particle
has a fixed value of $j_{3,i} $, which  is $\om$ independent and given by
the $K_i$ of the Nilsson label. The individual $j_{1,i}$ values of
the quasi particles are either calculated or extracted from differences
between the experimental curves $J_1(\om_1)$ 
( the quasi particle alignments ). The magnetic moments are approximated by
\beq
\vec \mu =g_K  \vec j,
\eeq 
where the gyromagnetic ratio $g_K$ is either calculated by means of the 
Nilsson model \cite{bm2} or taken from experiment. In TAC
 expression (\ref{bm1},\ref{mu}) the components of the magnetic moments
 are calculated, attenuating the  free spin magnetic
moment of the proton and neutron  by a constant factor.  

In ref. \cite{d}  an additional term is introduced which
 permits the  calculation of the
signature dependence of the  $B(M1)$ values
for the case that the $J_3$ component is generated by only one
\qp. It is not expected that this correction is also applicable  if  $J_3$ is
generated by many \qps, whereas the vector
model \cite{df} without the signature term also applies to this more
general case.

As discussed, the standard CSM becomes a good approximation of TAC if
for the active quasi particles\\
i) $j_{3,i}(\om,\th)$  can be approximated by the constant
$K_i$ and \\
ii) $j_{1,i}(\om,\th)$  can be approximated by $j_{1,i}(\om_1, 90^o)$.\\ 
Then 
\beq\label{e'csm}
e'_i(\om,\th)\approx e'_i(\om_1,90^o)-\om_3 K_i,
\eeq
where
$\om_1=\om \sin\th$ and $\om_3=\om \cos\th$ (cf. eqs. (\ref{scJintr}) 
and (\ref{h'nilsd})).
 The assumption $J_3\approx K$ is justified and the tilt angle is given
by $\cos \th = K/J_1$, which leads to the expressions of the vector model
\cite{df}. 

Fig. \ref{f:i1i3th} shows the \am components $j_1(\om,\th)$ and
$j_3(\om,\th)$ of some representative 
quasi particles. They are compared with the CSM values $j_{1,i}(\om_1, 90^o)$
 and $K_i$, respectively. The DAL quasi neutron E obeys 
i) and ii) rather well. The  orbital G shows
the behavior \mbox{$ j_1 \approx \frac{1}{2}\sin \th$} and 
\mbox{$j_3 \approx \frac{1}{2}\cos \th$},
which is characteristic for the  pseudo spin singlet. This is 
at variance with i) and ii). Although the contribution to the 
\am is small, the characteristic spacing of $\De e_{pso}=\om$ between the 
two pseudo spin partners is not obtained
 when $\th$ is substantially below $90^o$.
The intruder orbital A
shows in the range  $40^o<\th<60^o$ the typical FAL behavior, which  is 
fairly well reproduced by the approximations i) and ii). For larger
values of $\th$ it changes gradually into a state with a good signature.
The transition is accompanied by changes of $j_1$ and $j_3$ which
 are  at variance  with i) and ii).  For smaller values of $\th$ orbital A keeps its
FAL character and remains close to the approximations i) and ii).
The orbital B changes dramatically when $\th$ decreases from $90^o$, because
there is the quasi crossing with the 
down sloping orbital C (cf. fig. \ref{f:qn}). For  values of $\om$ smaller 
than displayed this crossing is
rather sharp. One can follow the FAL branch of B, which is well
approximated by i) and ii),  below the crossing.
For larger values of $\om$  the two orbitals strongly mix and a 
new $\th$ dependence emerges, which is shown fig. \ref{f:qnthom4}. 
Obviously such changes 
of the quasi particle  structure cannot be described by means of
 the traditional CSM
treatment basing on the assumptions i) and ii).

The use of $\om_1$ as the rotational parameter
in the CSM has the advantage that all configurations
can be constructed from one and the same quasi particle diagram
$e_i(\om_1,90^o)$, which is a great simplification and has lead to the 
popularity of this approach. Another pleasant feature is that the 
signature splitting appears in a gradual way. The disadvantage is
that it is only an approximation to the  TAC mean field solution,
 the latter being completely self-consistent and more accurate when 
$\th$ substantially deviates from $90^o$ \cite{fm}.
Sometimes  the differences    are only of quantitative nature, but
 there are many cases where  qualitatively different results are 
obtained.   Magnetic Rotation of weakly deformed 
nuclei
\cite{mr1,mr2,mr3,pblt,sn105,sn106,sb108,cd110,sn106l,cd109,cd108l,sm139,rn205,rb74} 
is a conspicuous example, which  will not be discussed
in this paper. In the following discussion of examples 
we shall point out the
differences between TAC  the standard CSM.

 High-$K$  bands, which are in experiment near yrast, appear 
in  the  CSM as 
relatively high lying configurations, embedded into the back ground of 
many configurations with low $K$. 
 In TAC  they are low lying configurations.
The reason can be seen in (\ref{e'csm}). CSM uses $e'_i(\om_1,90^o)$, 
which is
shifted up by $\om_3 K_i$ with respect to $e'_i(\om,\th)$ used in TAC. 
It is also noted that only 
the TAC mean field solutions  can be improved by means of RPA
corrections in a systematic way. The PAC
configurations corresponding to a finite $K$  are instable.

It is quite common to present the experimental branching ratios 
$\frac{B(M1)}{B(E2)}$
as effective values  of the ratio $\left|\frac{g_K-g_R}{Q_o}\right|$, which
would determine the branching ratio if the strong  coupling limit was
valid \cite {bm2}. This popular way of representing the data 
has the advantage that the ratio becomes  constant when approaching the
strong  coupling limit. 
One may convert 
the  ratios of $\frac{B(M1)}{B(E2)}$  calculated by means of TAC
into  effective ratios $\left|\frac{g_K-g_R}{Q_o}\right|$. The pertinent
relation 
\beq\label{gkgr}
\left|\frac{g_K-g_R}{Q_o}\right|=
\left( \frac{5(J^2-K^2)B(M1)}{16J^2B(E2)}\right)^\frac{1}{2}
\eeq  
is obtained from the expressions in section \ref{s:em} by 
making the assumption of strong coupling,
 $J_3=K$, $\mu_3=(g_K-g_R)K$ and $\mu_1=0$.
The square root of the branching ratio 
becomes the product of
 $\left|\frac{g_K-g_R}{Q_o}\right|$ and the inverse of the  
geometric factor on the right hand side of (\ref{gkgr}).
In order to avoid any misunderstanding it is noted that (\ref{gkgr})
is just a way to present the results of the exact TAC calculations, 
which do not make any strong coupling approximation. 


\section{Multi-quasi particle configurations near $^{174}$Hf}
\label{s:mqpc}
This section will explain how to construct   
multi  \qp configurations
in the TAC scheme and how to interpret them as rotational bands.   
  The nuclides  $^{174,175}$Hf, $^{175}$Ta and $^{174}$Lu
serve as examples.
The SCTAC scheme is used for the calculations. 
In order to simplify the discussion, 
the same  deformation parameters $\eps=0.258 $, $\eps_4=0.034$
and $\gamma=0^o$ are assumed for all the nuclides considered. The set
represents
an average  of equilibrium shapes   calculated 
for  several configurations and frequencies $\om$. The actual values
 scatter within the interval $0.25 <\eps <0.27$, but the differences 
in deformation  
do not change the discussed quantities in a substantial way.
We consider both the cases of no pairing and a constant pair field.
For the case of finite pairing we use
the prescription of the CSM \cite{bf}, which has turned out to give very
reasonable description of multi-quasi particle bands in the traditional PAC
scheme. Accordingly, 
 $\De_n=0.69~MeV $ and $\De_p=0.75~MeV $, which is 80\% 
of the experimental even -odd mass
difference. The chemical potentials
$\la_n$ and $\la_p$ are fixed to the values that give the correct particle
numbers for the ground state (\conf [0] at $\om=0$).  This scenario
provides a good description of configurations up to two excited quasi
particles and a frequency of about $0.35~MeV$. It will be discussed in 
the subsections \ref{s:0qp} - \ref{s:2qn}.
For higher frequency and
more excited quasi particles we   take 
zero neutron pairing into consideration.    
Self-consistency for $\De$ is invoked along the lines 
described in sect. \ref{s:dela} in order to decide where the transition
to zero pairing is located. Subsection \ref{s:p8m0} discusses this regime.

\subsection{Construction of multi quasi particle
configurations}\label{s:con} 

For zero pairing the  configurations are
generated by filling up the lowest $Z$ and $N$ single particle 
levels and then making particle-hole excitations.

In the case of pairing the configurations are constructed from the 
quasi particle Routhians. The quasi particle spectrum is symmetric with respect
to $e'=0$ and the double dimensional occupation scheme, 
discussed for the PAC solutions
in ref \cite{bf}, is applied: If a quasi particle state $i$ is occupied, its 
conjugate partner $ i^+$ must be free.
In contrast to the PAC case, 
  the conjugate states in general do not have opposite signature, which
is only for $\th=90^o$ a good quantum number. 

Diabatic tracing  
turns out  very practical for identifying 
the conjugate states. They always cross sharply because they are 
orthogonal.     
As in the PAC scheme, one must be careful in choosing the right particle
number parity when \qp trajectories cross the zero line. The most simple way
is to start at sufficiently low $\om$, where there is still a gap between
the positive and negative solutions. There it is clear how to excite an
odd or even number of \qps. Keeping the occupation
by diabatic tracing, the particle number parity of the
configuration is conserved.

In order to efficiently label the \confs  a compact notation 
which indicates the \qp composition is desirable. We follow the well-tried
practice of the CSM assigning letters to the \qp trajectories 
and quoting the excited \qps in parenthesis. The letters A, B, C,  D 
denote  positive parity \qns , E, F, G, H, ... 
negative parity \qns, a, b, c, d positive parity \qprs and e, f, g, h  ,...
negative parity \qprs. 

The letter code becomes to some extend ambiguous
when the structure of the \qps  strongly changes
 with the frequency $\om$ and the tilt angle $\th$.
The positive parity $i_{13/2}$ orbitals  are most susceptible to the 
inertial forces. Fig. \ref{f:qn} 
 shows  the  complex pattern of \qp trajectories, which 
strongly interact with each other and interchange their character as
functions of $\om$ and $\th$. 
An example are the orbitals B and
C in fig. \ref{f:qn}.   As discussed already in sect. \ref{s:csm}, 
they quasi-cross each other
near $\th=60^o$. For $\om =0.2~MeV$, (not shown)  the crossing is still
rather sharp and it would be  natural to follow each trajectory diabatically, i. e.
for  $\th<60^o$ to call the upper trajectory B and the lower one C. 
 For $\om =0.4~MeV$ (see fig. \ref{f:qnthom4}) they interchange their
character very gradually.
Now it is more natural  to call the lower trajectory B and the higher C
throughout the mixing region.
Fig. \ref{f:qn9045} shows the \qn trajectories as functions of $\om$.
For $\th=45^o$ (lower panel) the crossings are rather sharp for most
trajectories. Here it is natural to keep the labels in a diabatic way, as
indicated. For $\th=90^o$ the crossings between the $i_{13/2}$ trajectories 
are much softer and the question arises of how to label them after the 
first quasi crossing. The suggested labeling  tries to
follow the \qp trajectories both in the $\om$ and the $\th$ direction such  
that the structural change is as gradual as possible.
It  connects the two 
diagrams $e'_i(\om,\th=45^o)$ and $e'_i(\om,\th=90^o)$
the most  natural way. 
via the $\th$ degree of freedom   at high frequency (cf. figs. \ref{f:qn}
and \ref{f:qnthom4}). This implies that the smooth crossing between A and
B$^+$ as function of $\om$ at $\th=90^o$  must be treated diabatically.

It should  be pointed out that the suggested labeling is
a compromise.  As discussed above,
for $\om=0.2~MeV$ the \qns  B and C cross sharply as functions of $\th$.
In the adopted labeling the lower of the two levels is B and the higher C.
It would be more natural to follow the structures in $\th$ direction
 diabatically through the crossing. But a relabeling that accounts for this
leads to problems a high $\om$, where the adopted labeling 
is most natural.  
The difficulty to label the strongly interacting \qp trajectories 
  in a simple way has a topological origin, which 
can be best
understood if one follows in one of the \qp diagram \ref{f:qp}, \ref{f:qn},
\ref{f:sp3}, \ref{f:sn3}  a trajectory on the $(\om,\th)$-path: 
$ (0,90^o) \rightarrow (0.3~MeV,90^o)  \rightarrow
(0.3~MeV,0^o) \rightarrow (0,0^o)$. For weakly interacting trajectories, as
most  with normal parity, one returns to the same \qp. For the intruder 
trajectories, as $i_{13/2}$ and $h_{11/2}$, this is not  always the case.      

Trying to keep the notation as  simple as possible
we shall   assign the   
low-$\om$ composition to a \conf and shall not  change it when a 
crossing is encountered. The structural change can be figured out from the
\qp diagrams.

\subsection{Elimination of spurious states}\label{s:spur}

Each configuration with an equilibrium angle $\th_o \not= 90^o$ is associated
with a $\De I=1$ rotational band (TAC solution), whereas each  configuration
with an equilibrium angle $\th_o = 90^o$ is associated
with a $\De I=2$ rotational band (PAC solution).  Of course, the number
of quantal states cannot abruptly double when the equilibrium angle
moves away from $90^o$. Hence, one has to be careful in
avoiding spurious states. This problem was studied in ref. \cite{fm}
for the model system of one and two quasi particles coupled to a rotor.
An elimination scheme has been suggested 
which is based on the following principle:
The number of TAC \confs must be the same as the number of PAC \confs
at $\th=90^o$, where they emanate from. This means, for each TAC minimum,
which is interpreted as a $\De I =1$ band composed of
 {\em two} $\De I =2$ sequences, one has to discard
one configuration. One finds this spurious state most easily by taking into
account that the function $E'(\th)$ is symmetric with respect to $90^o$.  
If one \conf  ($J_3 \approx K > 0$ ) has a minimum at $\th_o$ its mirror image
 ($ K < 0$ ) has  the minimum at $90^o-\th_o$. Tracing the function
$E'(\th)$ of the mirror image diabatically through $\th=90^o$,
one arrives at the spurious \conf that must be discarded.

The tilt angle $\th_o$ increases with  $\om$.  When it approaches $90^o$ one has to 
switch from the TAC to the PAC interpretation. This results in a discontinuity of 
the function $E'(\om)$  for the unfavored (upper) signature branch: In the TAC
scheme it is
is degenerate with the favored branch whereas in the PAC scheme
it is  the discarded \conf, which is now taken into account. 
The  angle for switching from TAC to PAC is to some extend
arbitrary. We have found it reasonable  to use the PAC interpretation 
when the equilibrium angle $\th_o > 80^o$. 
If several \qps combine into high-$K$
and low-$K$ \confs it is important to switch from TAC to PAC for {\em both} the 
high- and the low-$K$ \confs at the {\em same} $\om$.
 As discussed in \cite{fm} and in sect. \ref{s:1qp1qn} for a concrete
example, changing to PAC only for a part of the configurations results in 
highly nonorthogonal states.

As an example, let us consider the most simple case of the one 
\qn \confs denoted by [E]
and [F] in fig. \ref{f:Eth1qn}, which represent, respectively, the two 
branches $j_3 \approx 5/2$ and -5/2 of the DAL orbital [512]5/2. 
For $\om = 0.2$  and $0.3 ~MeV$, [E] has a minimum below $80^0$, which
is interpreted as a $\De I =1 $ band.
The diabatic continuation of [F] becomes the mirror image of [E] for $\th >
90^o$. Hence,   [F] is spurious  and  must be discarded. The kink - like minimum 
of the upper branch of [512]5/2 must also be disregarded.

The elimination rules are somewhat differently formulated
in ref. \cite{fm}. The reader might find this complementary formulation 
instructive. The proposed scheme has
been tested for the model system of one and two quasi particles coupled to a
rotor \cite{fm}. No spurious states have been found in the low lying spectrum 
after applying the elimination rules.

\subsection{Band heads}\label{s:head}

Generally, a band is a quasi particle configuration whose
\am increases with the rotational frequency $\om$. 
Its structure changes gradually with $\om$, such that it remains similar for
 adjacent quantal states  of the band. This is a natural 
definition which permits calculating both the start and the termination of
a band.  Since we restrict ourselves to
well deformed nuclei, we shall  discuss only the start in this paper. 

Fig. \ref{f:Ethe} illustrates how the 
\conf [E] in $^{175}$Hf starts.  
 The function $E'(\om,\th)$ has a minimum at $\th=0$ (and 180$^o$) 
for $\om$ below the band head  frequency of $\om_h=0.08~MeV$.
In this range of $\om$ the band has not yet started, because \am does not
depend on $\om$, being  $J=K$. 
 The band actually starts at $\om_h$ when
the equilibrium value  $\th_o$ becomes finite, i. e. when
 the minimum
of $E'(\th)$ at $\th=0^o$ turns over  into a maximum and there appears  
a minimum at $\th_o>0^o$. That is, the  frequency  $\om_h$ where
\beq 
 \frac{\partial^2E'(\om_h,\th)}{\partial \th ^2}\mid_{\th=0}=0
\eeq
 has the 
physical meaning of the rotational frequency of the band head. 
Fig. \ref{f:thom1qn} shows the equilibrium  angles $\th_o$
 for several one \qn bands. The
bands heads lie where $\th_o$ bifurcates from the zero line. 

The experimental
band head frequency is the energy of the first transition \mbox{ $I=K+1\rightarrow
I=K$}. It should be compared  with the frequency where TAC gives
$J(\om)=K+1$, which
is somewhat larger than $\om_h$. In some of the figures (e. g. 
\ref{f:1qp1qn})  this frequency is
indicated by a fat dot.
In most of the  figures the calculated curves start with the first grid 
point for which 
$\th_o > 0$, i. e. $\om_h$ is only determined with the accuracy of 
$\De \om = 0.05~MeV$, which is the step used in the calculations.

One may distinguish between strong coupling behavior 
and more complex response near the band head. 
Strong coupling behavior corresponds to $J_3=K=const.$ and $J_1=\om_1{\cal J}$,
where ${\cal J}$ is the moment of inertia of the collective rotation.
In this case one has
\beq
J=\om{\cal J},~~\th_o = \arccos (\frac{K}{\om {\cal J}})~~
 for ~~\om>\om_h=\frac{K}{{\cal J}}.
\eeq 
 Axial nuclei are close to the strong coupling limit near the band head 
 if only DAL quasi particles are excited. The \conf [E] 
illustrated  in figs.  \ref{f:Eth1qn} - \ref{f:thom1qn} is of his type.     
 At a first glance, one might
expect that the TAC approximation, which treats 
the orientation angle $\th$ in a static
way, becomes a bad approximation near the band head,
 because the $E'(\th)$ is very flat there.
  The model studies \cite{fm} demonstrated that this is not 
the case. In fact, 
the wave function becomes narrow at
the band head, because  $J_3$ approaches the  good
quantum number $K$. One may interpret this as follows.
The mass parameter associated with the zero point motion  in
$\th$ increases faster
than the curvature near the minimum,
 $\partial^2E'(\om_h,\th)/\partial \th ^2 \mid_{\th=\th_o}$.

A more complex situation  is encountered when one or more
\qps  easily  align with the rotational axis.
 Configuration [A] in fig. \ref{f:thom1qn} is an example. 
 The band  starts significantly earlier than expected for 
a strongly coupled 7/2 band with a jump of $\th_o$.   
For cases like this 
 ref. \cite{fm} found that  TAC approximates 
 the quantal particle rotor calculation   less well near the band head,   
but  becomes again a very good approximation for higher frequency.
 
\subsection{The zero quasi particle configuration}\label{s:0qp}

In $^{174}$Hf, the
 lowest configuration at $\th=90^o$ is the vacuum [0] with all negative
levels occupied, which is the ground (g-) configuration at 
low $\om$. Around  $\om=0.30~ MeV$, the neutron system   gradually changes 
into s-configuration, which  is  seen in figs. \ref{f:qn} and \ref{f:qn9045}
(upper panel) as the 
quasi crossings of trajectories A  with B$^+$ and B with  A$^+$
(AB crossing). Near
these crossings  the $\th$ dependence of the trajectories is complicated. 
We shall return to the interpretation of this region in sect. \ref{s:2qn}.   
 
First, let us  discuss the proton system, which does not have such a
crossing in the considered frequency interval.
At $\th=90^o$, the configuration [0] has  the 
character of the g-band. It keeps this character when $\th$ decreases, 
provided the occupation is followed diabatically, i. e. the crossings 
between the $\pi=+$ trajectories at $\th=22^o$ and the $\pi=-$ trajectories
at $\th=8^o$ are ignored.  
  It becomes  the ground state for $\th=0$,  because
the wave function does not depend on $\om$ for  this orientation.
The ground state  is not the lowest configuration at  $\th=0$, because
a number of  quasi
particles have crossed the zero line and crossed each other.
 This example demonstrates the advantage of diabatic tracing, which
automatically finds [0] when starting from either small $\om$ and $\th=0^o$
or $\th$ close to $90^o$, where [0] is  the lowest \conf. 
The neutron system has an analogous structure for $\om < 0.25~MeV$, where
where [0] has the character of the g-band. Fig. \ref{f:qn} shows the \qn
trajectories at $\om =0.3~MeV$. It is seen that the \conf 
[0], which has a mixed  g- and  s- character  $\th=90^o$, 
becomes the ground state at $\th =0^o$, where it is no longer the lowest
\conf.

Fig. \ref{f:Eth2qn+} shows the total Routhian $E'(\om=0.2~MeV,\th)$ of 
the combined proton and neutron \confs [0]. 
Its $\th$ dependence
 reflects the g-band character: The \am is   collective, i. e.
\beq
E'(\om,\th)\approx-\frac{\om_1^2}{2}{\cal J}=-\frac{(\om\sin\th)^2}{2}{\cal J}. 
\eeq
The minimum lies a $\th=90^o$, the signature is $\al=0$,
corresponding to the even spin g -band. 
For $\om =0.3~MeV$ the level repulsion near between A and B$^+$ modifies the 
slightly the $\th$ dependence of the total Routhian.

In order to make a first
qualitative estimate of the tilt angle for 
 multi \qp configurations one has to add  this zero \qp Routhian to the
sum of the 
Routhians $e'_i(\th)$ of the excited \qps, which can be
 taken from the \qp diagrams.

For  $\om>0.35~MeV$ the \qp vacuum  has the character of the 
s-configuration at $\th=90^o$. With decreasing $\th$,  it changes 
into the t-configuration 
which becomes the $K^\pi=8^+$ configuration $[7/2^+,9/2^]$ at  $\th=0^o$. 
We shall discuss these changes and their consequences in sect. \ref{s:2qn},
 together with the two \qn excitations of positive parity.

\subsection{One quasi neutron configurations}\label{s:1qn}

They are generated by adding one quasi neutron to the \conf [0]. 
Fig. \ref{f:Eth1qn} shows their total Routhians $E'(\om,\th)$.

The \confs [G] and [H] have  $\th_o=90^o$ for all $\om$.
They are interpreted as $\De I =2$ bands.
They represent the two signatures $\alpha = \pm 1/2$
 of the pseudo spin singlet 
[521]1/2. Since the pseudo spin is decoupled (cf. 
sect. \ref{s:read}), \mbox{$E'(\th)=E'_{[0]}(\th)+const \pm \om/2$}  
and $\th_o=90^o$. 
 
For 
$\om=0.2$ and $0.3~ MeV$ the \confs [A] and [E] have   minima 
at $90^o>\th_o>0^o$. They are interpreted as
 $\De I =1$ bands ($K^\pi= 7/2^+$ and $5/2^-$).
The \confs [B] and [F] are the continuations of [A] and [E]
 reflected through $\th=90^o$. Accordingly they are discarded as spurious states
together with the kink at $90^o$. 
Then the condition is satisfied,
that the number of states
is the same as for the PAC interpretation at $\th=90^o$.      
For $\om=0.4~ MeV$  the minima of [A] and [E] have moved above $80^o$.
 Now
 we change to the PAC interpretation and refer to the
calculations at  $\th=90^o$.
{\em Both}
 [A] and [B] are interpreted as $\De I =2$ bands. They form 
the signature pair $(\pi, ~\alpha)=(+,~\pm1/2)$.
The \confs [E] and [F] represent
the two $\De I =2$ bands combining to the signature pair $(-,~\pm 1/2)$.
 
Fig. \ref{f:1qn} shows the calculated total Routhians. The change from the 
TAC interpretation to the PAC is seen as the sudden onset of the signature
splitting. The width of the transition region is determined by the size of
the calculation grid in $\om$, which is $0.05~MeV$ and does not bear
any physical relevance. As discussed above, the TAC approach is not 
able to describe the smooth transition from broken to conserved signature symmetry. 
In order to describe the gradual onset of the signature
splitting one has to go beyond the pure mean field theory 
\cite{onishi1,onishi2,doenau}.     

Since the orbital
E obeys the DAL coupling, the \conf  [E] is expected to be close
 to the strong coupling limit. 
 Fig.
\ref{f:thom1qn} shows that the $5/2^-$ band starts at $\om_h=0.09~MeV$ near the
strong coupling estimate $K/{\cal J}$. 
Also for higher $\om$ the tilt angle $\th_o$ remains
  close to the
 strong coupling value. The $7/2^+$ band
 starts at $\om_h=0.02~MeV$, below the $5/2^-$ band [E] and much below
the strong coupling estimate for the $K=7/2$ band. This indicates a
substantial deviation from  strong coupling. In fact,  fig.
\ref{f:thom1qn} shows that $\th_o$ jumps to a finite value at a
low frequency,  corresponding to the rapid transition from the DAL to
FAL coupling at the band head.

Fig. \ref{f:1qn} also shows the experimental Routhians \cite {hf174}.
The relative position of the Routhians as well as their slopes (i. e.
the \am) are reasonably well reproduced by the calculation. The frequency
of the first transition is  well described too. In particular the 
low value of $\om_h$ for 
the $7/2^+$ band indicates that the FAL coupling is 
seen in the experiment.

In the TAC calculation the \conf [C] 
 starts at  $\om_h=0.19 ~MeV$ with $J=6.1$ and $\th_o=60^o$
(cf. fig. \ref{f:Eth1qn}).
It represents the $K=9/2^+$ orbital. The \conf [D] is
discarded because it is the continuation of the mirror image of [C]. 
The minimum rapidly moves towards $90^o$, due to the strong admixture of
 $i_{13/2}$ components with low $K$.  After $\om=0.25~MeV$
one has to change to the PAC
interpretation, where both
[C] and [D] are interpreted as the signature pair $(+,~\pm1/2)$.
Experimentally, only the transition $I=9/2\rightarrow 11/2$ is seen
at $\om=0.153~MeV$, which is lower than the calculated value of $\om_h$.

The panel  $\om=0.2~MeV$ in fig. \ref{f:Eth1qn} shows that the
minimum of [C] is very shallow. For $\om=0.19~MeV$
 it becomes a shoulder.  In contrast to the experiment,
 there is no $\th >0$ solution
for lower frequency.
  Here, a limitation of the TAC approximation
is encountered. The tilt angle $\th_o$ is found in a static way by searching
for the minimum of the Routhian $E'(\th)$. This is an approximation to 
studying the dynamics of the $\th$ degree of freedom.
The static TAC treatment is expected to give good results as long as there
exists a  certain convex region around $\th_o$.
 Then $\th$ will execute symmetric oscillations and
averaging over them will  result in values close to the  ones 
for $\th_o$.
The model studies in ref. \cite{fm} have demonstrated this
 for the lowest configurations.
It is clear  for a curve like [C] that averaging over the 
the collective wave function in $\th$ 
needs not to give values close to the ones obtained for $\th_o$, 
in particular, when  the minimum has become a shoulder. Then the dynamics
of $\th$ must be explicitly calculated.  
Refs. \cite{onishi1,onishi2,doenau} have addressed
this problem in the frame work of  the Generator Coordinate Method.

Situations like the discussed one become more likely if one considers 
excited \confs. As seen in fig. \ref{f:Eth1qn}, the flat behavior of
[C] may be thought as the consequence of interaction (repulsion) with
the \confs below and above. This problem is not special to the 
orientation degree of freedom. Analog restrictions of the
static HFB approximation are encountered when it is used to calculate the shape
of excited configurations.

Since the $J_3<4.5$ for all the one \qn \confs 
  the tilt angle  $\th_o$ rapidly increases with the frequency.  
As seen in fig. \ref{f:thom1qn},  $\th_o > 60^o$ for $\om>0.2~MeV$.
 Accordingly, the \qp trajectories become  similar to 
the  ones at $\th=90^o$. One recognizes the familiar CSM pattern 
of band crossings.
The $\pi=-$ bands show the AB crossing and the 
$\pi=+$ bands the delayed BC crossing, because AB is blocked
(cf. e. g. \cite{bf}). 

\subsection{One quasi proton \confs }\label{s:1qp}

The lowest proton \confs are generated by occupying the 
orbitals e, a and c in fig. \ref{f:qp}. They are
all of DAL type and $\th_o < 80^o$. Accordingly, the
\confs  [e], [a] and [c] are interpreted  as the  $\De I=1$ 
 bands with $K^\pi=7/2^-,~7/2^+$ and $5/2^+$, respectively.
The \confs [f], [b] and [d] are discarded.
The \conf [g] has always $\th_o=90^o$, as can be expected from
fig. \ref{f:qp}. It is interpreted as the 
$\De I=2$ band $(-,1/2)$, i. e. the favored signature sequence
of the $h_{9/2}$ orbital. 
Fig. \ref{f:1qp} shows the calculated and
the experimental Routhians in $^{175}_{73}$Ta$_{102}$.
All bands show the neutron AB crossing
at $\om=0.3 ~MeV$.  The TAC calculation for the
 \conf [g] gives too high energy and
shows  too early the neutron AB crossing. This is a well known problem
of the  $h_{9/2}$ band which has been discussed in the 
literature. The discrepancies can partially be attributed to a larger
deformation. 
Since these questions have been addressed before \cite{fal,h92}
and are not at the focus of this
paper we have not tried to improve the agreement by optimizing 
the deformation.

\subsection{One quasi proton one quasi neutron \confs}
\label{s:1qp1qn}
The Routhians $E'(\th$) for the 
 four combinations of the  \qprs a and  b emanating from 
Nilsson states [404]7/2  with
the   \qns E and F emanating from [512]5/2 are shown 
in fig. \ref{f:Eth1qp1qn}.   
They are nearly degenerate at $\th=90^o$. 
The \conf [aE] has its minimum at $\th_o =35^o$ and  represents the 
$\De I=1$ band   $K^\pi=6^-$. The \conf [aF] has its minimum at  $\th_o =78^o$ 
and represents the 
$\De I=1$ band   $K^\pi=1^-$. The other two \confs [bF] and [bE]
continue the mirror images of [aE] and [aF]. 
 They are discarded as spurious
states. Both $\De I=1$ bands are seen  in $^{174}$Lu.
Fig. \ref{f:1qp1qn} shows that  separation and the slope of the bands is 
well reproduced by the TAC calculation.
Another bundle of Routhians are four combinations of a and b with A and B,
emanating from the neutron states [633]7/2. The \conf [aA] represents
the $\De I=1$ band   $K^\pi=7^+$. The \conf [bB] has  no minimum,
only the kink at $\th=90^o$. It continues the mirror image of [aA] and
is discarded. The \confs [aB] and [bA]
have both their minimum at $\th_o=90^o$. One is interpreted as a 
the $\De I=1$ band   $K^\pi=0^+$ and the other one is discarded. 
It does not matter which is chosen, because the two \confs differ only by
 their orientation ($|aB>=e^{-i\pi \hat J_2}|bA>$). To be definite we choose [aB].  

When   two \qps of the DAL type are combined into  a low-$K$ and a 
high-$K$ \conf, one must 
switch from the TAC to the PAC interpretation 
for both configurations {\em simultaneously}. 
As discussed in detail in ref. \cite{fm}, interpreting one
\conf as PAC and the other one as TAC  makes them highly nonorthogonal:
The TAC solution is of the type $\psi_{K_1}\psi_{K_2}$ whereas the
PAC solution is of the type $(\psi_{K_1}\pm \psi_{-K_1})(\psi_{K_2}\pm
\psi_{-K_2})/2$.     This has the following consequences for our example:

i) One cannot take [aB] at its minimum at $\th=90^o$, because there it is of
 PAC type with good signature and thus  nonorthogonal to the $K=7$ \conf
[aA]. It suffices to take the \conf at a somewhat smaller tilt angle,  
 for example  at $\th=85^o$, where [aB] has changed to the $K=0$ 
\conf. Energywise
this  makes barely a difference. However it is important for the 
calculation of the $B(M1)$ values, which 
according to ({\ref{bm1}, \ref{mu}) are zero for a PAC \conf.

ii) Since the $K^\pi =1^-$ \conf [aF]
 has a larger
 $\th_o$ than the  $6^-$ combination [aE], 
its minimum approaches $90^o$
at a lower value of $\om$. However,  one must keep the TAC interpretation
as long as [aE] is tilted. As discussed for [aB], 
one may take $\th=85^o$.        
In fact, substantial $M1$ transitions are seen in this band, 
which are illustrated by fig. \ref{f:bm1lu174}.   

Fig. \ref{f:1qp1qn} compares the experimental Routhians in
 $^{174}_{71}$Lu$_{103}$ with the TAC calculations. 
The relative positions and
slopes are well reproduced.
Within the frequency range, all \confs are of TAC type. 
For the experimental $K^\pi =0^+$ band [aB] the
two  signatures are separated (signature splitting). This is at variance with
the calculations, which assign a $\De I = 1$ band (cf. preceding paragraph).
The discrepancy is attributed to the residual interaction, which may lead
to signature dependent correlations in the low-$K$ bands, as well as to 
the zero point motion in $\th$.

It is noted that in $_{73}^{180}$Ta$_{107}$ the $K^\pi=9^-$ and $0^-$ bands
 composed of the \qn [624]9/2 and the \qpr [514]9/2 {\em both} do not
show signature splitting \cite{ta180} and both have 
a substantial M1 transitions. This is  consistent with 
the  predictions  by  TAC. 

The orbitals H and G emanate from the pseudo spin singlet
[521]1/2 (cf. subsection \ref{s:read}). The $K^\pi =4^-$ \conf [aG] and the
$K^\pi =3^-$ \conf [aH] correspond to the parallel and anti parallel
orientation  of the pseudo spin with respect to the rotational axis.
 Accordingly, the distance
between the $4^-$ and $3^-$ bands  is equal to
$\om$ in the TAC calculation. The experimental distance somewhat deviates 
from this value, what
can be seen as evidence for a pseudo spin dependence of the 
proton neutron interaction.

\subsection{Two quasi neutron excitations}\label{s:2qn}

Fig. \ref{f:Eth2qn+} shows the Routhians $E'(\om,\th)$
of the lowest positive parity configurations.
Let us start with $\om=0.2~MeV$.
 For $\th>60^o$, the first three    \confs are [0],  [AB]  and  [AC], which  
 have the character of g-,  s- and  t-
configurations, respectively. For both the s- and the t- \confs 
the $j_1$ components of the two \qns add up to a large
value of $J_1$. In the case of the s-\conf the two $j_3$ components are
opposite in sign, resulting in $J_3\approx 0$, whereas for the 
t-\conf the two $j_3$ components add up to a large $J_3$.
The structure of  the t- and s-bands was first   discussed in   
\cite{tac,re181}, where illustrations can be found. 

For $\om=0.2~MeV$, the \confs [AB] and [AC] change order at
$\th=60^o$. 
For $\om=0.3~MeV$, they mix strongly around $\th =65^o$,
 interchanging their character. This reflects the quasi crossing between the
orbital B and C, which can be seen in figs. \ref{f:qn} and \ref{f:i1i3th}. 
 For $\om=0.4~MeV$, the crossing feature has disappeared. The orbitals
 B and C are now substantially different from what they were at 
low frequency. The labeling of the \qn trajectories
suggested  in figs. \ref{f:qn9045}, 
always assigns [AB] to the lower and [AC] to higher \conf.
For   $\om=0.2~MeV$ this results in an abrupt exchange of the structure 
 when the \confs cross at $\th=60^o$.        
[AB]  takes the character of a t-band for  $\th <60^o$ (which changes into
the $K^\pi=8+$ state at $\th=0$). Their sudden exchange becomes a smooth
transition at high frequency.

 The   change of \conf [AB] from  the s- to the t- structure  
reflects the strong response of the $i_{13/2}$ \qns to the inertial forces, 
which depend on the orientation of the rotational axis.  In fig.
 \ref{f:i1i3th} it is seen for the \qn B  as the rapid 
change of  $j_3$  from negative to positive values near $60^o$.
This is an example of the complex rotational
response which cannot be guessed within the traditional PAC scheme.

 As seen in fig. \ref{f:Eth2qn+}, [AB] has its minimum at
$\th=90^o$. It has the signature $\al=0$ and is interpreted as the even spin
s - band. It becomes yrast after the  AB crossing at $\om=0.3 MeV$.
The \conf [AC] has also its minimum at $\th=90^o$. Since its signature is 
$\al=1$ it represents an odd spin band. The t-character of [AB] is 
only explored when additional  \qps of DAL type
 change $\th$ to smaller values. This
will be discussed in subsection \ref{s:p8m}.
 For neutron numbers  $N \geq 106$ the
 t-\conf   is more favored,   
becoming  a stable minimum. The   pure two \qn t-band
is seen  for example in 
$^{180}_{74}$W$_{106}$\cite{w180}  and $^{182}_{76}$Os$_{106}$ \cite{os182}.  

 Now we consider 
the combinations of the $i_{13/2}$ orbitals A, B, C, D
 with E and F, emanating from
the Nilsson state [512]5/2. Fig. \ref{f:Eth2qn-} shows the Routhians.
In the lower bundle the \qns E and F are combined with A and B,
 emanating from [633]7/2.
The \conf [AE] is the $K^\pi=6^-$ band and [AF] the $1^-$ band, both being
$\De I=1$ sequences.
The \confs [BE] and [BF] are discarded. 
At $\om=0.4 MeV$ the minimum of [AE] has
moved to $80^o$ and we switch to the PAC interpretation. Now,
[AE] and [BF] have $(\pi,\al)=(-,1)$, i. e. they
 represent two odd spin bands,
 and  [AF] and [BE] have $(\pi,\al)=(-,0)$, i.e. they represent
two even spin bands.
Fig. \ref{f:2qna} compares the calculation with the experimental bands in 
$^{174}_{72}$Hf$_{102}$. Whereas the $6^-$ band is seen as a 
$\De I =1 $ sequence, as expected, 
the experimental $1^-$ band shows a substantial signature splitting.   
The discrepancy is attributed to the residual interaction. For the low-$K$
negative parity \confs the octupole correlations are important. 
Usually they are stronger for the $(\pi,\al)=(-,1)$ bands than for
the $(-,0)$ bands \cite{oct}. This  can explain 
why the experimental 
$(-,1)$ sequence has a  low energy relative to 
the experimental $(-,0)$ but also
relative to the TAC calculation.   

The interpretation of the bundle formed  combining the
\qns E and F with C and D is less straightforward. The \conf [DE]
has a TAC minimum for $\om < 0.25 MeV$. 
The component $J_3 \approx 2$, i. e. it represents the $2^-$ band 
(cf. fig \ref{f:2qna}).
Some small signature splitting is seen in experiment.

 For $\om \ge 0.3 MeV$ we find $\th\approx 80^o$ for  [DE] and [CE]. 
Hence,    the PAC interpretation is applied to the 
whole bundle. Accordingly, the TAC \conf [DE] splits into the PAC
 \confs [DE] and [DF], which have $(\pi,\al)=(-,0)$ and $(-,1)$.
The TAC \conf [CE] splits into the PAC \confs [CE] and [CF],
which have $(-,1)$ and $(-,0)$, giving rise to two
$\De I= 2$ sequences with odd and even spin, respectively.
As seen in fig. \ref{f:2qna}, there is a $(-,0)$ band observed, which can be 
interpreted as [CF]. The odd spin band  [CE] should be nearby. It
is not given in the experimental level scheme, but ref. \cite{hf174} reports
two unplaced $\De I= 2$ sequences. 

 The interpretation of the
low frequency part of [CE] and [CF] has the same problems as 
discussed for the \conf [C] in sect. \ref{s:1qn}. As seen in fig.
\ref{f:Eth2qn-} for $\om=0.2MeV$, [CE] is rather flat. It has a minimum
at $\th=0^o$, which is due to the $K=9/2$ component in C. In addition it has
 a shoulder around $\th=60^o$ due  the $K=5/2$
component, which is hardly visible. However 
 the condition of uniform rotation (\ref{scJintr}) has
solutions $\th=58^o$ and $66^o$ for $\om=0.2~MeV$ and $0.25MeV$, respectively.
These points are included in fig.  \ref{f:2qna}. 
The experimental $(-,0)$ band, which we assign to [CF], is seen to substantial
lower frequency than $\om=0.2$, below which no TAC solution is found.
This is another  example for the limitations of the TAC approach, which 
treats the orientation angle $\th$  as a static variable.
The static approach is expected to work best 
if the function $E(\th)$ is relatively
symmetric around the minimum. Then, 
the fluctuations of $\th$ are also expected to be symmetric and
the contribution linear in $\th-\th_o$ will average out.

 The combinations of the $i_{13/2}$ \qns with the [514]7/2 orbitals
will not be discussed, because the results of the calculations are 
 similar to the ones obtained for the  $5/2^-$
orbitals and there is no experimental information about them. 

The
combinations of the \qn E with G and H, which emanate from the 
pseudo spin singlet [521]1/2, are shown in 
fig. \ref{f:2qng}. 
The situation is analogous to  $^{174}_{71}$Lu$_{103}$, where
the same orbitals combine with the \qprs a and b. 
The decoupled pseudo spin just adds or subtracts one half unit of
\am to the total \am. This means,
\bea \label{pseudo}
E'_{[EG]}=E'_{[E]}-\om/2 +const,\nonumber \\
E'_{[EH]}=E'_{[E]}+\om/2 +const.
\eea   
 The Routhians of [AG] and [AH] are related to the 
Routhian of [A] in a similar way (cf. figs. \ref{f:1qn} and \ref{f:2qng}).
The \confs [AG] and [EG] are observed as the $K^\pi=4^-$ and $3^+$ bands.
It would be interesting to observe the \confs [AH]$3^-$ and [EH]$2^+$
with  the pseudo spin being anti parallely oriented with respect to the 
rotational axis. Then one could  investigate to what extend the 
the residual interaction depends on the pseudo spin orientation.

\subsection{Quasi neutron \confs at a large tilt}\label{s:p8m}
 
The strong coupling estimate for the tilt angle
\mbox{ $\th = \arccos{(K/I)}$}
shows that the number of steps in $I$ needed to change $\th$ from $0^o$(
band head)  to $90^o$ increases with $K$.
Hence, the high-$K$ bands offer  the possibility to
explore the \qp spectrum at an   angle substantially 
different from 0$^o$ and 90$^o$. The bands with low or moderate
$K$, discussed in the preceding sections, only permit a sketchy view, because
the rotational axis  rapidly reorients from the
3- to the 1- axis. For $^{174,175}$Hf there is the family of 
bands
built on the $K^\pi=8^-$ proton \conf [ae]. They permit us studying the
 \qn \confs at a large tilt. These
high-$K$ bands will be discussed in terms of the \qn spectrum for $\th =
45^o$. Although the tilt angle varies along the bands, 
it stays below $60^o$ within the considered frequency range.
 Like  the \qp diagrams for $\th = 90^o$ (upper panel of fig. \ref{f:qn9045})
 give a first  orientation of the structure of the low-$K$ bands,
the  diagram  for $\th = 45^o$ (lower panel of fig. \ref{f:qn9045})
 permits qualitative interpretation of the \qn \confs in
 the [ae] family.    

 Comparing the two panels of fig. \ref{f:qn9045}
 one notices a  substantial reordering of the trajectories.
As a function of $\om$, they cross much more sharply for $\th=45^o$ than 
for $90^o$. This
reflects the  tilt of the rotational axis towards the symmetry axis, where
all levels cross sharply.       
The sharp crossings have the consequence that the zero \qn \conf
[ae]$8^-$ remains unperturbed up to the relatively high frequency of
$\om \approx 0.4 MeV$, where
the \qn A$^+$ interacts with the \qn C'. 
This is confirmed by the full TAC calculation
and the experiment shown in fig. \ref{f:p8m2qn}.  
The trajectory A$^+$ crosses  with a number of \qn trajectories  
before it interacts with C'. This means that there are several two \qn
\confs below [ae]$8^-$.   
The lowest of these, [aeAE],    is seen in $^{174}$Hf as the $14^+$ band. 
The lowest \conf of the opposite parity, [aeAB]$16^-$, has not yet been observed.
It is the \qn t - configuration, which is shifted to low energy by the
two DAL \qprs a and e.  Fig. \ref{f:p8m2qn} shows that
with increasing $\om$ it interchanges its character with [aeAC]. This
is the manifestation of the quasi crossing between B and C  seen in 
fig. \ref{f:qn} at $\th=65^o$.
 The contrast to the \qn spectrum 
at $\th \approx 90^o$ is noted. There, the  
configuration [AB] interchanges gradually character with the vacuum [0].
For substantially smaller $\th$, [AB] has a t - structure which is very
different from [0] ($J_3 \approx 8$ and 0, respectively) and, as a
consequence, couples only very weakly to [0]. 

The $18^+$ band [aeAEGI] is predicted too high relative to [aeAE]$14^+$.
As discussed in section \ref{s:p8m0}, the discrepancy is probably due   
a substantial reduction of the pair field by blocking the two \qns G and I.

The quasi crossing of  A$^+$  with  C', seen $\om \approx 0.4 MeV$
in the lower panel of fig. \ref{f:qn9045}, causes the down bend  
of the zero \qn \conf
[ae]$8^-$ seen at $\om_c=0.38~MeV$ in fig.  \ref{f:p8m2qn}. 
It is the AB crossing at the  tilt angle of $\th_o=60^o$. It
is delayed   as compared to the yrast band
where it appears at $\om_c=0.30~MeV$. Only part of the delay can be explained by 
geometry within the standard PAC scheme:
The projection of the angular frequency  on the 1 - axis is
 $\om_{c1}= \om_c \sin \th = \om_{c} \sin 60^o =  0.33~MeV$, which is 
larger than $\om_1=0.30~MeV$,
where the AB crossing appears in PAC. 

Let us now discuss the [ae] family  in the $N=103$ system
by means of the $\th=45^o$ \qn diagram (lower panel of \ref{f:qn9045}).
The one  \qn \conf  [aeA]$23/2^-$ is  lowest.
It is observed at about the right energy in $^{175}$Hf as the $23/2^-$ band. 
Since E lies above A  
the \conf [aeE] is expected at higher energy.  This confirmed by the full TAC
calculation shown  in  fig. \ref{f:p8m1qn}.  
In the $\th=45^o$ diagram \ref{f:qn9045},
the two \qn excitations EB and EI have negative energy above $\om =0.4 MeV$.
Fig. \ref{f:p8m1qn} demonstrates that after
 minimizing with respect to $\th$ the \confs [aeABE]$39/2^+$  and
[aeAEI]$35/2^-$ lie below [aeA]$23/2^-$. This  order of the
bands is seen in   $^{175}$Hf. 

Hence, the experimental high $K$ spectra clearly reflect
 the modification of the 
\qn spectrum with decreasing $\th$.

\subsection{Unpaired neutron \confs}\label{s:p8m0}
 
In this subsection we discuss, how to interpret the high-$K$ bands of the
[ae] family in terms of \confs of the unpaired neutron system.
Fig. \ref{f:sn45} shows the single neutron Routhians at 
$\th =45^o$. The labels of the states are chosen such that they are as close as
possible to the \qn neutron labels in fig. \ref{f:qn9045}.  
We consider  $N=102$.
 In order to keep the notation similar to the case of finite pairing, 
let us denote by [0] the yrast
\conf for $\om < 0.1 ~MeV$ ( below the crossing between the 
levels A and E ).
At $\th=90^o$ it 
  has the character of the s - \conf and at $\th=0^o$ it is the
 ground state. The particle hole excitations are constructed 
relative to this \conf [ae0]$8^-$. 
Above the AE - crossing, the yrast \conf is [aeA$^{-1}$E]$14^+$. 
The next higher \conf
is [aeA$^{-1}$G$^{-1}$EI]$18^+$. They form the two lowest bands of the [ae]
family. Both are shown in fig. \ref{f:p8m2sn} as results of the full
TAC calculation. The relative position and slopes compare well with the 
experiment shown in the upper panel of fig. \ref{f:p8m2qn}. Close by 
there is [aeA$^{-1}$G$^{-1}$BE]$19^-$, which has not yet been observed. 
Above $\om=0.35~MeV$ one expects  a number of \confs with higher $K$,
generated by lifting a neutron from the levels $9/2^-$ and C to
I and B. One example is [ae A$^{-1}$C$^{-1}$EI]$20^+$.

The unpaired \confs [aeA$^{-1}$E]$14^+$ and 
 [aeA$^{-1}$G$^{-1}$EI]$18^+$ 
are also shown in fig. \ref{f:p8m2qn} for a comparison with the calculation at 
finite neutron pairing.
There they are labeled in terms of the 
\qp notation as [aeAE]$14^+$ and [aeAEGI]$18^+$, respectively. The unpaired
\conf [aeA$^{-1}$E]$18^+$ lies significantly below the paired one. 
In this \conf   four \qns are blocked and the
description as unpaired neutron state is better (within the HFB scheme to which
we restrict here).   For  [aeAE]$14^+$, the paired calculation is 
favored at low and the unpaired at high $\om$.
The zero pairing calculation compares better with the experiment.
The zero pairing Routhians of both \confs lie too low as compared to the
yrast band [0]. This may be a consequence of the 
dynamical pair correlations which are not taken into account. 

Fig. \ref{f:bm1hf174} shows the branching ratios for
two bands [aeAE]$14^+$ and [aeAEGI]$18^+$. 	
In case of the $K^\pi=14^+$ band,
both the paired and the unpaired calculation give similar results.
In case of the $18^+$ band, the unpaired calculation shows a similar
 increase at low frequency as the experiment, however it underestimates
the experimental ratio. We have to underline here that the microscopic 
calculation of the magnetic transition probabilities by means of 
(\ref{bm1}, \ref{mu}) is expected to be less accurate than the popular
strong coupling estimates (if applicable),
 which are based on \qp g - factors that
are adjusted to the experiment.

The unpaired \conf [ae0]$8^-$ lies slightly 
below the paired \conf [ae]$8^-$ after
the latter has bend down. It has the character of the neutron s - \conf. 
That is the pairfield becomes small when  the pair of $i_{13/2}$ \qns
is broken in the AB crossing.  

The unpaired \conf [aeA$^{-1}$B]$8^-$ lies above the two paired bands 
[aeAC] and [aeAB], which interchange character. Thus the interpretation
 in terms of a \qp structure is favored. It will be interesting to see      
if the experimental bands shows the paired or unpaired pattern, which are
markedly different.

As seen in fig. \ref{f:sn45}, the lowest \conf in the $N=103$ system is 
[ae A$^{-1}$EI]$35/2^-$, which is shown in fig. \ref{f:p8m1qn} as  
[aeAEI]$35/2^-, \De=0$. It lies below the paired \conf [aeAEI]. This is 
expected because three \qns are blocked, destroying the static pair field.
The relatively high band head frequency  in experiment is consistent with the
zero pairing calculation. In the calculation with  finite pairing the band 
starts at a much lower
 frequency.  The \conf [ae A$^{-1}$EF]$23/2^-$ is shown in fig. \ref{f:p8m1qn} as  
[aeA]$23/2^-, \De=0$. It lies above the paired band [aeA]$23/2^-$ for most
of the frequency range. 
Finite pairing is favored for the one \qn band.
 The zero pairing solution wins only at high frequency where a band crossing
 (down bend) is encountered.
The band  [aeABI]$39/2^+$ corresponds in 
the unpaired scheme  to [ae A$^{-1}$BI]$39/2^+$. This \conf, shown as
[aeABI]$39/2^+, \De=0$ in fig. \ref{f:p8m1qn}, lies {\em above}  the finite
 pairing calculation. This somewhat surprising result (static pairing for the 
three  \qn \conf) is understood from fig. \ref{f:sn45}. In contrast to 
[ae A$^{-1}$EI]$35/2^-$, where the EF is blocked for  pair scattering,
for  [ae A$^{-1}$BI]$39/2^+$ the pair BB' blocked. Since BB' is
much further away from the Fermi - surface than EF, blocking is much  
less effective.

For higher frequency it becomes favorable to generate \am by exciting
 \qprs. As seen in fig. \ref{f:qp}, the next pair at large tilt is [cg].
In fact, the \conf [aecgAEGI]21$^-$ appears  at the yrast line. In the TAC 
calculation the paired four \qpr \conf [aecg] has a lower energy than the 
unpaired.
It is not unexpected that the TAC calculation gives a too high energy,
because already the one \qp \conf [g] is predicted too high by TAC (cf. sect.
\ref{s:1qp}).

\section{Rules for TAC}\label{s:rules}

Let us summarize  the  experience gained in applying the TAC approach
to the analysis of rotational spectra in 
axial well deformed nuclei. It is assumed
that for $\th=0$ the rotational axis coincides with the symmetry axis.
We shall use the form of  rules.
\begin{enumerate}
\item In order to construct the configurations use the particle - hole scheme
in the case of zero pairing. For finite pairing use the \qp occupation scheme.
The \qp levels appear in  pairs of conjugate levels, one of which is
occupied and the other is empty. Only for $\th=90^o$
the conjugate levels  have opposite signature.
\item In order to check if a  \conf corresponds  to even or odd
 particle number  trace it  diabatically back to low $\om$, where
 gap between the \qp levels of positive and negative energy exists.
\item  In searching the equilibrium orientation
$\th_o$ try to stay within structurally the same configuration. Use diabatic
tracing. 
\item If there are avoided crossings between the levels, diabatic tracing
may end up in an unwanted configuration. Usually this shows
up as an  irregularity of the calculated quantities as functions of $\om$.
In such cases one has to resort to the \qp diagrams
and manually reassign the desired \conf.
\item Reduce the chance of unwanted    \conf changes  by 
choosing the start angle $\th_s$ of the diabatic 
tracing    close a to the expected equilibrium
angles $\th_o$.
Draw the \qp diagram $e'_i(\om,\th_s)$.
Never start diabatic tracing at $\th_s=90^o$, use $85^o$.
\item The crossings between \qp \confs represent crossings between real 
bands, but the description of the mixing region itself is incorrect.
\item Rotational bands correspond to a function $J(\om)$ which
 increases with $\om$.
So long as $\th_o=0$ the band has not  started yet. The band head lies at
the frequency where  a minimum at $\th_0\not=0$ appears.
\item Solutions with $\th_o<80^o$ are of the TAC type. They represent $\De I=1$
 bands. The two signature partners $(\pi, ~\al)$ and $(\pi, \al +1)$ are
degenerate. 
\item   Solutions with $\th_o>80^o$ are of the PAC type. 
They represent $\De I=2$ bands of given $(\pi,~\al)$, i. e.
 $I=\al~mod ~2$.      
The signature $\al$ is given by its value at $\th=90^o$.
\item When  the tilt angle $\th_o(\om)$
becomes larger than
 $80^o$ the  change from the TAC to the PAC interpretation results in
 unphysical jumps of the energy distance between signature partners  and
of other quantities. The experimental quantities show a gradual transition 
between  the two  cases, which cannot be calculated by TAC.
\item Since the number of $\De I=2$ bands of given $(\pi,~\al)$ must be the 
same in the PAC and TAC interpretation, one half of the TAC \confs
is spurious. 
\item Each \conf in the region $\th<90^o$ 
 has its mirror image $E'(\pi-\th)= E'(\th)$ in the region $\th>90^o$.
The diabatic continuation of the mirror image of an adopted configuration
into the region $\th<90^o$ is spurious and must be discarded.
\item The spurious \confs   have  minima or kinks only at $\th=90^o$.
 If such a \conf has another  minimum for $\th<80^o$ this must be considered 
as physical.  Then \conf has changed its 
character with $\th$,  being no longer spurious.
\item For a strong tilt ($0^o \ll \th_o \ll 90^o$), the spurious \conf are usually high
in energy and  do not interfere.
\item In the case of  multi \qp \confs there are  bundles of \confs emanating
from $\th=90^0$, each of which has  its own $\th_o$.
Only when lowest $\th_o$  has reached $80^o$
one must  change from the TAC to the PAC interpretation 
simultaneously for the whole bundle. 
\item The  intra band matrix elements 
 of the electromagnetic operators can be calculated. PAC solutions provide 
their signature dependence.  TAC solutions 
give only the  average over both signatures.     
\end{enumerate}
  
The rules can  be applied to triaxial nuclei with  few obvious
modifications. Rule 5 must be complemented by: Do not use $\th_s=0^o$, start
at $5^o$.   
Only the first, general  statement of rule 7 remains valid. 
The second which assumes that  
 $\th=0$ is a  symmetry axis does not apply.
The low frequency behavior of triaxial nuclei  has not yet been studied, 
except the investigation of a model case in \cite{mftr}. 
PAC solutions are possible for all three principal axes and 
TAC solutions are possible in all   three planes
spanned by the principal axes. Accordingly one must extend the search.
The simplest way is  letting $\gamma$ vary from 120$^o$ to -60$^o$ 
(cf. tab. 1). If the rotational axis does not lie in one of the three planes
spanned by the principal axes the rotation becomes chiral.  
The consequences of this 3D geometry are
discussed in \cite{rmp} and \cite{mftr}.

\section{Conclusions}\label{s:conc}

The semiclassic concept of uniform rotation about an axis that is 
tilted with respect to the principal axes of the deformed density distribution
leads to a mean field theory which describes energies
and intra band electro-magnetic matrix elements of
 $\De I =1 $ bands in a quantitative way. The orientation of the rotational
axis turns out to be as good a collective degree of freedom as 
the familiar shape degrees of freedom are.  
The tendency of high spin particles to align with the  {\em rotational} axis, 
which in general does not have the direction of one of the principal axes of
the deformed mean field, is a concept that permits to explain    
many features
of high-$K$  bands from a new  perspective.

The tilted solutions do not  have
the familiar $C_2$ symmetry, which appears when 
 rotational axis coincides with one of the  principal axes.
The lower symmetry results in the  loss of the signature
quantum number, which manifests itself by the appearance of one $\Delta I
=1$ band instead 
 two separate $\De I=2$ sequences of opposite signature. The  
transverse magnetic dipole moment, which determines the rate of 
magnetic dipole transitions, plays the role of an
order parameter. For tilted solutions it has a finite expectation value,
which
  may become quite large, because it is the sum
of contributions of several \qps. For rotation about a principal axis the 
expectation value of the  transverse magnetic dipole moment is zero. The
magnetic transition probability is given by a matrix element between
two different \qp \confs, which is of single particle order.   

Being a mean field approach,
tilted axis cranking is not capable of describing
the transition from a tilted to a principal axis solution
in a correct way,
because this involves the transition from a broken to a restored 
discrete symmetry. The signature dependence of the energy and other quantities
appears in a sudden, unphysical way when switching from the broken symmetry
to the conserved symmetry interpretation. Still, one can  guess from the 
calculations at which rotational frequency
 the signature effects are expected and how strong they 
should be.  

The  breaking of the $C_2$ symmetry leads to the appearance of
spurious states. An elimination  method is suggested. After applying it
the calculated sequence of  
the first  bands  above the yrast line which agrees with the observed one
for the studied examples. No spurious  states remain in the near yrast
region.  

It is the strength of the tilted axis cranking approach that
treating  many excited  \qps is no more complicated 
than treating few. In order to demonstrate the application of the method
we  studied \confs with up to four excited \qprs and four excited \qns
in the nuclides with $N=102,103$ and $Z=71,72,73$. The calculated energies
and branching ratios agree with the experimental values within an accuracy
that is 
typical for microscopic mean field calculations. In particular, it is found
that the order and structure of the high-$K$ bands can be qualitatively
understood in terms  configurations  constructed from  \qp levels, which are
calculated as functions of the rotational frequency $\om$ at a fixed tilt
angle of about $45^o$. 
    
The regime of quenched static pairing is encountered in the multi \qp bands
of high seniority. Since the change of  the pair field is not at the focus
of this paper, it was treated in a rough way by comparing the paired
with unpaired solutions and choosing the one with the lower energy.
This schematic approach turned out quite practicable for a 
first analysis of the high-$K$ band structure. A  refined description 
of pairing within the tilted cranking model, which includes dynamical  
pair correlations, will be given elsewhere \cite{almehet}.     

The study of high-$K$ bands at the largest frequencies attainable is a 
interesting problem of nuclear physics.
 More systematic
  studies than the present are expected to reveal the response of the single
particle levels to the tilt of the rotational axis. 
A particularly  interesting  question is
how the $K$ quantum number is eroded with increasing rotational frequency. 
Tilted axis cranking is the proper 
mean field theory to address this question. It is also the appropriate   
starting point for theories that go beyond the mean field, like RPA.

\section{Acknowledgment}
 
I should like to thank F. D\"onau and Jing-ye Zhang for carefully reading
the manuscript.  The work was
partially carried out under the Grant DE-FG02-95ER40934.

\newpage
\onecolumn

\begin{figure}
\mbox{\psfig{file=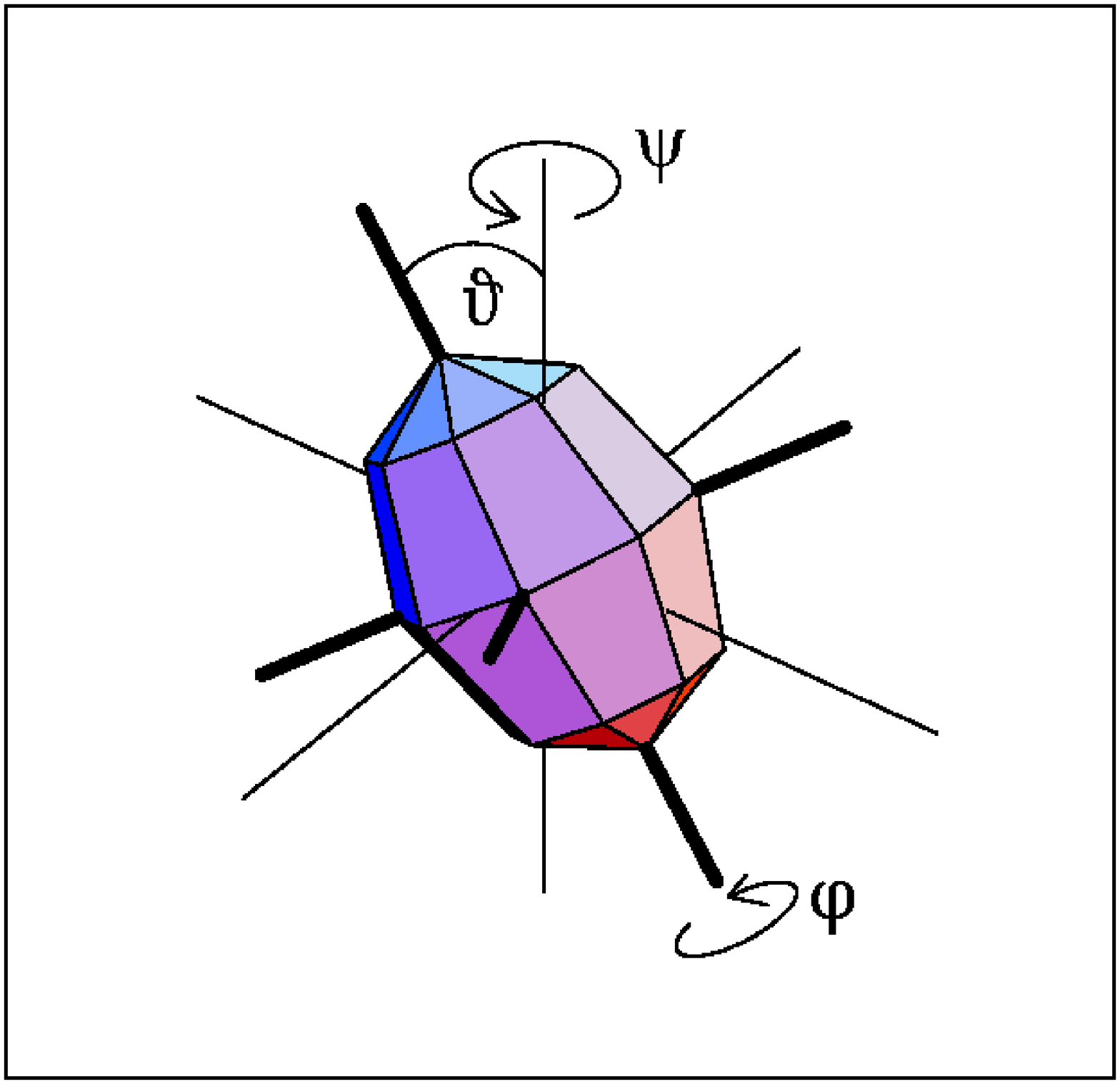,width=16cm}}
\caption{\label{f:angles} The Euler angles specifying the orientation of
the triaxial density distribution in the laboratory frame. A polyeder
shape is shown, which makes the geometry better visible. The principal axes
1, 2, 3 are fat and the laboratory axes x, y, z are thin.} 
\end{figure}

\begin{figure}
\mbox{\psfig{file=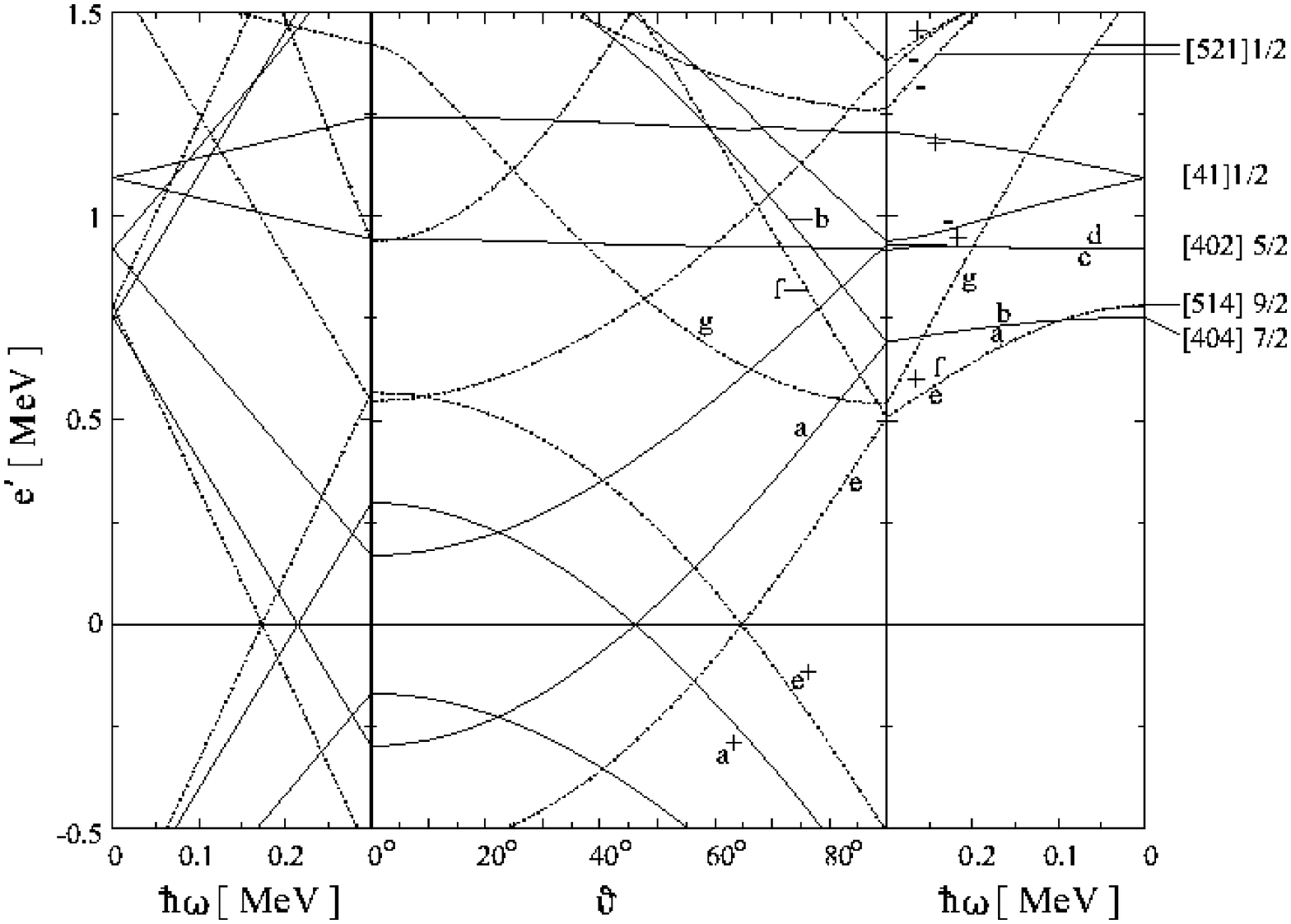,width=18cm}}
\caption{\label{f:qp}
Quasi proton energies for $\eps=0.258,~\eps_4=0.034, ~\De=0.75 MeV$.
The chemical potential $\la$ is adjusted to have $<Z>=72$ at $\om=0$.
Full lines: positive parity, dashed-dotted lines: negative parity. For $\th=90^o$
the signature is indicated by $\pm$ standing for $\al=\pm 1/2$.}
\end{figure}
\begin{figure}
\mbox{\psfig{file=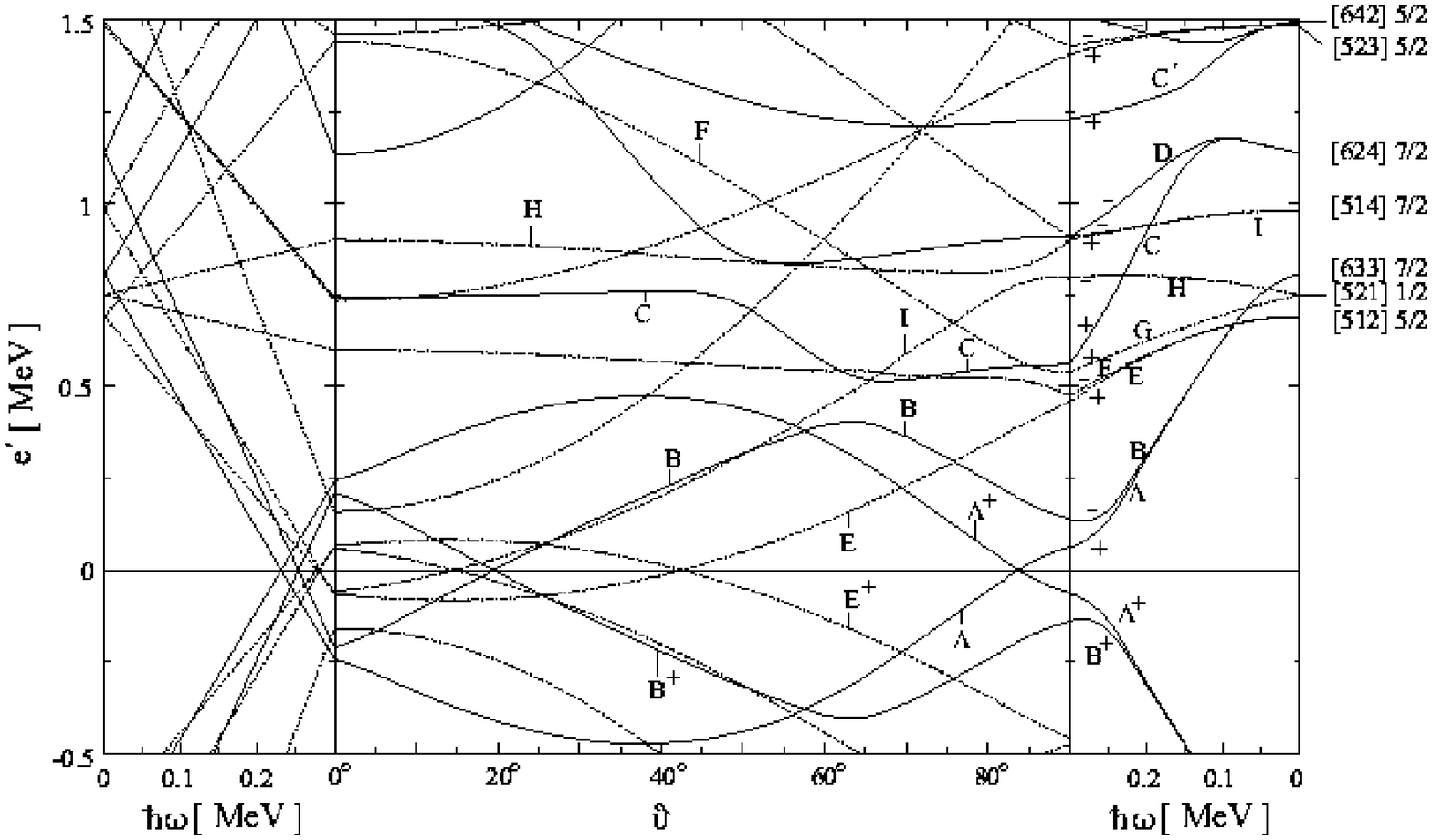,width=18cm}}
\caption{\label{f:qn}
Quasi neutron energies for $\eps=0.258,~\eps_4=0.034, ~\De=0.69 MeV$.
The chemical potential $\la$ is adjusted to have $<N>=102$ at $\om=0$.
Full lines: positive parity, dashed-dotted lines: negative parity. For $\th=90^o$
the signature is indicated by $\pm$ standing for $\al=\pm 1/2$.}
\end{figure}

\begin{figure}
\mbox{\psfig{file=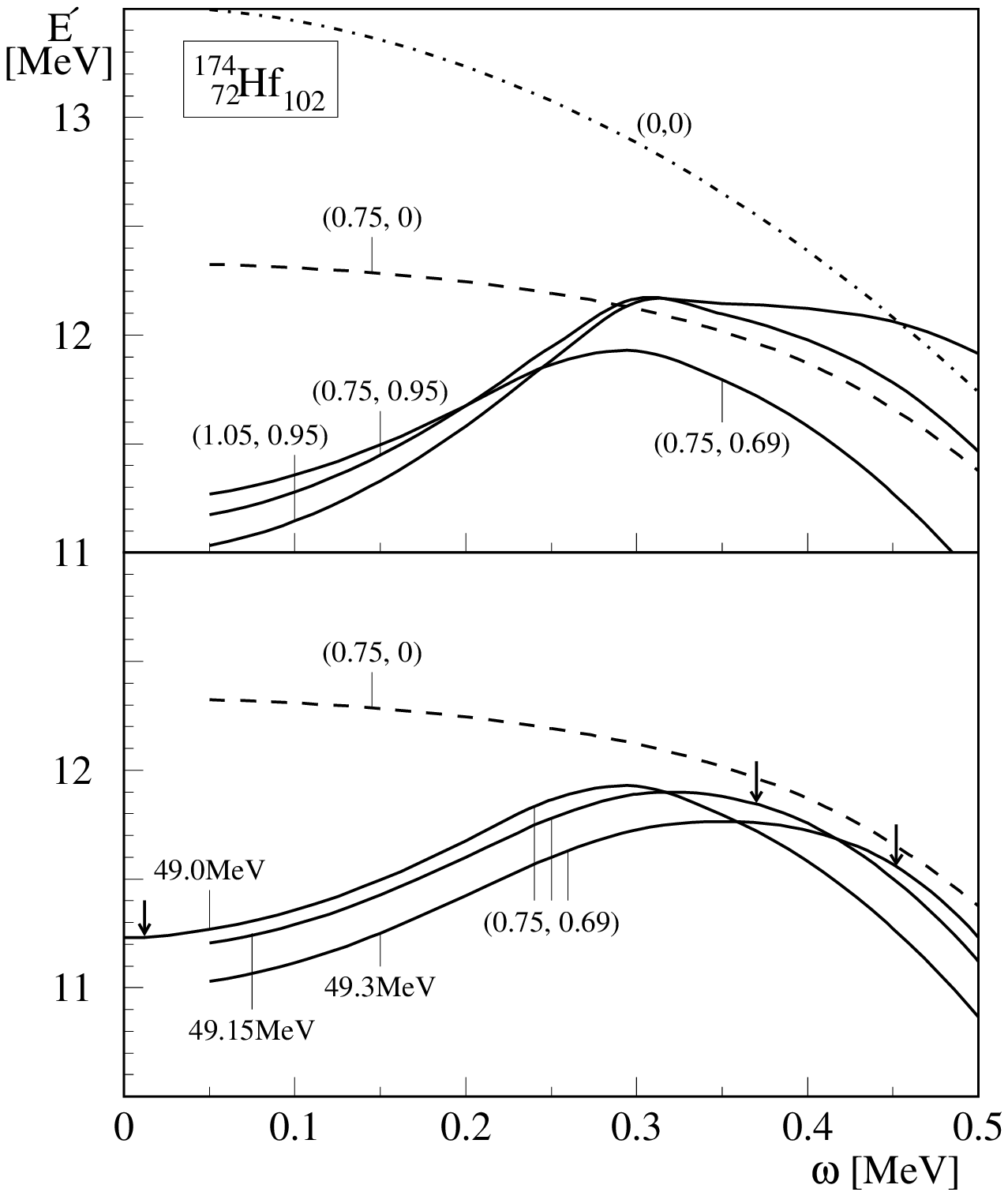,width=14cm}}
\caption{\label{f:dela}
Routhians of the configuration [0] for different values of the pair gaps
(upper panel) and of the chemical potential $\la$ (lower panel).   
The quantity  $E'+\la_pZ+\la_nN$ is displayed.
The pairs of numbers give $(\De_p,\De_n)$  in $MeV$. The values  
$\la_p=43.95~MeV$ and  $\la_n=49.0~MeV$, 
corresponding to $<Z>=72$ and $<N>=102$ for $\om=0$, 
are used in the upper panel. In the lower panel $\la_p=43.95~MeV$ and
$\la_n$ is explicitly indicated in $MeV$. The arrows show at
which $\om$ the condition
$<N>=102$ is satisfied.} 
\end{figure}

 \begin{figure}
\mbox{\psfig{file=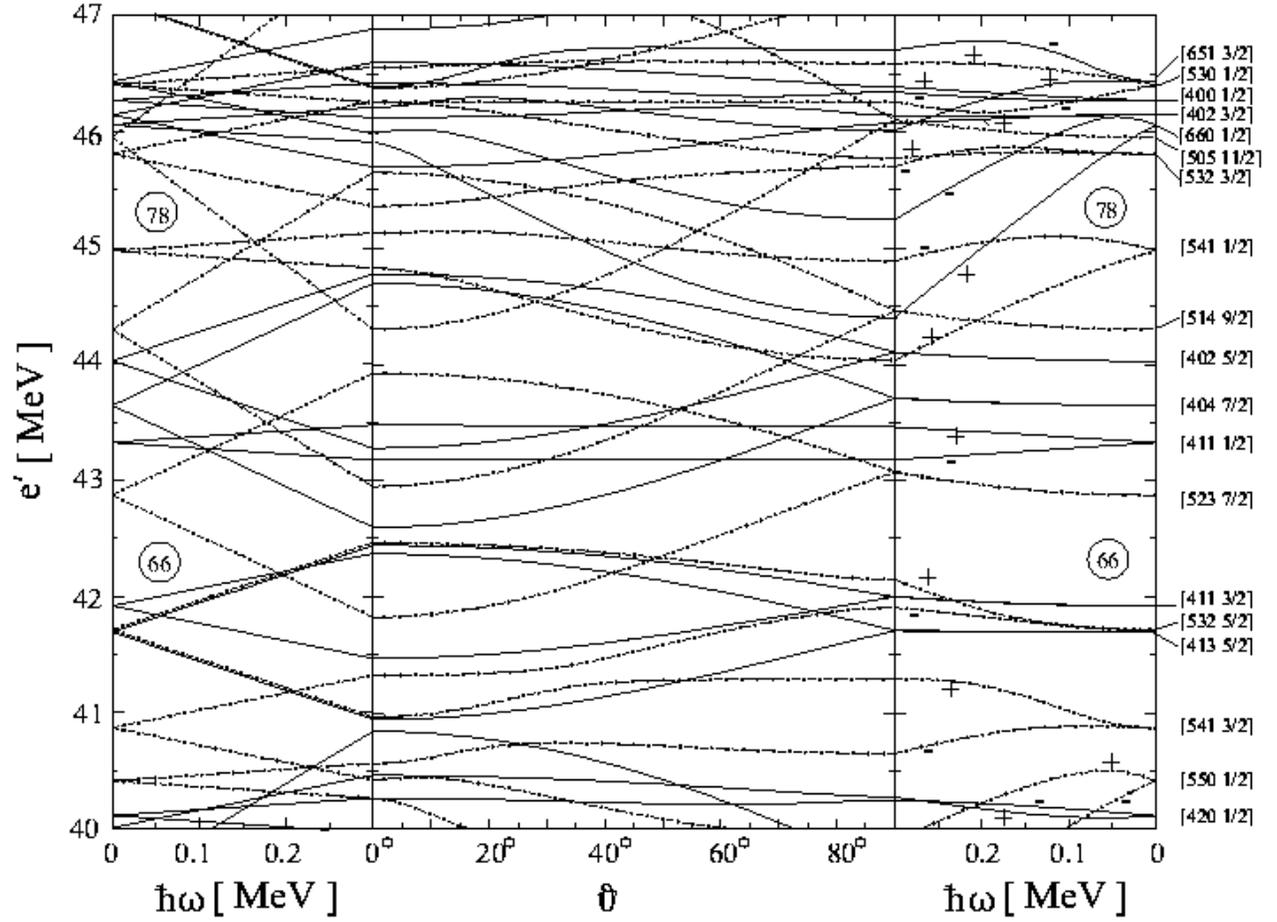,width=18cm}}
\caption{\label{f:sp3}
Single proton energies at intermediate frequency for $\eps=0.24,~\eps_4=0$.
Full lines: positive parity, dashed-dotted lines: negative parity. For $\th=90^o$
the signature is indicated by $\pm$ standing for $\al=\pm 1/2$.
On the right hand side the Nilsson labels are given, which are
relevant for $\om=0$. }
\end{figure}

 \begin{figure}
\mbox{\psfig{file=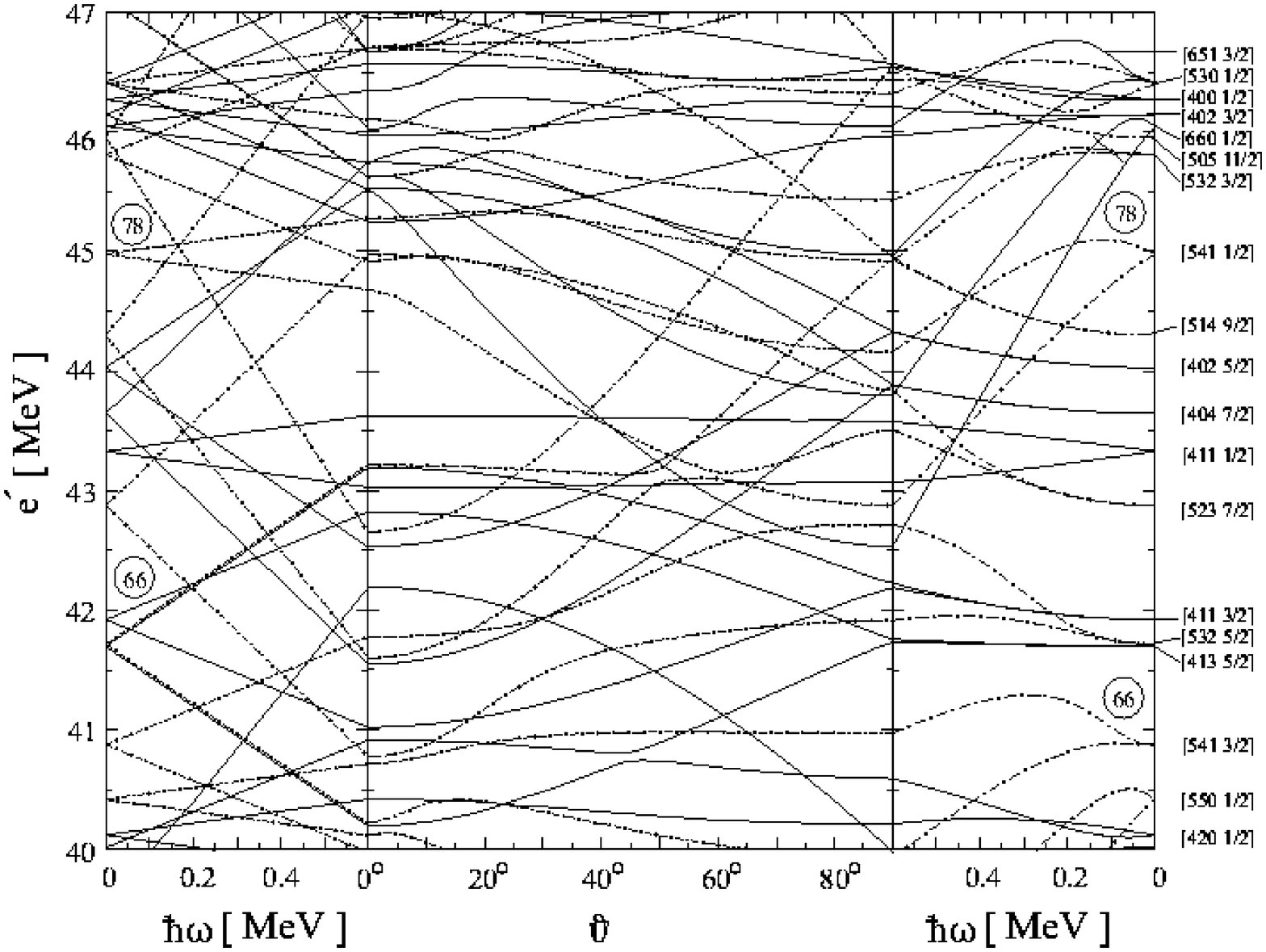,width=18cm}}
\caption{\label{f:sp6}
Single proton energies at high frequency for $\eps=0.24,~\eps_4=0$.
Full lines: positive parity, dashed-dotted lines: negative parity.}
\end{figure}
\newpage

\begin{figure}
\mbox{\psfig{file=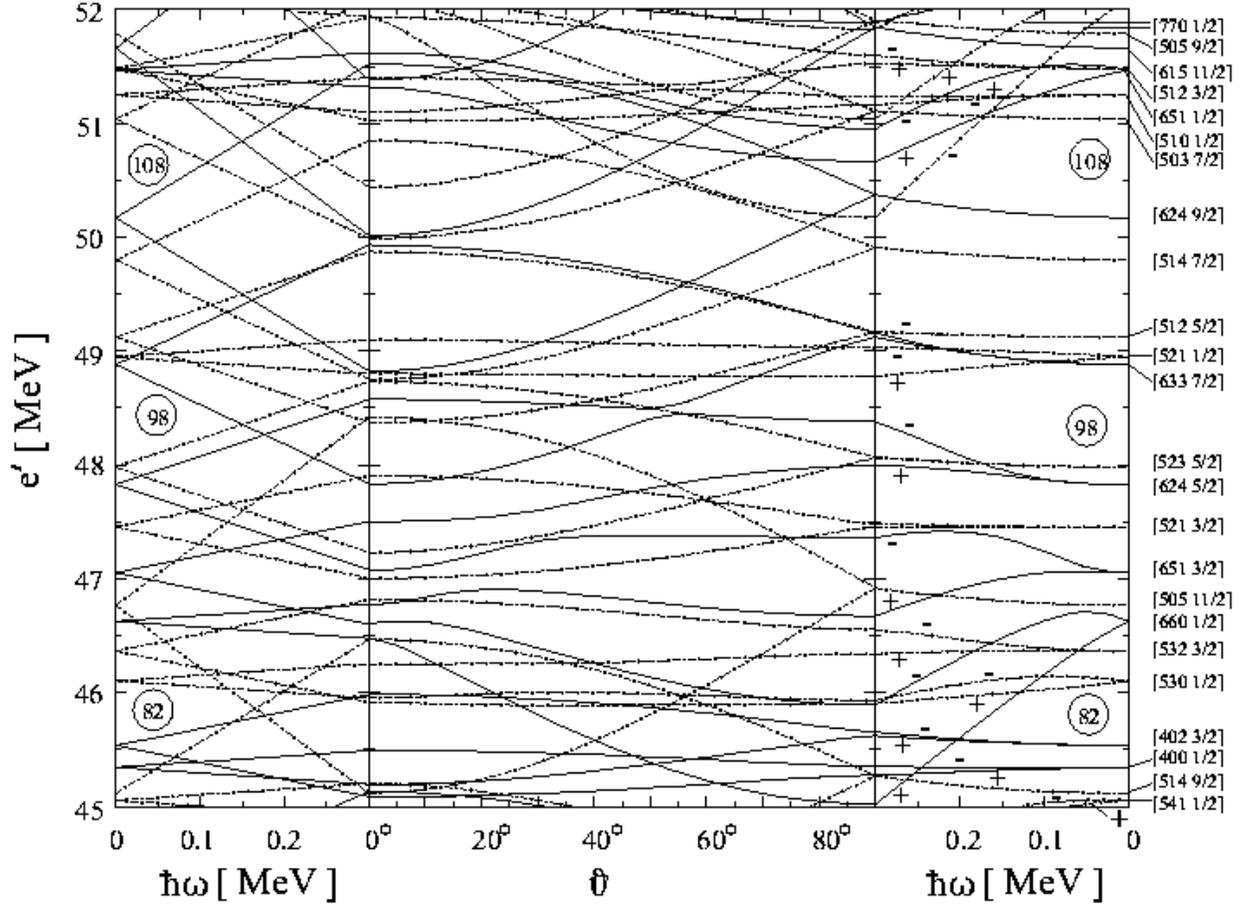,width=18cm}}
\caption{\label{f:sn3}
Single neutron energies at intermediate frequency for $\eps=0.24,~\eps_4=0$.
Full lines: positive parity, dashed-dotted lines: negative parity. For $\th=90^o$
the signature is indicated by $\pm$ standing for $\al=\pm 1/2$.}
\end{figure}
\newpage
\begin{figure}
\mbox{\psfig{file=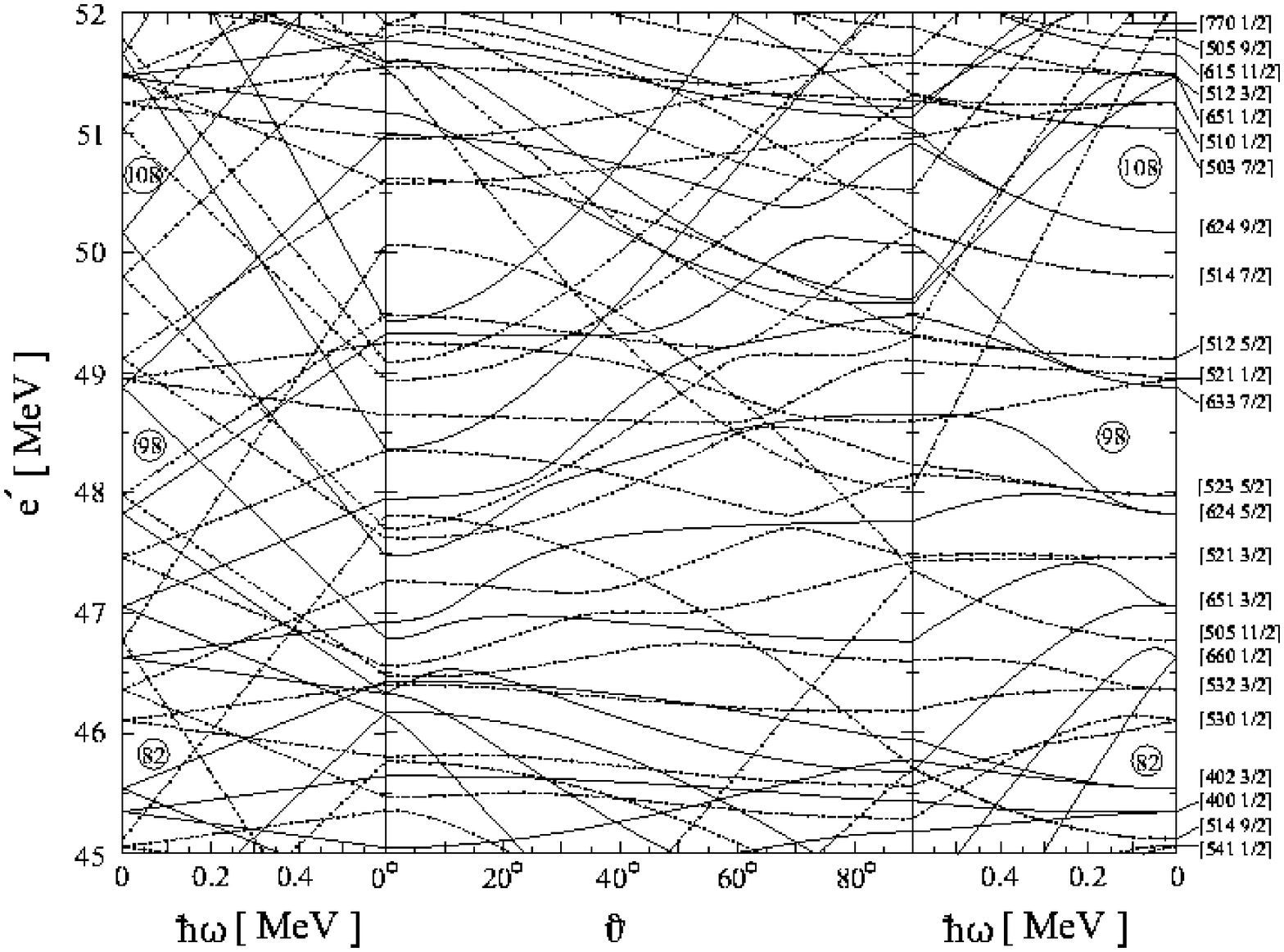,width=18cm}}
\caption{\label{f:sn6}
Single neutron energies at high frequency for $\eps=0.24,~\eps_4=0$.
Full lines: positive parity, dashed-dotted: lines negative parity.}
\end{figure}
\newpage

\begin{figure}
\mbox{\psfig{file=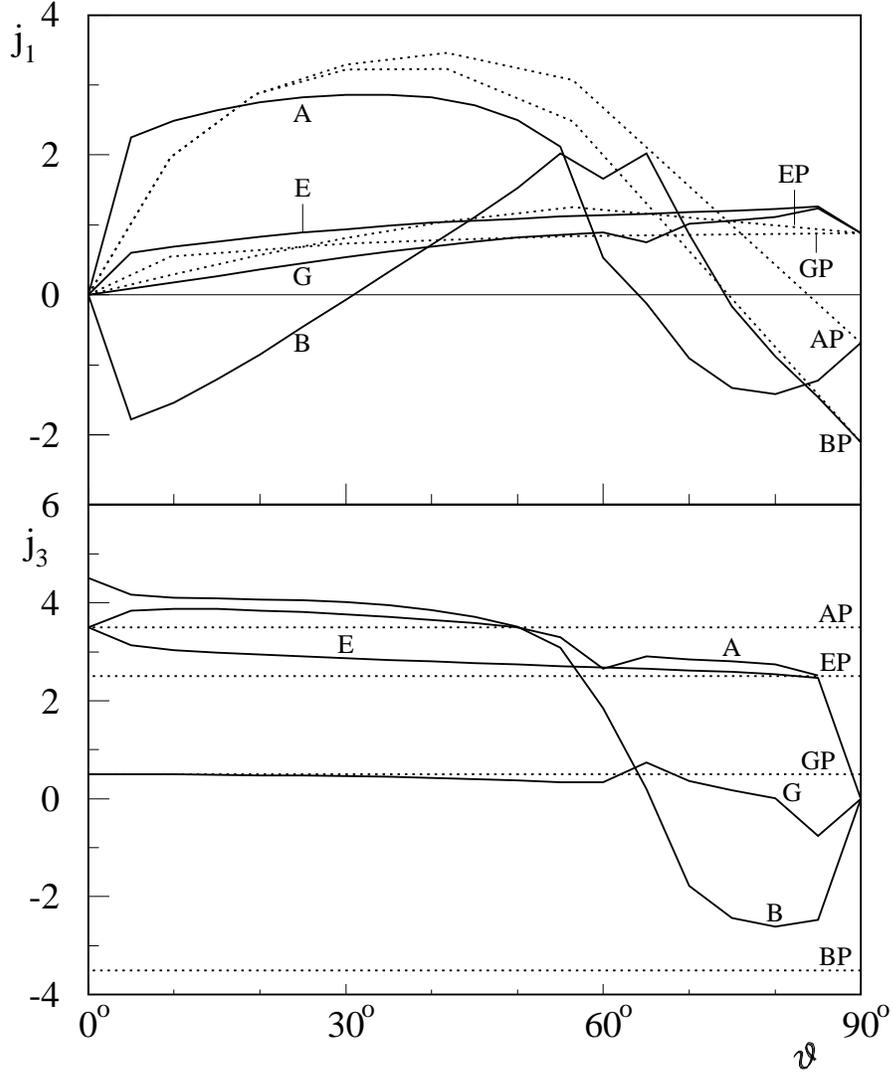,width=14cm}}
\caption{\label{f:i1i3th}
Angular momentum components of  different \qns as
function of the tilt angle $\th$ at $\om=0.3~MeV$.
They are labeled by A, B, ... as in fig.
\protect\ref{f:qn}. The dotted lines  show 
the values $j_3=K$ and $j_1=j_1( \om=0.3~MeV\sin \th, \th=90^o$),
which are used in the standard CSM treatment of bands with finite
$K$. They are labeled by AP, BP, ... The parameters of the calculation are
same as in fig. \protect \ref{f:qp} and \protect \ref{f:qn}.}
\end{figure}

\begin{figure}
\mbox{\psfig{file=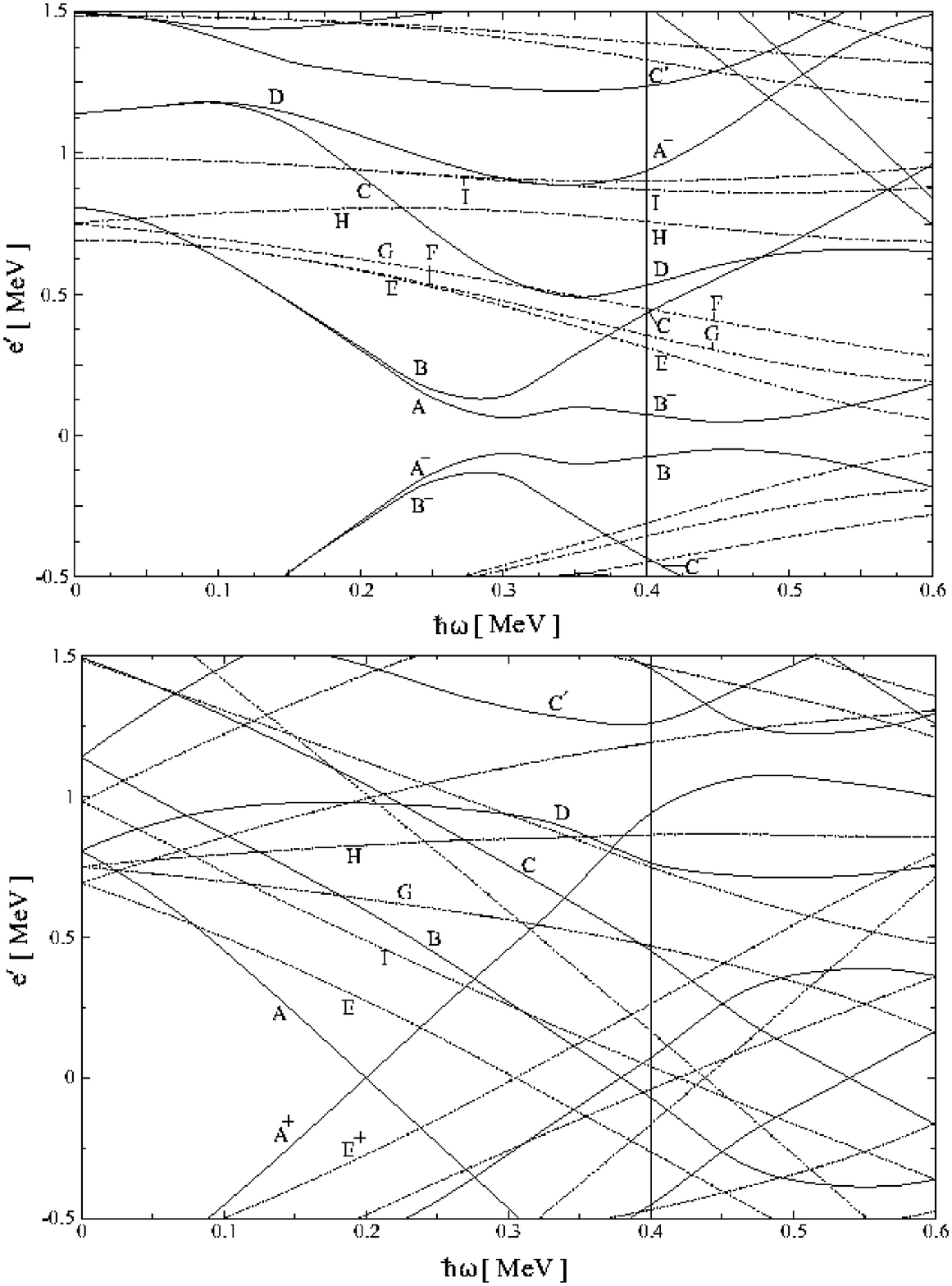,width=16cm}}
\caption{\label{f:qn9045}
 Quasi neutron energies for $N\approx 102$ and $\th=90^o$ (upper panel)
and $\th=45^o$ (lower panel). 
The parameters are $\eps=0.258,~\eps_4=0.034,
~\De=0.69 MeV$.
The chemical potential $\la$ is adjusted to $<N>=102$ at $\om=0$ .}
\end{figure}

\begin{figure}
\mbox{\psfig{file=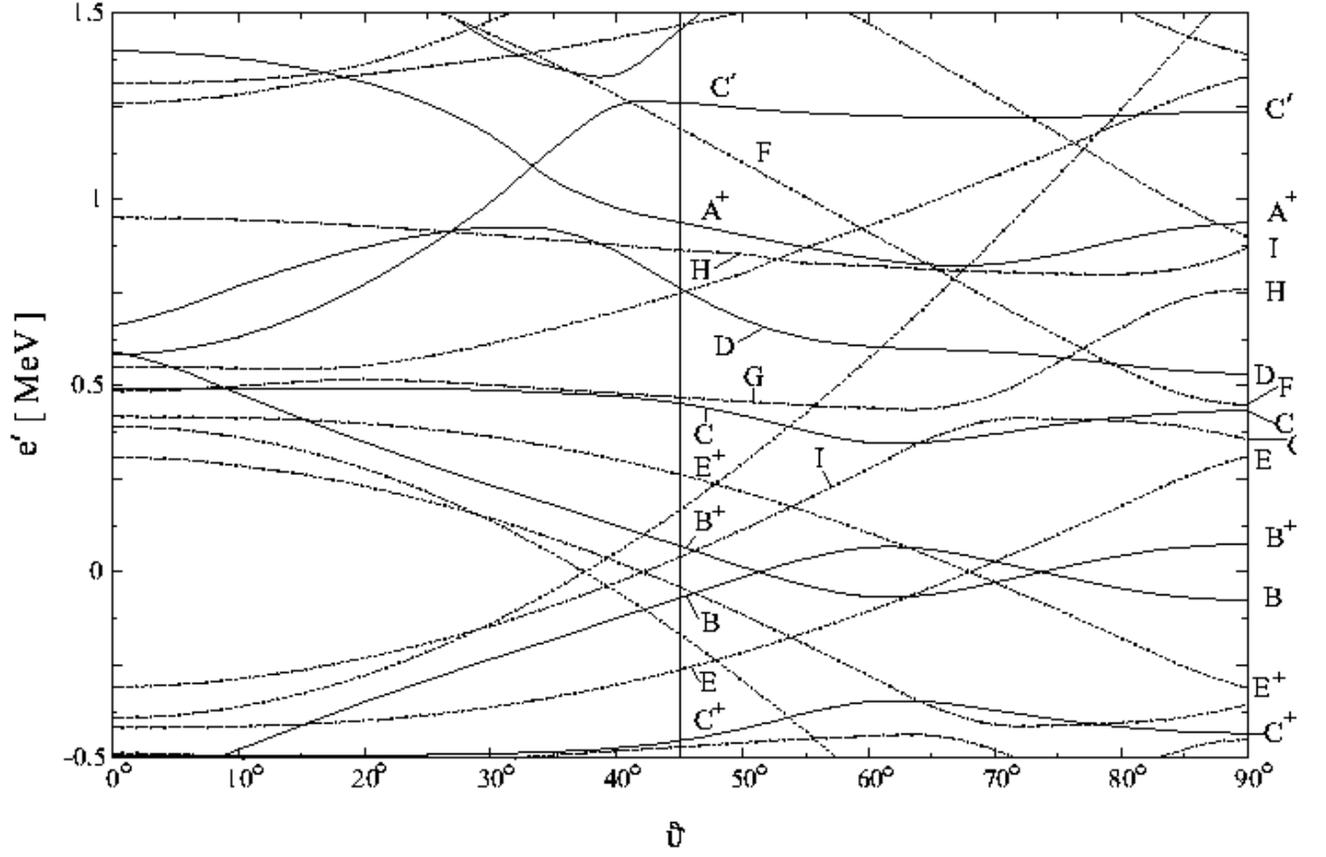,width=18cm}}
\caption{\label{f:qnthom4}
Quasi neutron energies for $\eps=0.258,~\eps_4=0.034, ~\De=0.69 MeV$
and $\om=0.4~MeV$.
The chemical potential $\la$ is adjusted to have $<N>=102$ at $\om=0$.
Full lines: positive parity, dashed-dotted lines: negative parity.
}
\end{figure}

\begin{figure}
\mbox{\psfig{file=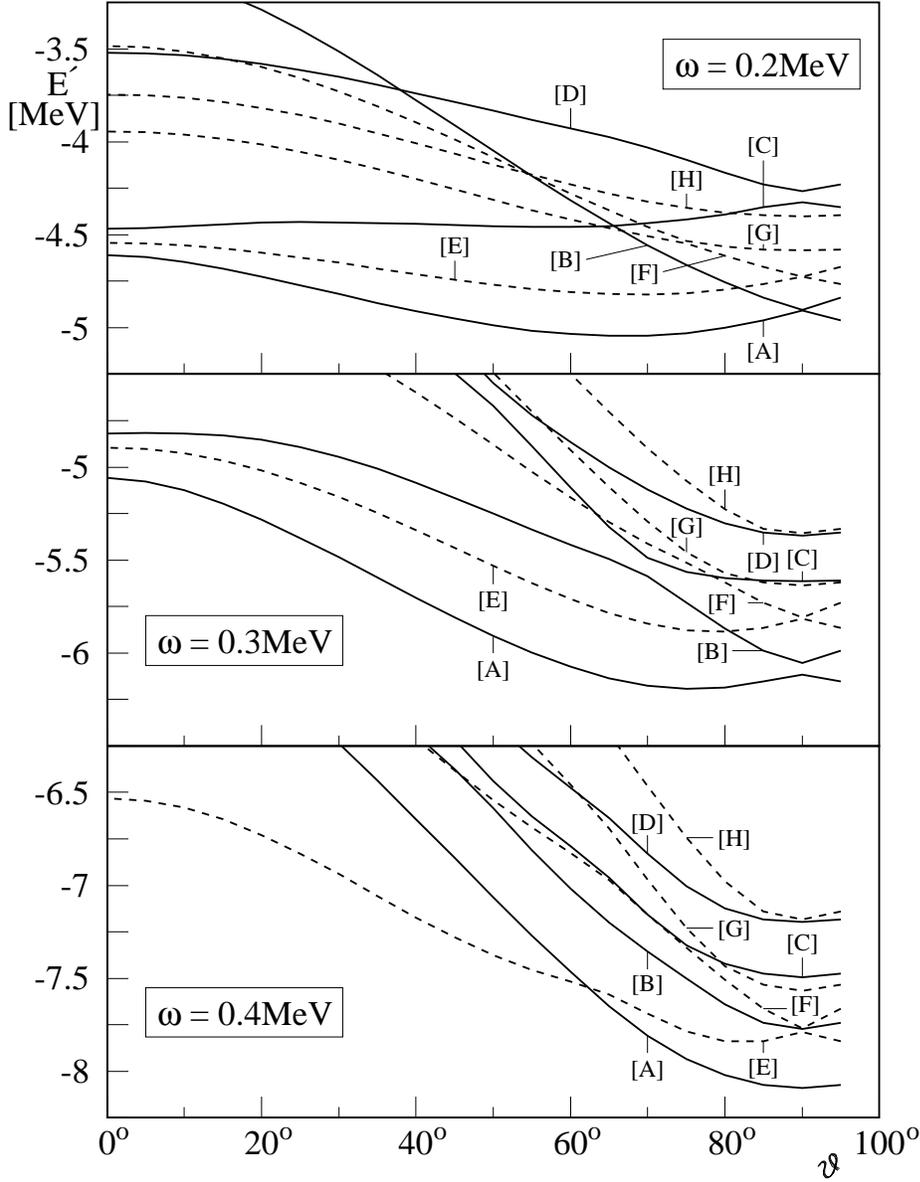,width=14cm}}
\caption{\label{f:Eth1qn}
Total Routhian as function of the tilt angle for the one quasi neutron
configurations in $^{175}_{72}$Hf$_{103}$. }
\end{figure}

\bef
\mbox{\psfig{file=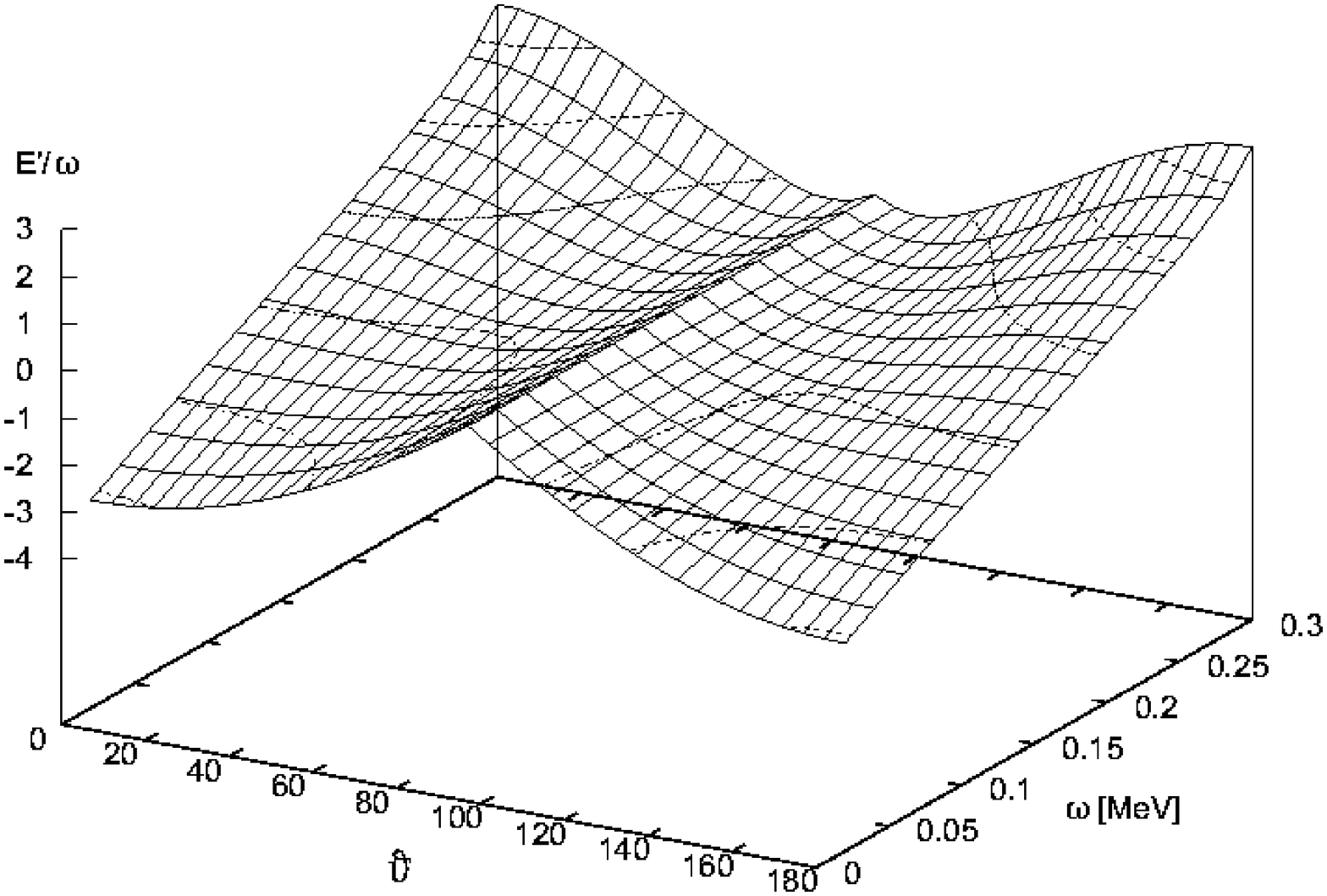,width=18cm}}
\caption{\label{f:Ethe}
Total Routhian $E'(\om,\th)/\om$ as function of the tilt angle for the  one quasi neutron
configuration [E] in $^{175}_{72}$Hf$_{103}$. The gridlines correspond to
the steps $\De \th =5^o$ and  $\De \om =0.02~MeV$, starting with $\om=0.02~MeV$.
A  constant is added such that $E'(\om,\th=90^o)/\om=0$.   }
\end{figure}

\begin{figure}
\mbox{\psfig{file=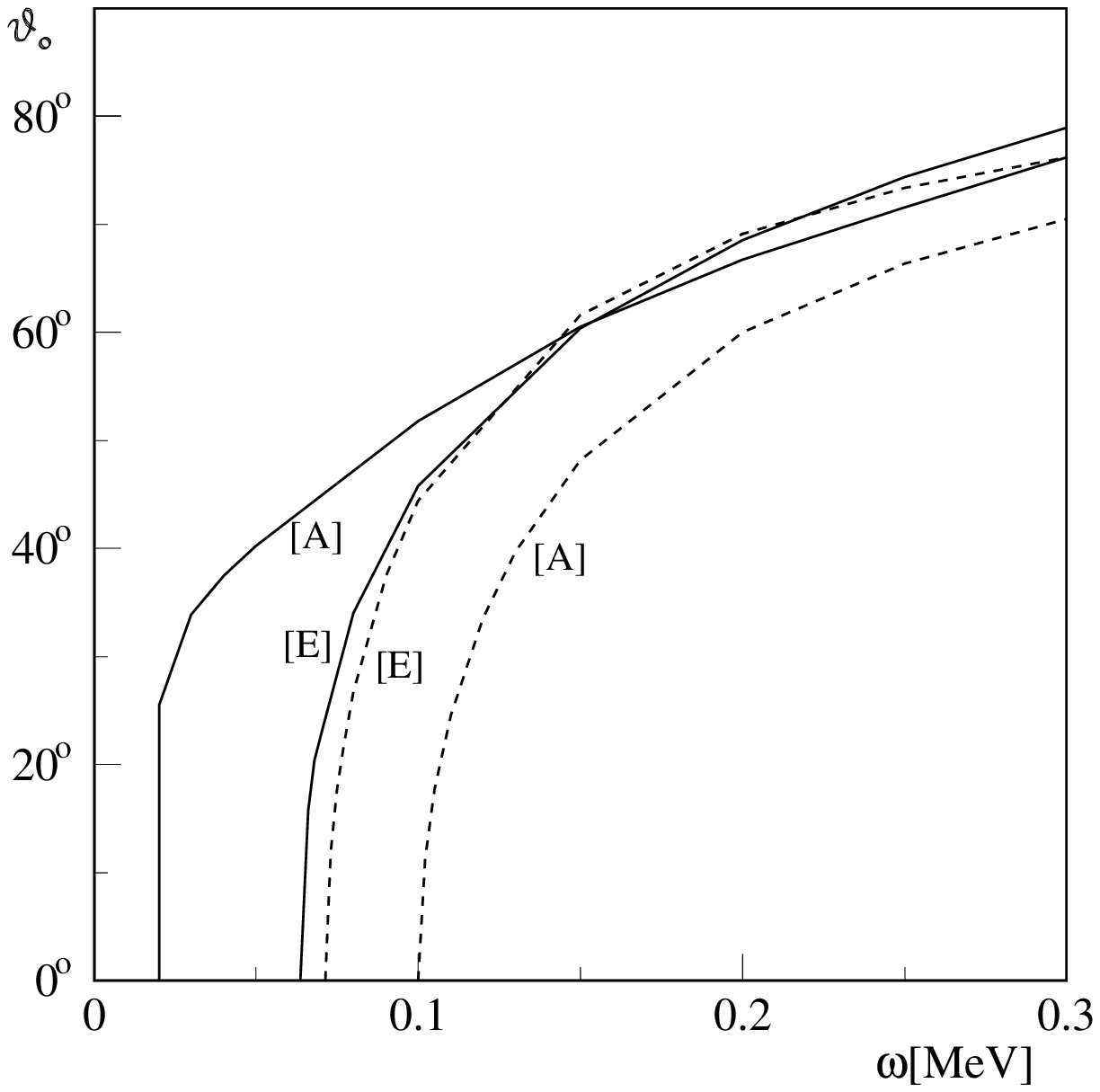,width=14cm}}
\caption{\label{f:thom1qn}
Equilibrium tilt angle $\th_o$  as function of the rotational frequency $\om$ for
the one \qn \confs [A] and [E].
The full lines display the TAC result. The dashed lines show the strong
coupling estimate,  where  the moment of inertia
${\cal J}=35~MeV^{-1}$  is used, which is the value of $J/\om$ for the \conf
[E].}
\end{figure}

\begin{figure}
\vspace*{-3cm}
\mbox{\psfig{file=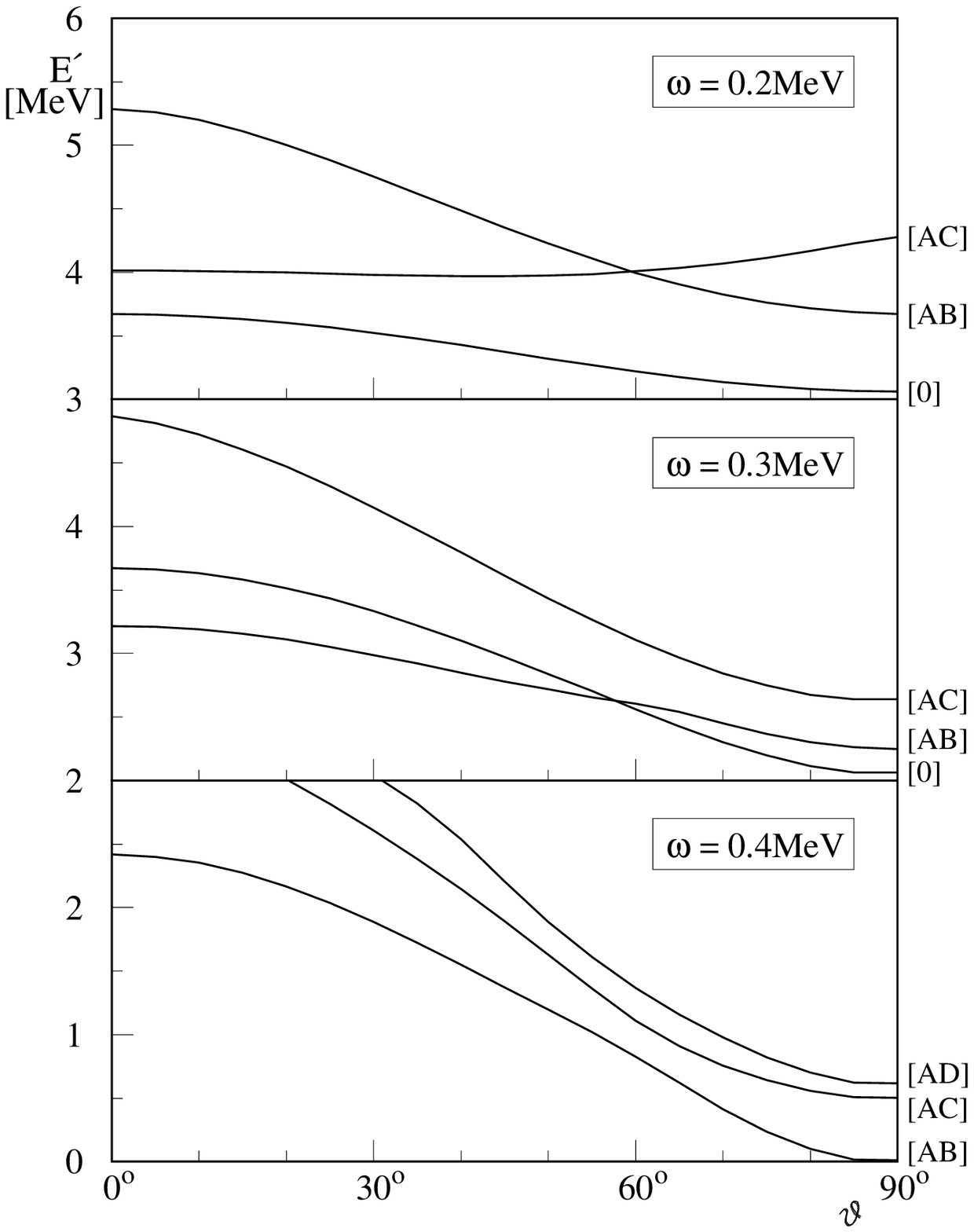,width=14cm}}
\caption{\label{f:Eth2qn+}
Total Routhians as function of the tilt angle for the lowest positive parity
\qn configurations in $^{174}_{72}$Hf$_{102}$. }
\end{figure}

\begin{figure}
\vspace*{-3cm}
\mbox{\psfig{file=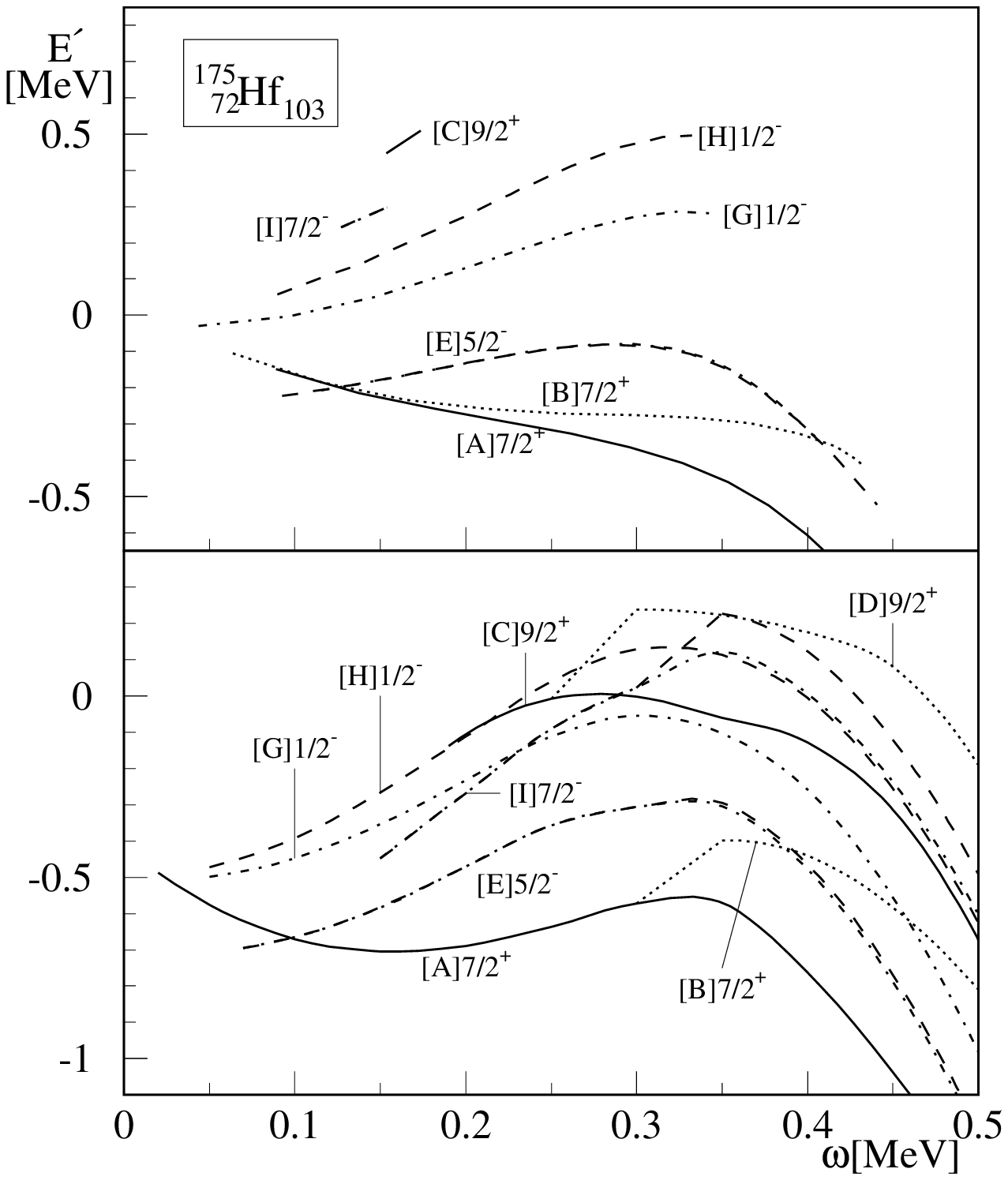,width=14cm}}
\caption{\label{f:1qn} Total Routhians of the lowest one quasi neutron
\confs in  $^{175}_{72}$Hf$_{103}$.\\
The following legend applies to all figures of this type and  contains 
more information than needed for this figure.\\
The line convention for the parity and signature $(\pi,~\al)$ is as
follows.\\
Odd $A$: full (+, 1/2), dotted (+, -1/2), dashed dotted (-, 1/2), and
dashed (-, -1/2).\\ 
Even $A$: 
full (+, 0), dotted (+, 1), dashed dotted (-, 0), and
dashed (-, 1).\\
Upper panel: Experimental Routhians  from the data in 
\protect\cite{tab,hf174}.
If the bands do not show signature splitting 
the Routhians are calculated by means of eqs. 
(\protect\ref{omexp1}-\protect\ref{E'exp1}) if there is a signature splitting 
eqs. (\protect\ref{omexp1}-\protect\ref{E'exp1}) are used.
The configuration is given as [X]$K^\pi$ where X stands
for  the letter code explained in the text
 and  $K^\pi$ gives the parity  $\pi$ and the $K$ value of
the band head.\\ 
Lower panel: Results of the TAC calculations. For TAC solutions $(\th_o \leq
80^o)$ the 
two lines representing the signatures are on top of each other. If they
are separated the solution is of PAC type  $(\th_o > 80^o)$. The last TAC
point and the first PAC point are connected by a straight line. The grid
is $\De \om =0.05~ MeV$.
  The band head is only determined with the
accuracy of  $\De \om$. For several bands the frequency of the first
transition ($J(\om)=K+1$) is indicated by a fat dot.\\ 
A rigid rotor reference with 
${\cal J}=50~ MeV^{-1}$ is subtracted from all Routhians.}
\end{figure}

\begin{figure}
\mbox{\psfig{file=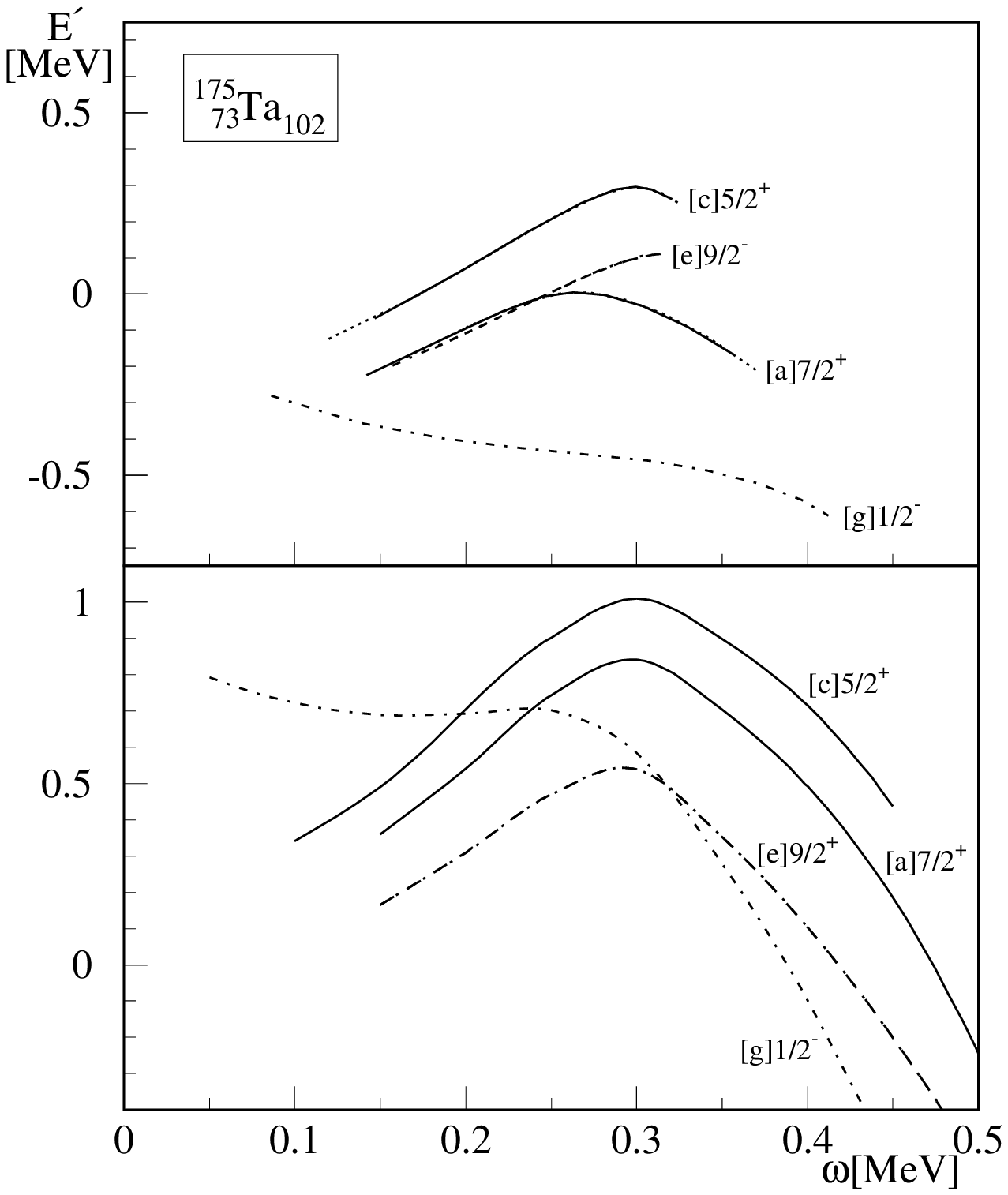,width=14cm}}
\caption{\label{f:1qp} Total Routhians of the lowest one quasi proton
\confs in in $^{175}_{73}$Ta$_{102}$. \\
Upper panel: Experimental Routhians 
from the data in \protect\cite{tab,ta175}.
Lower panel: TAC calculations.
Cf. caption of fig. \protect\ref{f:1qn}.} 
\end{figure}

\begin{figure}
\mbox{\psfig{file=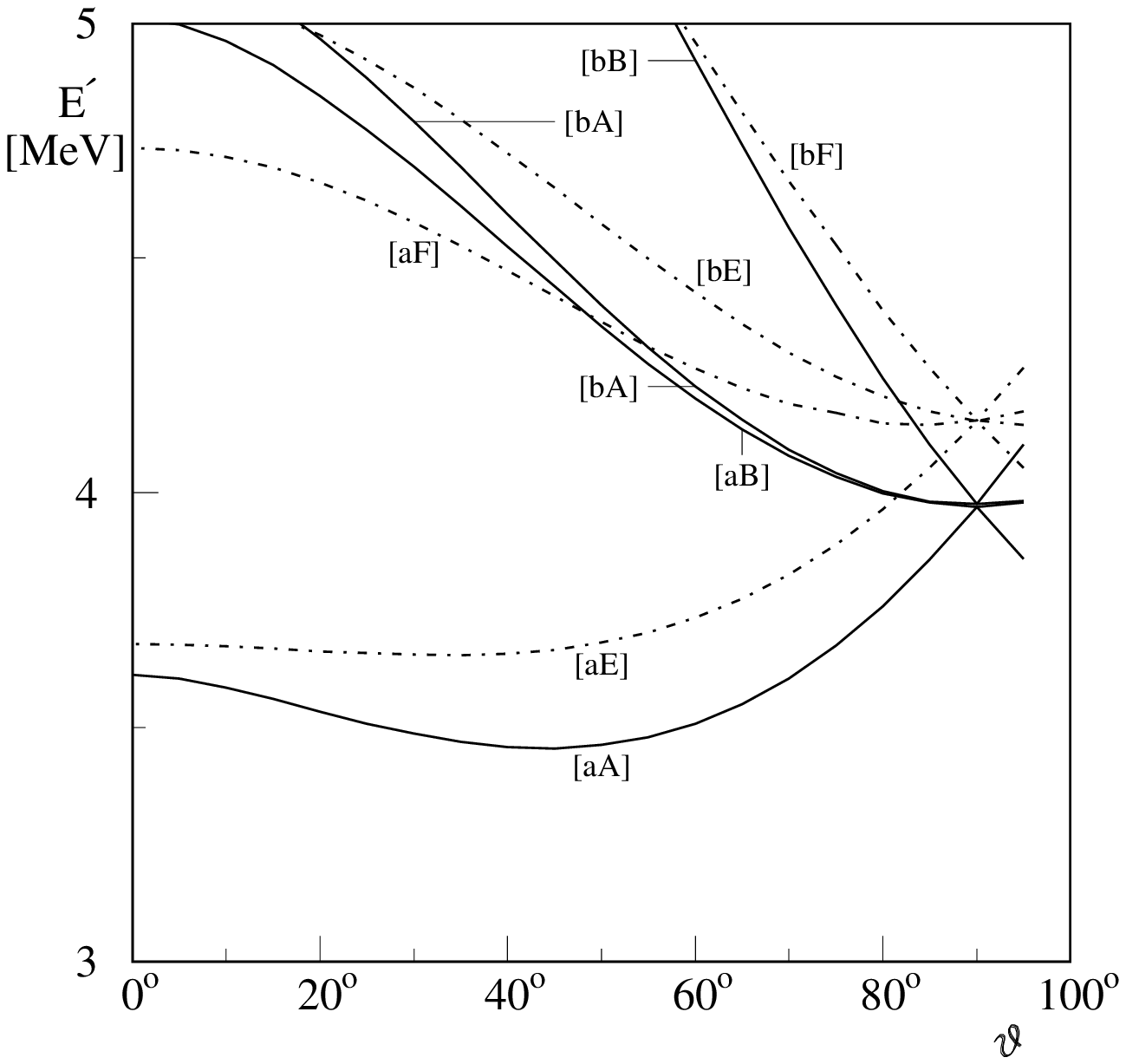,width=14cm}}
\caption{\label{f:Eth1qp1qn}
Total Routhians at $\om=0.2~MeV$ as functions of the tilt angle for 
 the lowest configurations in $^{174}_{71}$Lu$_{103}$. }
\end{figure}

\begin{figure}
\mbox{\psfig{file=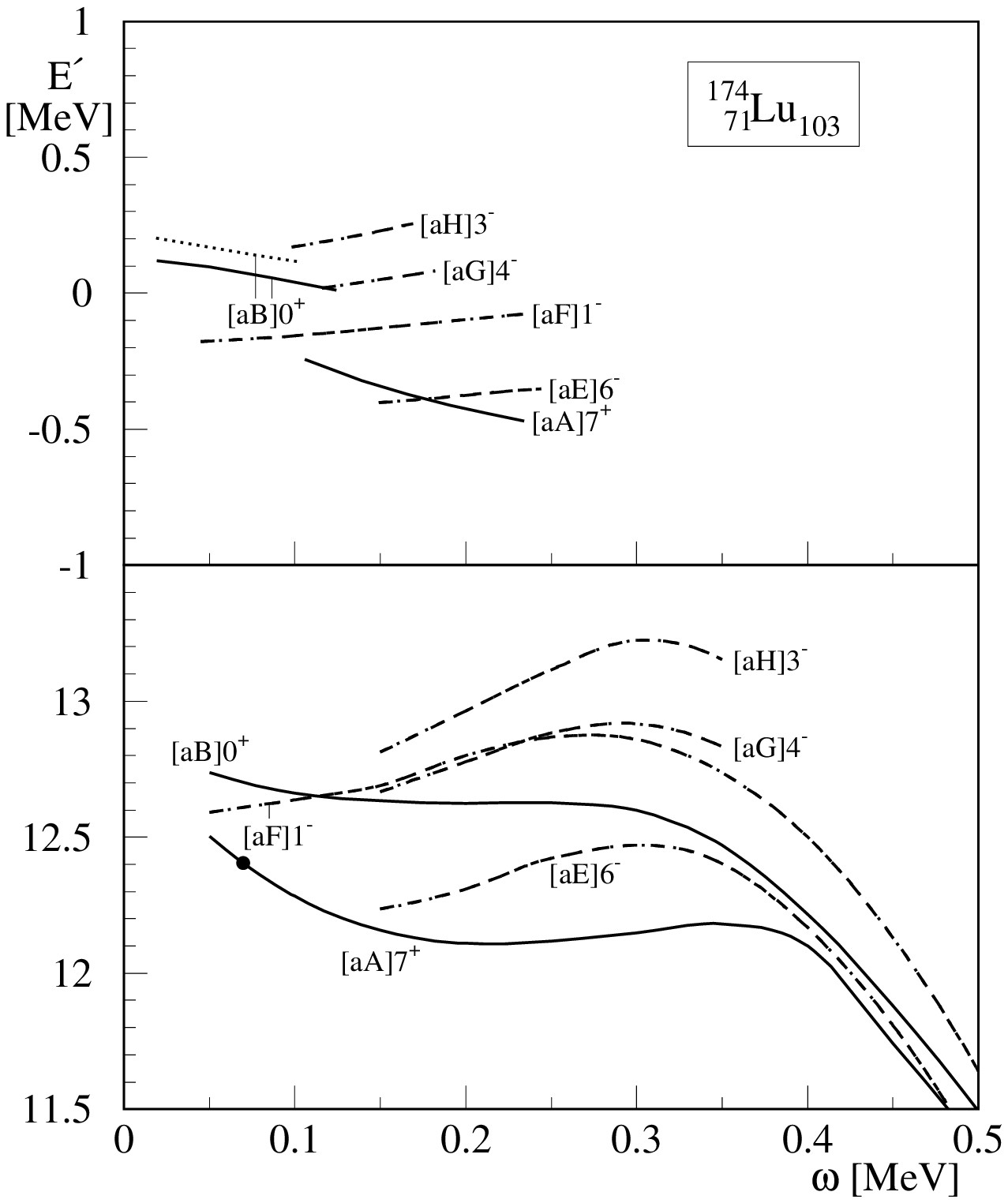,width=14cm}}
\caption{\label{f:1qp1qn} Total Routhians of the lowest one quasi proton one
quasi neutron 
\confs in $^{174}_{71}$Lu$_{103}$. \\
Upper panel: Experimental Routhians 
from the data in \protect\cite{lu174}.
Lower panel: TAC calculations.
Cf. caption of fig. \protect\ref{f:1qn}.} 
\end{figure}

\begin{figure}
\mbox{\psfig{file=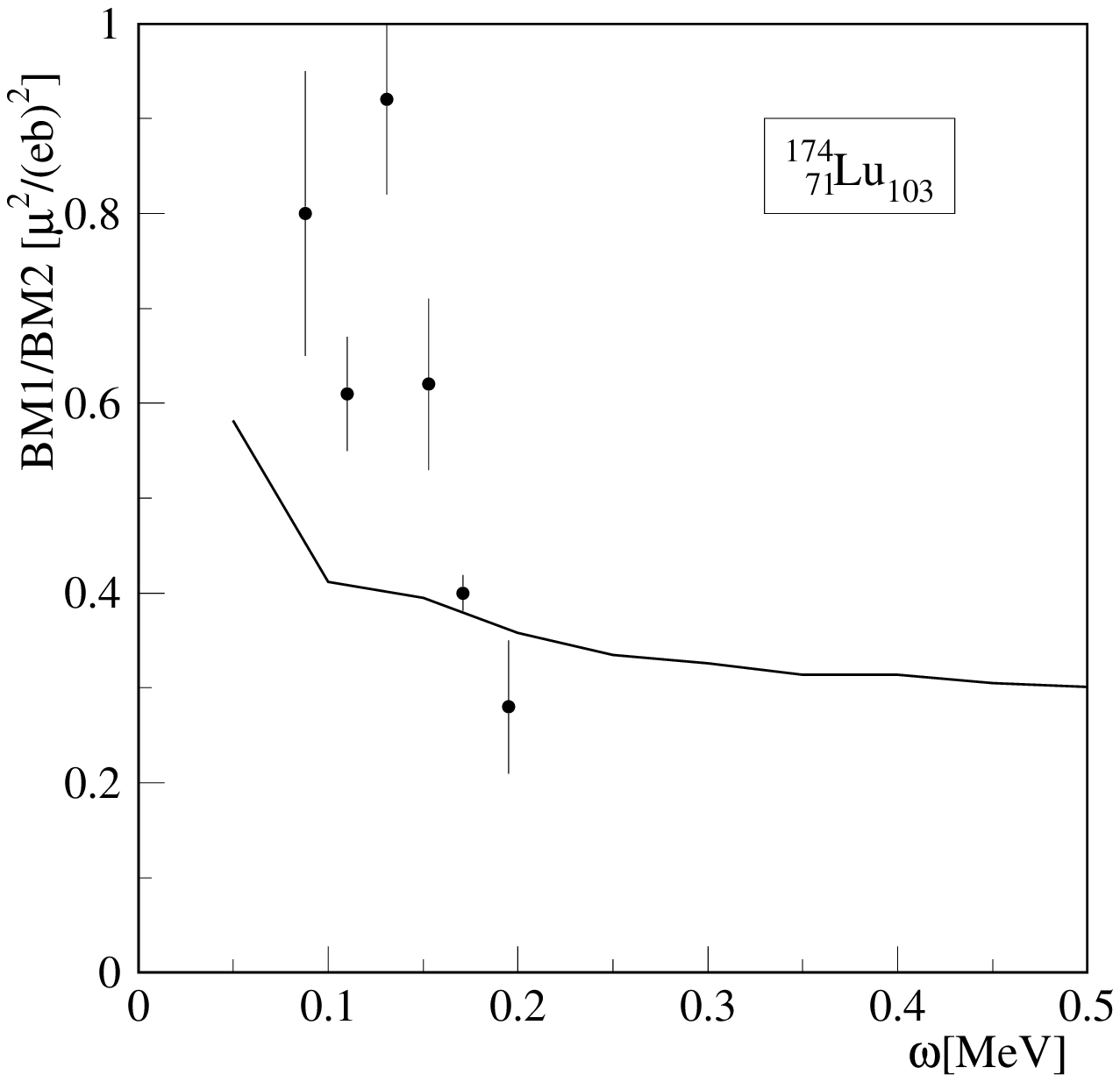,width=14cm}}
\caption{\label{f:bm1lu174} Branching ratios of the one quasi proton one
quasi neutron 
\conf  [aF]$_{1^-}$ in  $^{174}_{71}$Lu$_{103}$. \\
Experimental values from \protect\cite{lu174}.
Full line: TAC calculations.}
\end{figure}

\begin{figure}
\mbox{\psfig{file=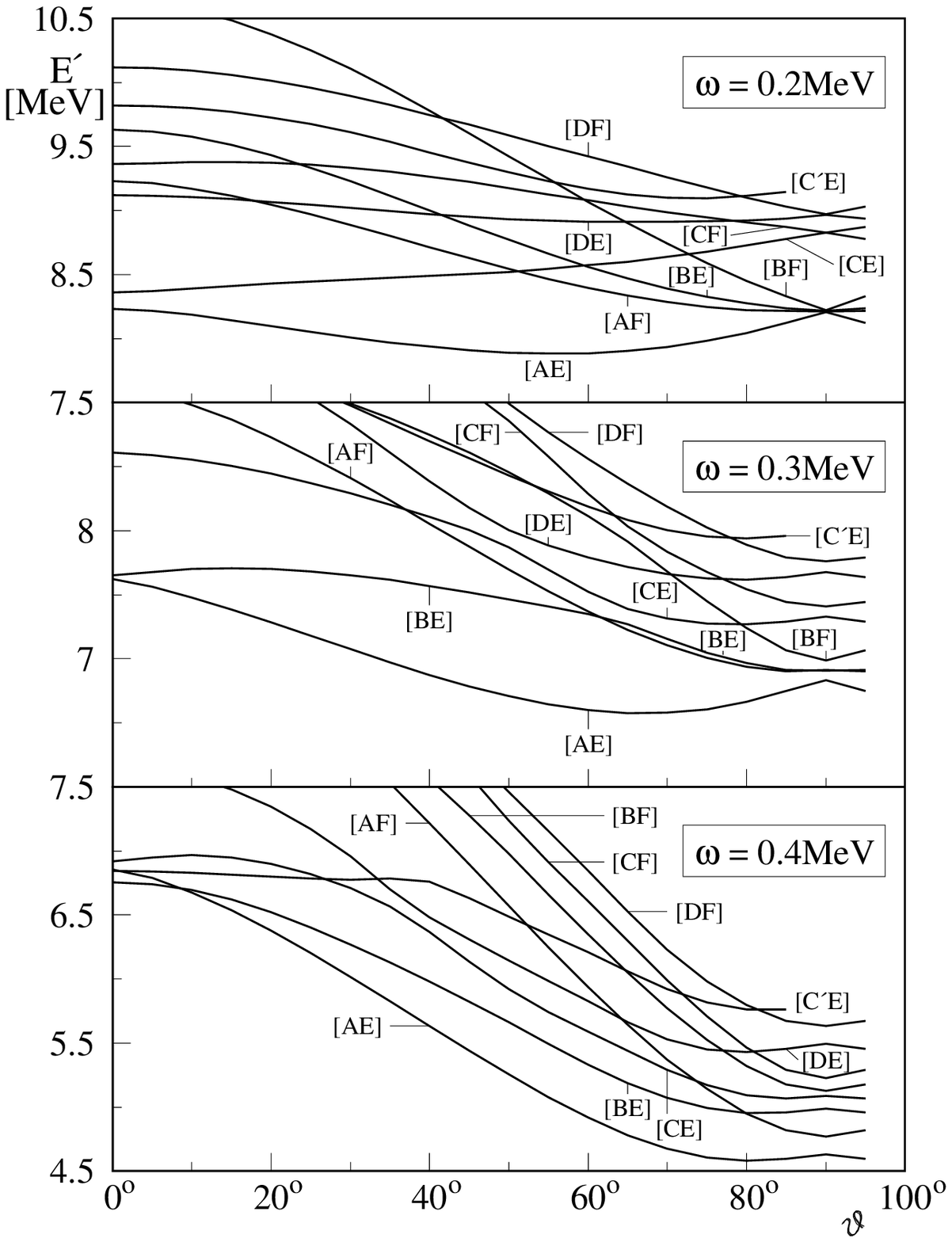,width=14cm}}
\caption{\label{f:Eth2qn-}
Total Routhians as functions of the tilt angle for negative parity
lowest two  \qn configurations in $^{174}_{72}$Hf$_{102}$. }
\end{figure}

\begin{figure}
\mbox{\psfig{file=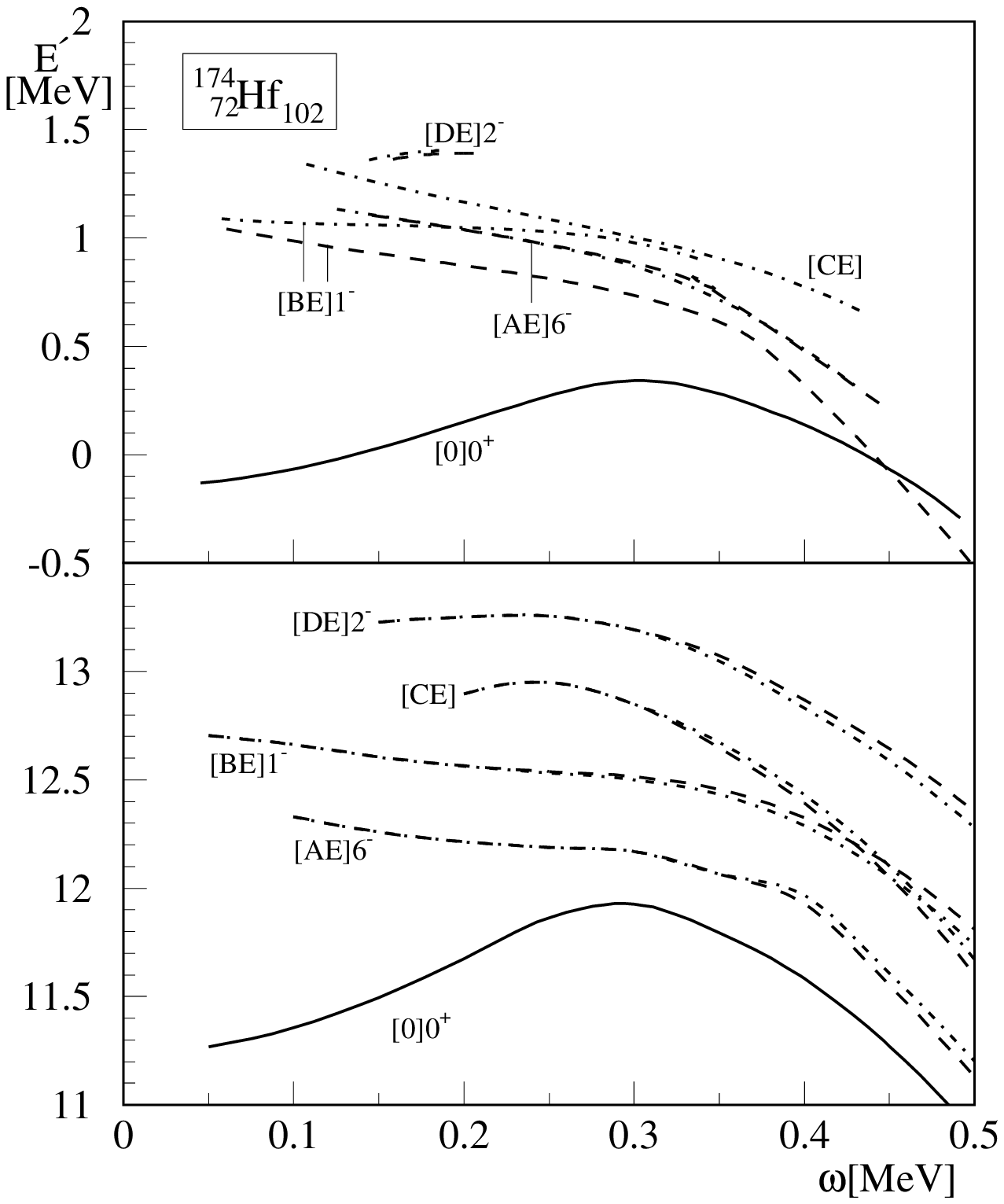,width=14cm}}
\caption{\label{f:2qna} Total Routhians of the lowest negative parity two
quasi neutron 
\confs in  $^{174}_{72}$Hf$_{104}$. 
Upper panel: Experimental Routhians 
from the data in \protect\cite{hf174}.
Lower panel: TAC calculations.
Cf. caption of fig. \protect\ref{f:1qn}.} 
\end{figure}

\begin{figure}
\mbox{\psfig{file=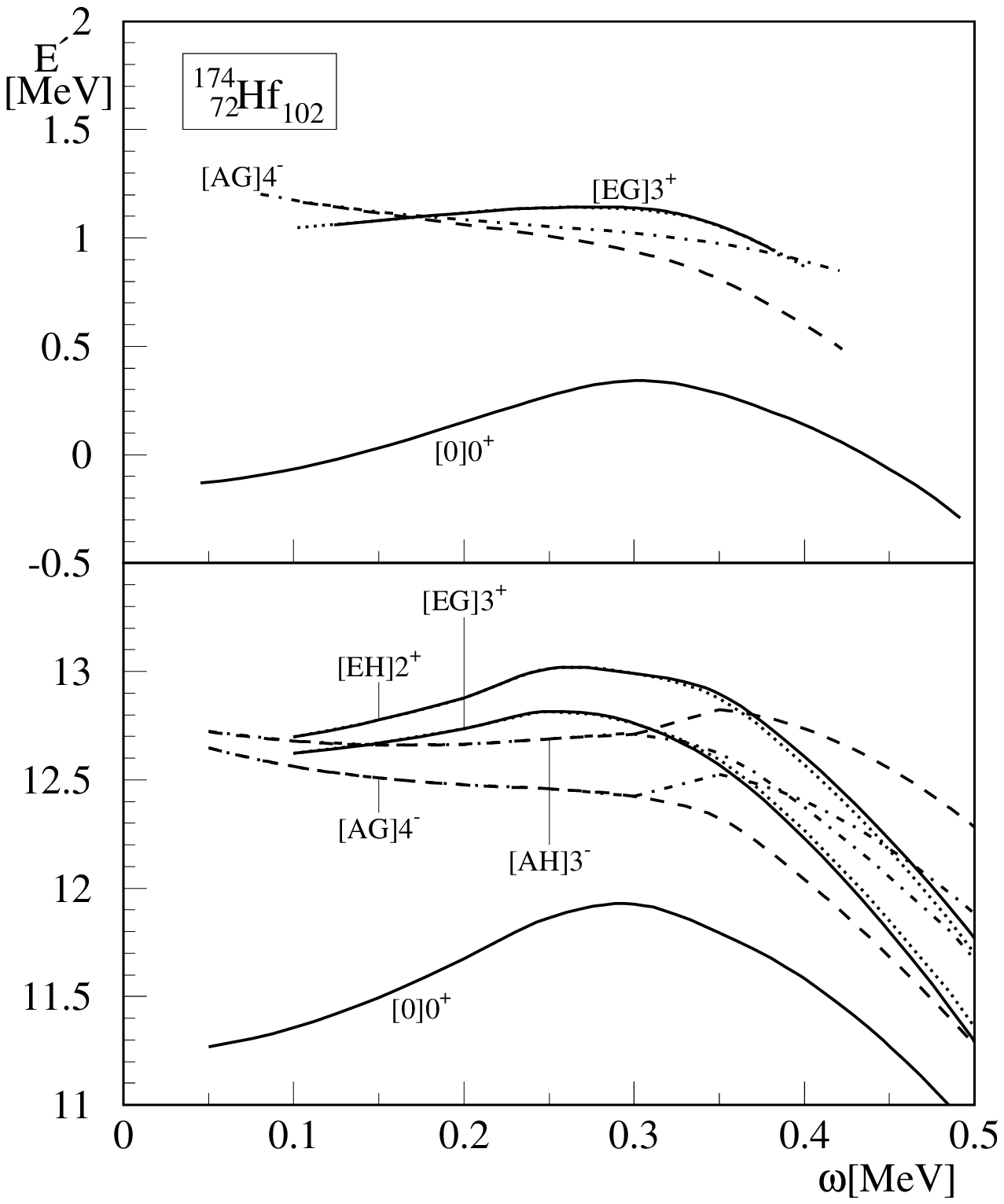,width=14cm}}
\caption{
\label{f:2qng} Total Routhians of the two
quasi neutron 
\confs containing the pseudo spin singlet \qns G and H
 in  $^{174}_{72}$Hf$_{104}$. 
Upper panel: Experimental Routhians 
from the data in \protect\cite{hf174}.
Lower panel: TAC calculations.
Cf. caption of fig. \protect\ref{f:1qn}.} 
\end{figure}

\begin{figure}
\mbox{\psfig{file=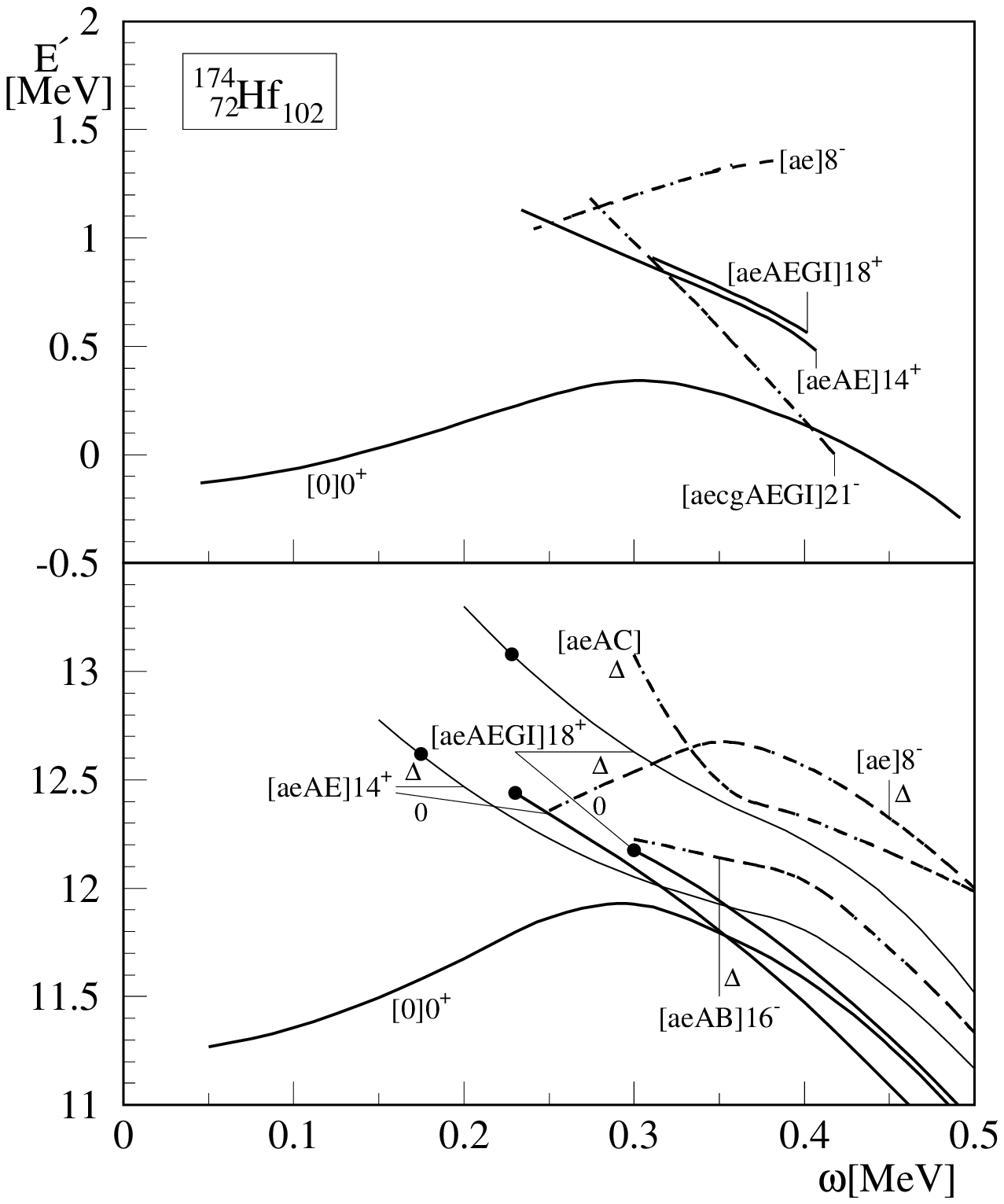,width=14cm}}
\caption{\label{f:p8m2qn} Total Routhians of  the lowest \qn
\confs in $^{174}_{72}$Hf$_{102}$ built on the $K^\pi=8^-$ two \qpr
excitation. \\
Upper panel: Experimental Routhians 
from the data in \protect\cite{hf174}.
Lower panel: TAC calculations. The labels $\De$ and 0 denote
$\De_n=0.69~MeV$ and 0, respectively.
Cf. caption of fig. \protect\ref{f:1qn}.} 
\end{figure}
\begin{figure}
\mbox{\psfig{file=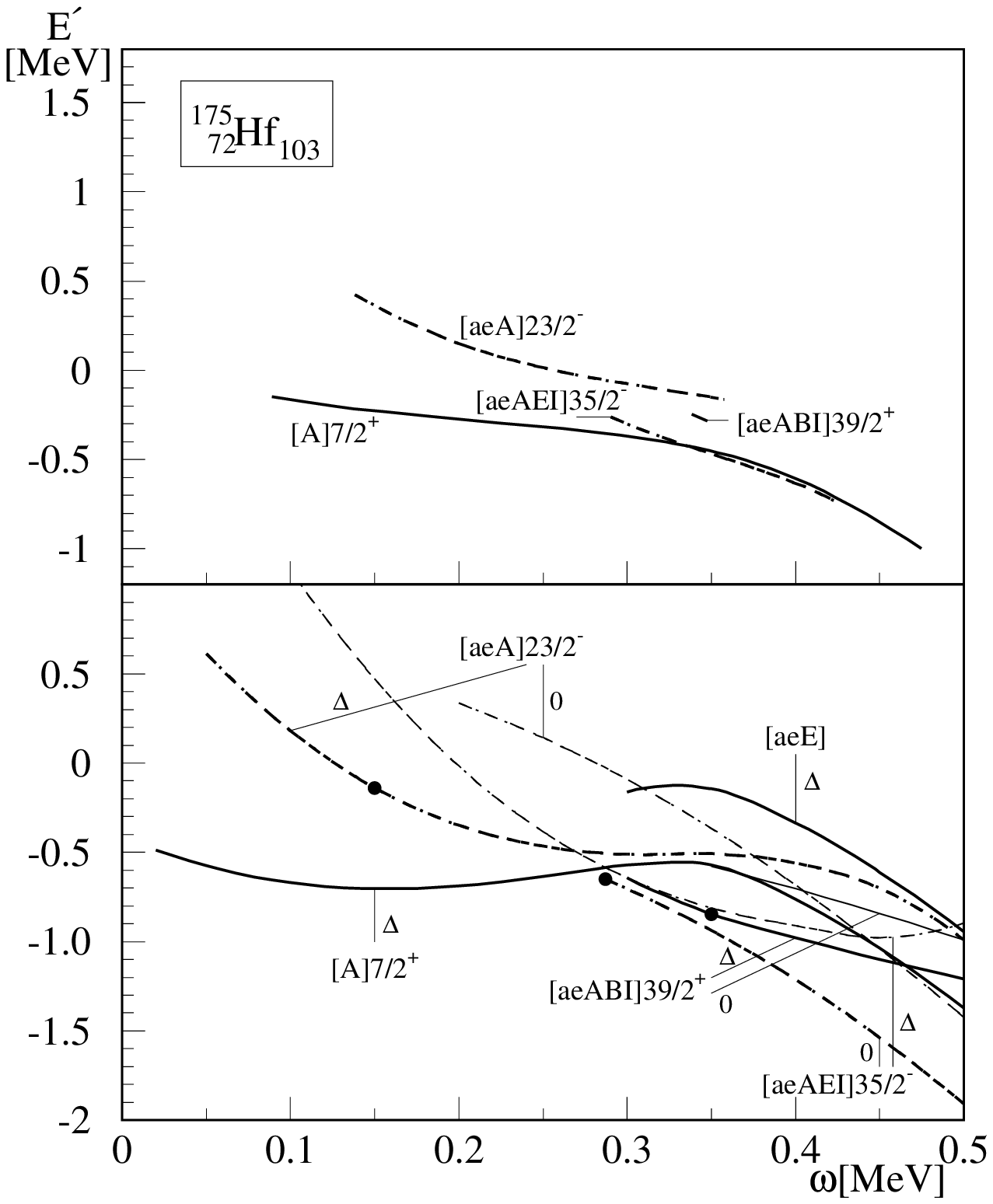,width=14cm}}
\caption{\label{f:p8m1qn} Total Routhians of  the lowest \qn
\confs in $^{175}_{72}$Hf$_{103}$ built on the $K^\pi=8^-$ two \qpr
excitation. \\
Upper panel: Experimental Routhians 
from the data in \protect\cite{hf174}.
Lower panel: TAC calculations. The labels $\De$ and 0 denote
 $\De_n=0.69~MeV$ and 0, respectively.
Cf. caption of fig. \protect\ref{f:1qn}.} 
\end{figure}

\begin{figure}
\mbox{\psfig{file=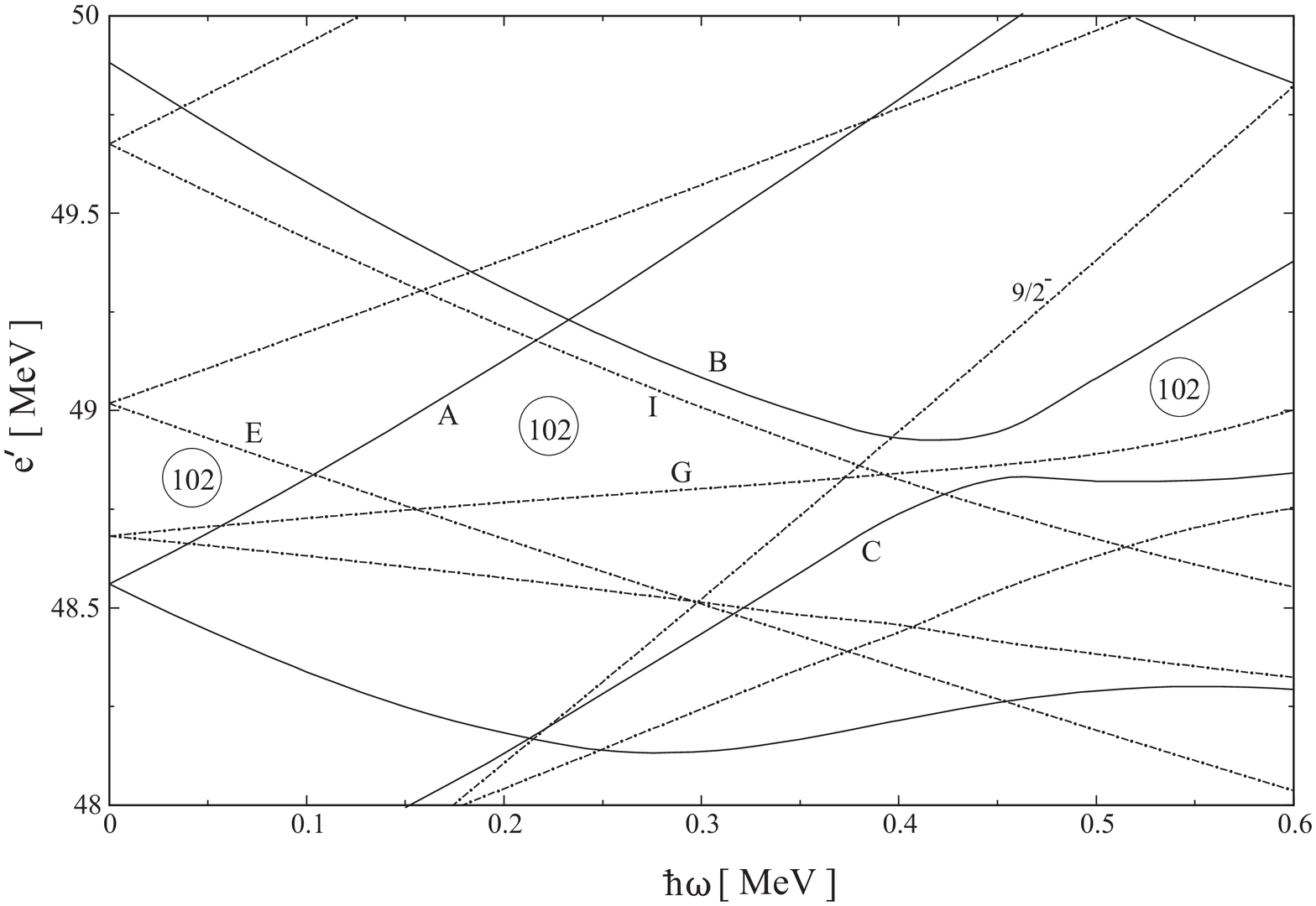,width=14cm}}
\caption{\label{f:sn45}
Single neutron energies for $\th=45^o,~\eps=0.258,~\eps_4=0.034$.}
\end{figure}

\begin{figure}
\mbox{\psfig{file=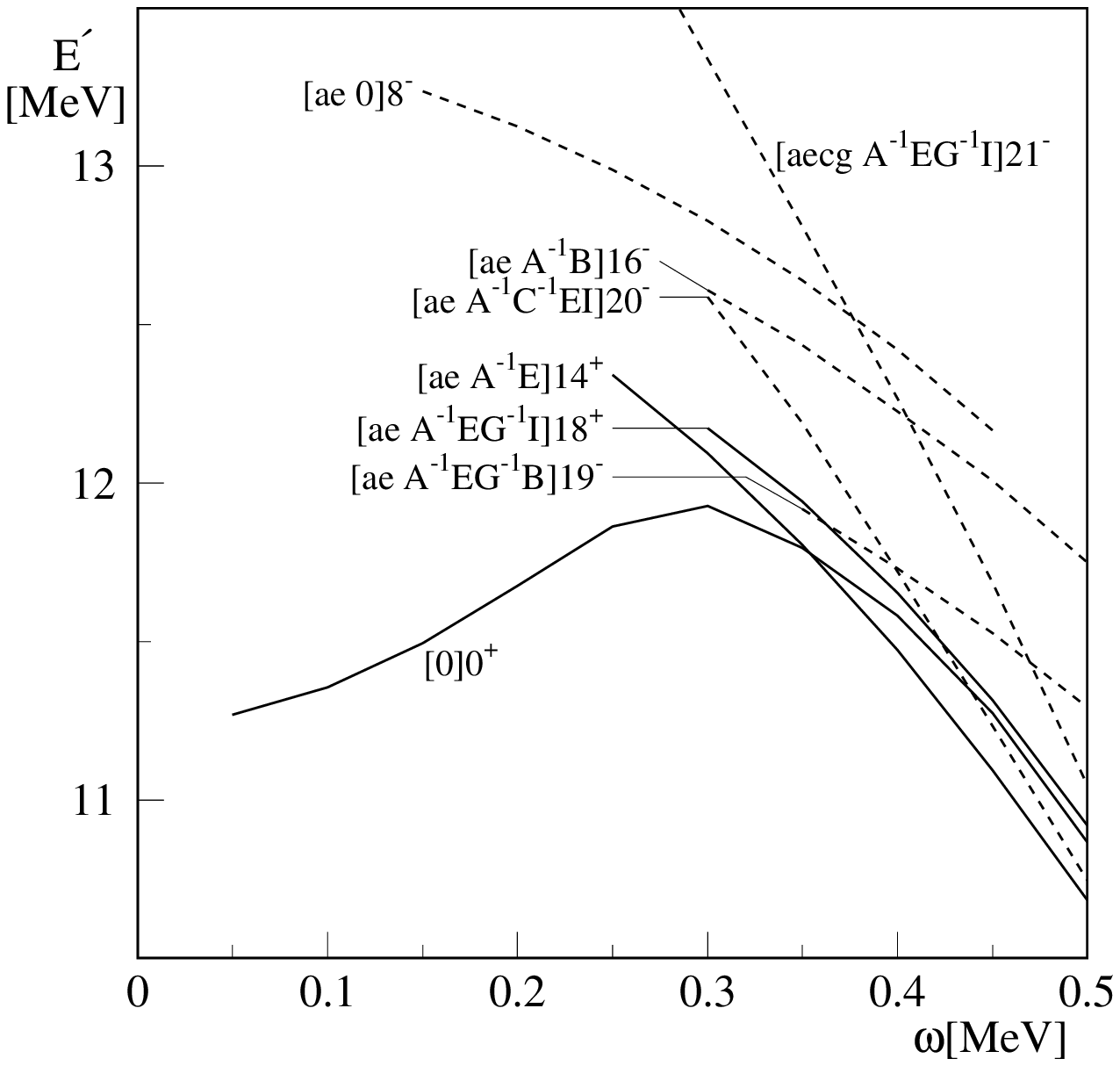,width=14cm}}
\caption{\label{f:p8m2sn} Total Routhians of the lowest  neutron
\confs in $^{174}_{72}$Hf$_{102}$ built on the $K^\pi=8^-$ two \qpr
excitation. The TAC calculations assume zero neutron pairing.
Cf. caption of fig. \protect\ref{f:1qn}.} 
\end{figure}

\begin{figure}
\mbox{\psfig{file=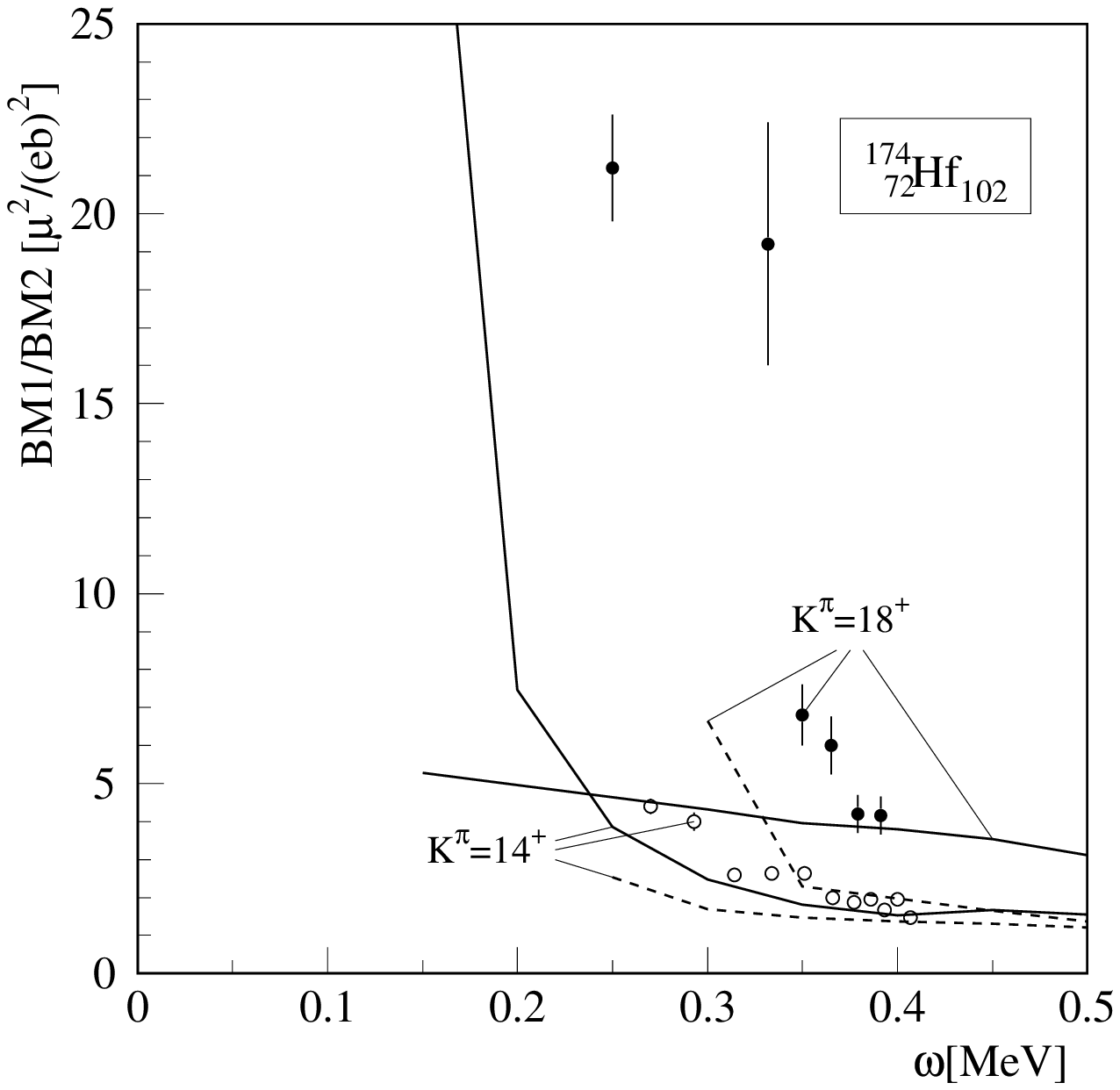,width=14cm}}
\caption{\label{f:bm1hf174} Branching ratios of the 
\conf  [aeAE]$_{14^+}$ and  [aeAEGI]$_{18^+}$ in  $^{174}_{72}$Hf$_{102}$. \\
Experimental values from \protect\cite{hf174}.
Full lines: TAC calculations with neutron pairing dashed lines with
zero neutron pairing.}
\end{figure}
\newpage

\begin{table}
\begin{tabular}{|cccccccccccccc|}
$\gamma$&$-270^o$&&$-180^o$&&$-120^o$&&$-60^o$&&$0^o$&&$60^o$&&$120^o$\\
\hline
shape&p&t&o&t&p&t&o&t&p&t&o&t&p\\
\hline
1-axis&s&i&l&l&\,l$^*$&l&l&i&s&s&\,s$^*$&s&s\\
\hline
3-axis&s&s&\,s$^*$&s&s&i&l&l&\,l$^*$&l&l&i&s
\end{tabular}
\caption{\label{table} Association of 
the principal axes of the triaxial potential with the axes 1 and 3.
The shapes are labeled by p (prolate), o (oblate), and t (triaxial).
The axes are labeled by l (long), i (intermediate), and s (short).
The star indicates a symmetry axis.  }   
\end{table}

\end{document}